\documentclass{aa}

\usepackage{graphicx}
\usepackage{amsmath,amsfonts,amssymb}
\usepackage{txfonts}
\usepackage{color}
\usepackage{natbib}
\usepackage{float}
\usepackage{dblfloatfix}
\usepackage{afterpage}
\usepackage{ifthen}
\usepackage[morefloats=12]{morefloats}
\usepackage{placeins}
\usepackage{multicol}
\bibpunct{(}{)}{;}{a}{}{,}
\usepackage[switch]{lineno}
\definecolor{linkcolor}{rgb}{0.6,0,0}
\definecolor{citecolor}{rgb}{0,0,0.75}
\definecolor{urlcolor}{rgb}{0.12,0.46,0.7}
\usepackage[breaklinks, colorlinks, urlcolor=urlcolor,
    linkcolor=linkcolor,citecolor=citecolor,pdfencoding=auto]{hyperref}
\hypersetup{linktocpage}
\usepackage{bold-extra}

\def\setsymbol#1#2{\expandafter\def\csname #1\endcsname{#2}}
\def\getsymbol#1{\csname #1\endcsname}

\def\Planck{\textit{Planck}}

\newbox\tablebox    \newdimen\tablewidth
\def\leaderfil{\leaders\hbox to 5pt{\hss.\hss}\hfil}
\def\endPlancktable{\tablewidth=\columnwidth 
    $$\hss\copy\tablebox\hss$$
    \vskip-\lastskip\vskip -2pt}
\def\endPlancktablewide{\tablewidth=\textwidth 
    $$\hss\copy\tablebox\hss$$
    \vskip-\lastskip\vskip -2pt}
\def\tablenote#1 #2\par{\begingroup \parindent=0.8em
    \abovedisplayshortskip=0pt\belowdisplayshortskip=0pt
    \noindent
    $$\hss\vbox{\hsize\tablewidth \hangindent=\parindent \hangafter=1 \noindent
    \hbox to \parindent{$^#1$\hss}\strut#2\strut\par}\hss$$
    \endgroup}

\def\nside{\ifmmode N_{\mathrm{side}}\else $N_{\mathrm{side}}$\fi}
\def\L2{\ifmmode L_2\else $L_2$\fi}

\def\DeltaT{\ifmmode \Delta T\else $\Delta T$\fi}
\def\deltat{\ifmmode \Delta t\else $\Delta t$\fi}
\def\fknee{\ifmmode f_{\rm knee}\else $f_{\rm knee}$\fi}
\def\Fmax{\ifmmode F_{\rm max}\else $F_{\rm max}$\fi}
\def\solar{\ifmmode{\rm M}_{\mathord\odot}\else${\rm M}_{\mathord\odot}$\fi}
\def\Msolar{\ifmmode{\rm M}_{\mathord\odot}\else${\rm M}_{\mathord\odot}$\fi}
\def\Lsolar{\ifmmode{\rm L}_{\mathord\odot}\else${\rm L}_{\mathord\odot}$\fi}
\def\inv{\ifmmode^{-1}\else$^{-1}$\fi}
\def\mo{\ifmmode^{-1}\else$^{-1}$\fi}
\def\sup#1{\ifmmode ^{\rm #1}\else $^{\rm #1}$\fi}
\def\expo#1{\ifmmode \times 10^{#1}\else $\times 10^{#1}$\fi}
\def\,{\thinspace}
\def\lsim{\mathrel{\raise .4ex\hbox{\rlap{$<$}\lower 1.2ex\hbox{$\sim$}}}}
\def\gsim{\mathrel{\raise .4ex\hbox{\rlap{$>$}\lower 1.2ex\hbox{$\sim$}}}}

\def\simprop{\mathrel{\raise .4ex\hbox{\rlap{$\propto$}\lower 1.2ex\hbox{$\sim$}}}}
\def\deg{\ifmmode^\circ\else$^\circ$\fi}
\def\pdeg{\ifmmode $\setbox0=\hbox{$^{\circ}$}\rlap{\hskip.11\wd0 .}$^{\circ}
          \else \setbox0=\hbox{$^{\circ}$}\rlap{\hskip.11\wd0 .}$^{\circ}$\fi}
\def\arcs{\ifmmode {^{\scriptstyle\prime\prime}}
          \else $^{\scriptstyle\prime\prime}$\fi}
\def\arcm{\ifmmode {^{\scriptstyle\prime}}
          \else $^{\scriptstyle\prime}$\fi}
\newdimen\sa  \newdimen\sb
\def\parcs{\sa=.07em \sb=.03em
     \ifmmode \hbox{\rlap{.}}^{\scriptstyle\prime\kern -\sb\prime}\hbox{\kern -\sa}
     \else \rlap{.}$^{\scriptstyle\prime\kern -\sb\prime}$\kern -\sa\fi}
\def\parcm{\sa=.08em \sb=.03em
     \ifmmode \hbox{\rlap{.}\kern\sa}^{\scriptstyle\prime}\hbox{\kern-\sb}
     \else \rlap{.}\kern\sa$^{\scriptstyle\prime}$\kern-\sb\fi}
\def\ra[#1 #2 #3.#4]{#1\sup{h}#2\sup{m}#3\sup{s}\llap.#4}
\def\dec[#1 #2 #3.#4]{#1\deg#2\arcm#3\arcs\llap.#4}
\def\deco[#1 #2 #3]{#1\deg#2\arcm#3\arcs}
\def\rra[#1 #2]{#1\sup{h}#2\sup{m}}

\def\dots{\relax\ifmmode \ldots\else $\ldots$\fi}
\def\WHzsr{\ifmmode \,\mathrm{W\,Hz\mo\,sr\mo}\else \,W\,Hz\mo\,sr\mo\fi}
\def\MHz{\ifmmode \,\mathrm{MHz}\else \,MHz\fi}
\def\GHz{\ifmmode \,\mathrm{GHz}\else \,GHz\fi}
\def\mKs{\ifmmode \,\mathrm{mK\,s}^{1/2}\else \,mK\,s$^{1/2}$\fi}
\def\muKs{\ifmmode \,\mu\mathrm{K\,s}^{1/2}\else \,$\mu$K\,s$^{1/2}$\fi}
\def\muKRJs{\ifmmode \,\mu\mathrm{K_{RJ}}\,\mathrm{s}^{1/2}\else \,$\mu$K$_{\mathrm{RJ}}$\,s$^{1/2}$\fi}
\def\muKRJ{\ifmmode \,\mu\mathrm{K_{RJ}}\else \,$\mu$K$_{\mathrm{RJ}}$\fi}
\def\muKCMB{\ifmmode \,\mu\mathrm{K_{CMB}}\else \,$\mu$K$_{\mathrm{CMB}}$\fi}
\def\KRJ{\ifmmode \,\mathrm{K_{RJ}}\else \,K$_{\mathrm{RJ}}$\fi}
\def\KCMB{\ifmmode \,\mathrm{K_{CMB}}\else \,K$_{\mathrm{CMB}}$\fi}
\def\muKHz{\ifmmode \,\mu\mathrm{K\,Hz}^{-1/2}\else \,$\mu$K\,Hz$^{-1/2}$\fi}
\def\MJysr{\ifmmode \,\mathrm{MJy\,sr\mo}\else \,MJy\,sr\mo\fi}
\def\MJysrmK{\ifmmode \,\mathrm{MJy\,sr\mo}\,\mathrm{mK}_{\mathrm{ CMB}}\mo\else \,MJy\,sr\mo\,mK$_{\mathrm{CMB}}\mo$\fi}
\def\microns{\ifmmode \,\mu\mathrm{m}\else \,$\mu$m\fi}

\def\muK{\ifmmode \,\mu\mathrm{K}\else \,$\mu$\hbox{K}\fi}
\def\microK{\ifmmode \,\mu\mathrm{K}\else \,$\mu$\hbox{K}\fi}
\def\muW{\ifmmode \,\mu\mathrm{W}\else \,$\mu$\hbox{W}\fi}
\def\kms{\ifmmode \,\mathrm{km\,s}^{-1}\else \,km\,s$^{-1}$\fi}
\def\kmsMpc{\ifmmode \,\kms\,\mathrm{Mpc\mo}\else \,\kms\,Mpc\mo\fi}
\def\cmisq{\ifmmode \,\mathrm{cm}^{-2}\else $\,\mathrm{cm}^{-2}$\fi}
\def\Acmb{\ifmmode a_\mathrm{CMB}\else $a_{\mathrm{CMB}}$\fi}
\def\Aquad{\ifmmode a_\mathrm{quad}\else $a_{\mathrm{quad}}$\fi}
\def\Asynch{\ifmmode a_\mathrm{s}\else $a_{\mathrm{s}}$\fi}
\def\Asrc{\ifmmode a_\mathrm{src}\else $a_{\mathrm{src}}$\fi}
\def\Adust{\ifmmode a_\mathrm{d}\else $a_{\mathrm{d}}$\fi}
\def\Asdust{\ifmmode a_\mathrm{sd}\else $a_{\mathrm{sd}}$\fi}
\def\Aame{\ifmmode a_\mathrm{ame}\else $a_{\mathrm{ame}}$\fi}
\def\Aco{\ifmmode a_\mathrm{CO}\else $a_{\mathrm{CO}}$\fi}
\def\AcoOne{\ifmmode a_\mathrm{CO10}\else $a_{\mathrm{CO10}}$\fi}
\def\AcoTwo{\ifmmode a_\mathrm{CO21}\else $a_{\mathrm{CO21}}$\fi}
\def\AcoThree{\ifmmode a_\mathrm{CO32}\else $a_{\mathrm{CO32}}$\fi}
\def\Aff{\ifmmode a_\mathrm{ff}\else $a_{\mathrm{ff}}$\fi}
\def\Tcmb{\ifmmode T_\mathrm{CMB}\else $T_{\mathrm{CMB}}$\fi}
\def\Tdust{\ifmmode T_\mathrm{d}\else $T_{\mathrm{d}}$\fi}
\def\scmb{\ifmmode s_\mathrm{CMB}\else $s_{\mathrm{CMB}}$\fi}
\def\squad{\ifmmode s_\mathrm{quad}\else $s_{\mathrm{quad}}$\fi}
\def\ssynch{\ifmmode s_\mathrm{s}\else $s_\mathrm{s}$\fi}
\def\sdust{\ifmmode s_\mathrm{d}\else $s_{\mathrm{d}}$\fi}
\def\ssdust{\ifmmode s_\mathrm{sd}\else $s_{\mathrm{sd}}$\fi}
\def\same{\ifmmode s_\mathrm{ame}\else $s_{\mathrm{ame}}$\fi}
\def\ssrc{\ifmmode s_\mathrm{src}\else $s_{\mathrm{src}}$\fi}
\def\sco{\ifmmode s_\mathrm{CO}\else $s_{\mathrm{CO}}$\fi}
\def\sff{\ifmmode s_\mathrm{ff}\else $s_{\mathrm{ff}}$\fi}
\def\gff{\ifmmode g_\mathrm{ff}\else $g_{\mathrm{ff}}$\fi}
\def\fsynch{\ifmmode f_\mathrm{s}\else $f_{\mathrm{s}}$\fi}
\def\fsd{\ifmmode f_\mathrm{sd}\else $f_{\mathrm{sd}}$\fi}
\def\fame{\ifmmode f_\mathrm{ame}\else $f_{\mathrm{ame}}$\fi}
\def\alphasrc{\ifmmode \alpha_\mathrm{src}\else $\alpha_{\mathrm{src}}$\fi}
\def\bdust{\ifmmode \beta_\mathrm{d}\else $\beta_{\mathrm{d}}$\fi}
\def\bsynch{\ifmmode \beta_\mathrm{s}\else $\beta_{\mathrm{s}}$\fi}
\def\bsun{\ifmmode \beta_\mathrm{sun}\else $\beta_{\mathrm{sun}}$\fi}
\def\nuzeros{\ifmmode \nu_{0,\mathrm{s}}\else $\nu_{0,\mathrm{s}}$\fi}
\def\nuzeroff{\ifmmode \nu_{0,\mathrm{ff}}\else $\nu_{0,\mathrm{ff}}$\fi}
\def\nuzerod{\ifmmode \nu_{0,\mathrm{d}}\else $\nu_{0,\mathrm{d}}$\fi}
\def\nuzeroame{\ifmmode \nu_{0,\mathrm{ame}}\else $\nu_{0,\mathrm{ame}}$\fi}
\def\nuzerosd{\ifmmode \nu_{0,\mathrm{}}\else $\nu_{0,\mathrm{sd}}$\fi}
\def\nuzerosrc{\ifmmode \nu_{0,\mathrm{src}}\else $\nu_{0,\mathrm{src}}$\fi}
\def\nup{\ifmmode \nu_{\mathrm{p}}\else $\nu_{\mathrm{p}}$\fi}
\def\alphasd{\ifmmode \alpha_{\mathrm{sd}}\else $\alpha_{\mathrm{sd}}$\fi}
\def\Te{\ifmmode T_{\mathrm{e}}\else $T_{\mathrm{e}}$\fi}
\def\lmax{\ifmmode \ell_{\mathrm{max}}\else $\ell_{\mathrm{max}}$\fi}
\def\NHI{\ifmmode N_{\mathrm{H\,\textsc i}}\else $N_{\mathrm{H\,\textsc i}}$\fi}
\def\chisq{\ifmmode \chi^2\else $\chi^2$\fi}

\def\kB{\ifmmode k_\mathrm{B}\else $k_{\mathrm{B}}$\fi}
\providecommand{\sorthelp}[1]{}

\def\WMAP{\emph{WMAP}}
\def\COBE{\emph{COBE}}

\def\healpix{\texttt{HEALPix}}
\def\commander{\texttt{Commander}}

\def\commanderthree{\texttt{Commander3}}

\renewcommand{\d}[0]{\vec{d}}

\newcommand{\A}[0]{\tens{A}}

\def\YtW{\ifmmode{\tens{Y}^t \tens{W}} \else {$\tens{Y}^t \tens{W}$}\fi}
\newcommand{\n}[0]{\vec{n}}

\newcommand{\s}[0]{\vec{s}}
\renewcommand{\a}[0]{\vec{a}}
\newcommand{\m}[0]{\vec{m}}

\newcommand{\B}[0]{\tens{B}}

\renewcommand{\L}[0]{\tens{L}}
\newcommand{\g}[0]{\vec{g}}
\newcommand{\N}[0]{\tens{N}}

\newcommand{\M}[0]{\tens{M}}
\newcommand{\iN}[0]{\tens{N}^{-1}}

\renewcommand{\S}[0]{\tens{S}}
\renewcommand{\r}[0]{\vec{r}}

\renewcommand{\P}[0]{\tens{P}}

\newcommand{\Dbp}[0]{\Delta_{\mathrm{bp}}}

\newcommand{\BP}{\textsc{BeyondPlanck}}

\newcommand{\npipe}[0]{\texttt{NPIPE}}

\newcommand{\phm}[0]{\phantom{-}}

\def\inv{^{-1}}

\begin{document}

\title{\bfseries{\scshape{BeyondPlanck}} XIII. Intensity foreground
  sampling,\\ degeneracies, and priors}

\newcommand{\oslo}[0]{1}
\newcommand{\milanoA}[0]{2}
\newcommand{\milanoB}[0]{3}
\newcommand{\milanoC}[0]{4}
\newcommand{\triesteB}[0]{5}
\newcommand{\planetek}[0]{6}
\newcommand{\princeton}[0]{7}
\newcommand{\jpl}[0]{8}
\newcommand{\helsinkiA}[0]{9}
\newcommand{\helsinkiB}[0]{10}
\newcommand{\nersc}[0]{11}
\newcommand{\haverford}[0]{12}
\newcommand{\mpa}[0]{13}
\newcommand{\triesteA}[0]{14}
\author{\small
K.~J.~Andersen\inst{\oslo}\thanks{Corresponding author: K.~J.~Andersen; \url{k.j.andersen@astro.uio.no}, \url{andersen.kristian.joten@gmail.com}}
\and
D.~Herman\inst{\oslo}
\and
\textcolor{black}{R.~Aurlien}\inst{\oslo}
\and
\textcolor{black}{R.~Banerji}\inst{\oslo}
\and
\textcolor{black}{A.~Basyrov}\inst{\oslo}
\and
M.~Bersanelli\inst{\milanoA, \milanoB, \milanoC}
\and
S.~Bertocco\inst{\triesteB}
\and
M.~Brilenkov\inst{\oslo}
\and
M.~Carbone\inst{\planetek}
\and
L.~P.~L.~Colombo\inst{\milanoA}
\and
H.~K.~Eriksen\inst{\oslo}
\and
J.~R.~Eskilt\inst{\oslo}
\and
\textcolor{black}{M.~K.~Foss}\inst{\oslo}
\and
C.~Franceschet\inst{\milanoA,\milanoC}
\and
\textcolor{black}{U.~Fuskeland}\inst{\oslo}
\and
S.~Galeotta\inst{\triesteB}
\and
M.~Galloway\inst{\oslo}
\and
S.~Gerakakis\inst{\planetek}
\and
E.~Gjerl{\o}w\inst{\oslo}
\and
\textcolor{black}{B.~Hensley}\inst{\princeton}
\and
M.~Iacobellis\inst{\planetek}
\and
M.~Ieronymaki\inst{\planetek}
\and
\textcolor{black}{H.~T.~Ihle}\inst{\oslo}
\and
J.~B.~Jewell\inst{\jpl}
\and
\textcolor{black}{A.~Karakci}\inst{\oslo}
\and
E.~Keih\"{a}nen\inst{\helsinkiA, \helsinkiB}
\and
R.~Keskitalo\inst{\nersc}
\and
J.~G.~S.~Lunde\inst{\oslo}
\and
G.~Maggio\inst{\triesteB}
\and
D.~Maino\inst{\milanoA, \milanoB, \milanoC}
\and
M.~Maris\inst{\triesteB}
\and
A.~Mennella\inst{\milanoA, \milanoB, \milanoC}
\and
S.~Paradiso\inst{\milanoA, \milanoC}
\and
B.~Partridge\inst{\haverford}
\and
M.~Reinecke\inst{\mpa}
\and
M.~San\inst{\oslo}
\and
N.-O.~Stutzer\inst{\oslo}
\and
A.-S.~Suur-Uski\inst{\helsinkiA, \helsinkiB}
\and
T.~L.~Svalheim\inst{\oslo}
\and
D.~Tavagnacco\inst{\triesteB, \triesteA}
\and
H.~Thommesen\inst{\oslo}
\and
D.~J.~Watts\inst{\oslo}
\and
I.~K.~Wehus\inst{\oslo}
\and
A.~Zacchei\inst{\triesteB}
}
\institute{\small
Institute of Theoretical Astrophysics, University of Oslo, Blindern, Oslo, Norway\goodbreak
\and
Dipartimento di Fisica, Universit\`{a} degli Studi di Milano, Via Celoria, 16, Milano, Italy\goodbreak
\and
INAF-IASF Milano, Via E. Bassini 15, Milano, Italy\goodbreak
\and
INFN, Sezione di Milano, Via Celoria 16, Milano, Italy\goodbreak
\and
INAF - Osservatorio Astronomico di Trieste, Via G.B. Tiepolo 11, Trieste, Italy\goodbreak
\and
Planetek Hellas, Leoforos Kifisias 44, Marousi 151 25, Greece\goodbreak
\and
Department of Astrophysical Sciences, Princeton University, Princeton, NJ 08544,
U.S.A.\goodbreak
\and
Jet Propulsion Laboratory, California Institute of Technology, 4800 Oak Grove Drive, Pasadena, California, U.S.A.\goodbreak
\and
Department of Physics, Gustaf H\"{a}llstr\"{o}min katu 2, University of Helsinki, Helsinki, Finland\goodbreak
\and
Helsinki Institute of Physics, Gustaf H\"{a}llstr\"{o}min katu 2, University of Helsinki, Helsinki, Finland\goodbreak
\and
Computational Cosmology Center, Lawrence Berkeley National Laboratory, Berkeley, California, U.S.A.\goodbreak
\and
Haverford College Astronomy Department, 370 Lancaster Avenue,
Haverford, Pennsylvania, U.S.A.\goodbreak
\and
Max-Planck-Institut f\"{u}r Astrophysik, Karl-Schwarzschild-Str. 1, 85741 Garching, Germany\goodbreak
\and
Dipartimento di Fisica, Universit\`{a} degli Studi di Trieste, via A. Valerio 2, Trieste, Italy\goodbreak
}

\abstract{We present the intensity foreground algorithms and model
  employed within the \BP\ analysis framework. The
    \BP\ analysis is aimed at integrating component separation and
    instrumental parameter sampling within a global framework, leading to
    complete end-to-end error propagation in the \Planck\ Low
    Frequency Instrument (LFI) data analysis.
    Given the scope of the \BP\ analysis, a limited set of
    data is included in the component separation process, leading to
    foreground parameter degeneracies. In order to properly constrain
    the Galactic foreground parameters, we improve upon the previous
    \commander\ component separation implementation by adding a suite
    of algorithmic techniques.  These algorithms are designed to
  improve the stability and computational efficiency for weakly
  constrained posterior distributions. These are: 1) joint foreground
  spectral parameter and amplitude sampling, building on ideas from
  \textsc{Miramare}; 2) component-based monopole determination; 3)
  joint spectral parameter and monopole sampling; and 4) application
  of informative spatial priors for component amplitude maps. We find
  that the only spectral parameter with a significant signal-to-noise
  ratio using the current \BP\ data set is the peak frequency of the
  anomalous microwave emission component, for which we find
  $\nu_{\mathrm{p}}=25.3\pm0.5\,$GHz; all others must be constrained
  through external priors. Future works will be aimed at integrating many
  more data sets into this analysis, both map and time-ordered based,
thereby gradually eliminating the currently observed
  degeneracies in a controlled manner with respect to both
  instrumental systematic effects and astrophysical degeneracies. When
  this happens, the simple LFI-oriented data model employed in the
  current work will need to be generalized to account for both a
  richer astrophysical model and additional instrumental effects. This work
  will be organized within the Open Science-based \textsc{Cosmoglobe}
  community effort.}

\keywords{ISM: general -- Cosmology: observations,
    cosmic microwave background, diffuse radiation -- Galaxy:
    general}

\maketitle


\section{Introduction}
\label{sec:introduction}

The cosmic microwave background (CMB) represents one of our best
sources for knowledge of the early universe. The intensity of the CMB
peaks within the microwave frequency range at about 161\,GHz \citep{mather:1994} and a long line of experiments have
targeted this frequency range since the CMB was first
discovered by \citet{penzias:1965}. However, there are many sources of
radiation in the microwave sky that
obscure our view of the CMB, both from within the Milky Way Galaxy and from distant
sources \citep[see, e.g.,][and references therein]{delabrouille2012,
  planck2016-l04}. Each of these sources must be modeled to high accuracy
in order to establish a clean estimate of the CMB sky.

A modeling of the Galactic foreground emission in the microwave frequency range
has been a vital venture in the CMB field as modern observations require higher accuracy
to properly characterize the fluctuations in the CMB \citep{planck2014-a12}. During the 
\Planck\ mission, several different component separation software were created and
compared to clean the microwave sky \citep{planck2016-l04}. Since the official 
\Planck\ analysis ended, considerable development has occurred within the Bayesian
\commander\ component separation software, implementing a comprehensive scheme which
feeds the results from component separation back into the time-domain analysis, resulting
in well defined posteriors \citep{bp01}.

The current paper outlines the additions to the component separation portion
of \commander, highlighting the new functionalities and their feasibility through a
set of simulations. The test case for these features is within the context of the 
\BP\ framework, which aims to reprocess the \Planck\ Low Frequency Instrument (LFI) data in an end-to-end
Bayesian approach. Here, we present the low-frequency foreground sky model and
component separation algorithms used for intensity analysis within the \BP\ 
framework, as applied to the \Planck\ \citep{planck2016-l01} LFI
\citep{planck2016-l02}. As the focus of the \BP\ analysis is limited 
to LFI, we limit the amount of ancillary data used to assist in the component
separation effort, aiming for LFI driven posteriors.

\BP\ is a novel end-to-end Bayesian CMB analysis
framework that builds on decades of experience gained within the
\Planck\ collaboration and its main defining feature is that
instrument characterization and calibration is performed jointly with
mapmaking and component separation, resulting in a single
statistically consistent model for the full data set. As such, the
foreground sky model plays a key role in the process, feeding directly
into many aspects of the analysis, from gain and correlated noise
estimation via leakage corrections and mapmaking to final component
maps, CMB estimates, and cosmological parameters. An overview of the
full process is provided in \citet{bp01} and its companion papers.

As the goal of the \BP\ project is to focus on end-to-end error
propagation with respect to the \Planck\ LFI data set, a limited set of
external data is concerned in order to ensure that the statistical weight
of LFI drives the results. The ancillary data included here are described
in Sect. \ref{sec:data}, and come in two flavors, namely as foreground 
amplitude priors and pixelized frequency maps to add constraining power to spectral index parameters.
In order to properly characterize the foregrounds within this limited data set,
a suite of algorithms are introduced to the \commander\ Gibbs sampling framework.
These algorithmic implementations are the focus of the current paper.
These algorithms are not exclusively useful to the minimal \BP\ data set
and they are widely applicable to future analyses.

The algorithms used in this paper derives most closely from a similar
\commander-based \citep{eriksen:2004,eriksen2008} Bayesian analysis of
the \Planck, \WMAP, and Haslam et al.\ data presented by
\citet{planck2014-a11}. The main differences between the two analyses
are as follows. First, in the previous analysis there was limited
feedback between the gain estimation and component separation, as only
a single overall absolute calibration factor was fitted during the
component separation process. In the current analysis, a full
time-dependent time-ordered data model is propagated throughout the
analysis. Second, the previous analysis provided only a single maximum-likelihood solution for each component, together with marginal
per-pixel uncertainties. In the present analysis, we provide a full
Monte Carlo ensemble of samples drawn from the full posterior
distribution, which represents full end-to-end propagation of all
uncertainties. Third, the computational framework used in the
\Planck\ 2015 analysis required uniform resolution across all
frequency bands, thus limiting the resolution to that of the
instrument with the poorest resolution. This was improved upon in the
\Planck\ 2018 analysis \citep{planck2016-l04}, where a new
computational framework that allows for different resolution and
smoothing scales was developed. However, the corresponding analysis
only included a single joint low-frequency component in intensity and
did not provide new estimates of the individual low-frequency
components. In the current analysis, we provide full-resolution
parameter maps for all components listed above, limited only by the
signal-to-noise ratio (S/N) of the data in question.

In this paper, we also describe the implementation of
four important new algorithmic
features into the \commander\ framework that are all designed to
improve sampling efficiency and stability for weakly constrained
posterior distributions. The first of these is a joint spectral
parameter and amplitude sampler that employs the marginal
spectral parameter posterior distribution to move quickly through the
multi-dimensional parameter space. This algorithm was first introduced
in the CMB literature by \citet{2009MNRAS.392..216S} and later implemented
in the \textsc{Miramare} component separation code by
\citet{stivoli:2010}. The main advantage of this approach is a
significantly reduced Markov chain Monte Carlo  (MCMC) correlation
length and overall lower computational costs as a result. In this paper,
we demonstrate this sampler on the peak frequency of the anomalous
microwave emission (AME) spectral energy density (SED). However,
the importance of this new step will increase significantly when
additional spectral parameters are explored in the future and will be vital in probing, for instance, the thermal dust spectral
index and temperature efficiently with the \Planck\ High Frequency
Instrument (HFI).

The second algorithmic improvement is component-based monopole
determination. As discussed at length by, for instance, \citet{planck2014-a12}
and \citet{wehus2014}, one of the most important challenges regarding
intensity-based spectral parameter estimation is an accurate
determination of the zero-levels at each frequency band; any error in
this will translate directly into a bias in spectral parameters, which
typically manifests itself as a spatial correlation between the
spectral parameter map and the corresponding component amplitude
map. At the same time, it is important to note that few (if any) CMB
experiments actually have sensitivity to the true sky monopole, and
they have therefore typically instead resorted to morphology-based
algorithms to determine meaningful zero-levels. In this paper, we
point out that this problem may be entirely circumvented by instead
focusing on determining the zero-levels of the astrophysical
component maps, rather than individual frequency\emph{}
maps. The resulting global sky model may then be used to determine the
frequency map offsets. This approach is significantly more transparent from a
physical point of view (e.g., the CMB temperature perturbation component may be assumed to have
an identically vanishing monopole), it automatically guarantees consistency between
the zero levels at different frequency channels and it gives zero
statistical weight to the frequency monopoles in the actual fitting
procedure.

Nevertheless, there is a formal degeneracy between the spectral
parameters and the frequency monopoles at each step in the algorithm
and to eliminate the Markov chain correlation length increase from
this degeneracy, we additionally implement a new joint spectral
parameter and monopole sampler as our third algorithmic improvement.

Finally, we generalize the concept of informative spatial
component map priors that was introduced by
\citet{planck2016-l04} and \citet{npipe} and use the results to apply informative
physical Gaussian spatial priors. This can be leveraged to significantly
reduce correlations between various sky components on small angular
scales and, in particular, degeneracies between AME, free-free and CMB
\citep{bp11} may be alleviated in this manner. Of course, the ideal
approach to resolve such degeneracies is not through the use of informative
priors, but rather by integrating additional data. The
current algorithms allow for a gradual and controlled introduction of
such data sets, without introducing pathological artifacts along the
way. 

The rest of the paper is organized as follows.  Section~\ref{sec:bp}
gives a short overview of the \BP\ framework, and sky
signal model. Section~\ref{sec:data}
describes the data set used within this analysis, the motivation for 
the data selection, and the simulation data used for algorithm validation. Section~\ref{sec:algorithms} describes the main
sampling algorithms for intensity foreground parameters. The main
results are summarized in Sect.~\ref{sec:results} and we present our conclusions
in Sect.~\ref{sec:conclusions}. We note that polarization-based
component separation results are discussed separately by \citet{bp14}.

\section{Overview of the \BP\ sampling framework}
\label{sec:bp}
The \BP\ project has implemented an
integrated end-to-end data analysis pipeline for CMB experiments \citep{bp01},
connecting all steps going from raw time-ordered data to cosmological
analysis in a self-consistent Bayesian framework, finally realizing ideas originally proposed almost 20 years ago by \citet{jewell2004} and \citet{wandelt2004}. This methodology allows us to
characterize degeneracies between instrumental and astrophysical
parameters in a statistically well-defined framework, with
uncertainties propagating consistently through all stages of the
pipeline. It also seamlessly connects low-level instrumental quantities
like gain \citep{bp07} and correlated noise \citep{bp06}, bandpasses
\citep{bp09}, and far sidelobes \citep{bp08} via Galactic parameters
such as the synchrotron amplitude and spectral index \citep[current
  paper and][]{bp14}, to the angular CMB power spectrum and
cosmological parameters \citep{bp11,bp12}. In this section, we provide
a brief review of the \BP\ sky model, data selection, and sampling
scheme; we refer to the various companion
papers for more details on the model.

\subsection{Data model and Gibbs chain}

\begin{table*}[tbh]
\centering
\caption{Frequency band summary for the \BP\ intensity analysis. }
\footnotesize
 \renewcommand{\arraystretch}{1.4}
 \setlength{\tabcolsep}{3pt}
 \begin{tabular}{p{2.0cm} r@{}l r@{}l r@{}l r@{}l r@{}l r@{}l l}
  \hline 
  \hline
  \multicolumn{1}{c}{\begin{tabular}{c} \textsc{Survey} \\ \, \\ \, \end{tabular} } &
  \multicolumn{2}{c}{\begin{tabular}{c} \textsc{Detector} \\ \textsc{label} \\ \, \end{tabular}} & \multicolumn{2}{c}{\begin{tabular}{c}\textsc{Central} \\ \textsc{frequency} \\ $[\,\mathrm{GHz}\,]$ \end{tabular} } & \multicolumn{2}{c}{\begin{tabular}{c}\textsc{Bandwidth} \\ \, \\ $[\,\mathrm{GHz}\,]$ \end{tabular} } & \multicolumn{2}{c}{\begin{tabular}{c} \textsc{Beam size} \\ (\,FWHM\,)  \\ $[\,\mathrm{arcmin}\,]$ \end{tabular}} & \multicolumn{2}{c}{\begin{tabular}{c} \healpix\ \\ \textsc{resolution}  \\ $(\,N_{\mathrm{side}}\,)$ \end{tabular}} & \multicolumn{2}{c}{\begin{tabular}{c} \textsc{Average} \\ \textsc{RMS}\tablefootmark{a}  \\ $[\,\mu\mathrm{K_{CMB}\,arcmin}\,]$ \end{tabular}} & \multicolumn{1}{c}{\textsc{Reference(s)}} \\
  \hline
  \Planck\ LFI  & \;\;\;\;\;\; 30& & \;\;\;\;\;\,28&.4 & \;\;\;\;\;\;\;5&.7 & \;\;\;\;\;\,32&.4 & \;\;\;\;\;\;\, 512 & &  \;\;\;\;\;\;\;179&.3 & \citet{planck2014-a05} \\
       & 44& & 44&.1 & 8&.8 & 27&.1 & 512 & & 213&.8 & \\
       & 70& & 70&.1 & 14&.0 & 13&.3 & 1024 & & 189&.0 &\vspace{0.3cm} \\
  \WMAP \dotfill& \textit{Ka}& & 33&.0 & 7&.0 & 40& & 512 & & 290&.2 & \citet{bennett2012} \\ 
       & \textit Q1& & 40&.6 & 8&.3 & 31& & 512 & & 400&.6 & \\
       & \textit Q2& & 40&.6 & 8&.3 & 31& & 512 & & 380&.0 & \\
       & \textit V1& & 60&.8 & 14&.0 & 21& & 512 & & 517&.2 & \\
       & \textit V2& & 60&.8 & 14&.0 & 21& & 512 & & 446&.6 &\vspace{0.3cm} \\
  \Planck\ HFI \dotfill& 857& & 857& & 249& & 10&.0\tablefootmark{b} & 1024 & & 5984 & & \citet{npipe} \\
  Haslam \dotfill& \dots& & 0&.408 & 0&\tablefootmark{c} & 60&\tablefootmark{d} & 512 & & 7&.886\tablefootmark{e} & \citet{haslam1982} \\
  \hline
 \end{tabular}
 \label{tab:data_survey_char}
 \endPlancktablewide
\tablefoot{\\
 \tablefoottext{a}{Average white noise rms without regularization noise. }\\
 \tablefoottext{b}{The native resolution of 857\,GHz is 4.64\arcmin\ FWHM
   \citep{planck2014-a08}, smoothed to 10\arcmin\ FWHM in this analysis. }\\
 \tablefoottext{c}{408\,MHz bandpass profile is assumed to be a $\delta$ function. }\\
 \tablefoottext{d}{The native resolution of 408\,MHz is 56\arcmin\ FWHM
   \citep{haslam1982}, smoothed to 60\arcmin\ FWHM in this analysis. }\\
 \tablefoottext{e}{Unit is $\mathrm{K_{CMB}\,arcmin}$. }\\
}
\end{table*}

In \BP,\, the most basic data sets are raw un-calibrated time-ordered
data (TOD), which are modeled as follows:
\begin{equation}
  \begin{split}
    d_{j,t} = g_{j,t}&\P_{tp,j}\left[\B^{\mathrm{symm}}_{pp',j}\sum_{c}
      \M_{cj}(\beta_{p'}, \Dbp^{j})a^c_{p'}  + \B^{\mathrm{asymm}}_{pp',j}\left(s^{\mathrm{orb}}_{j,t}  
      + s^{\mathrm{fsl}}_{j,t}\right)\right] + \\
    + &s^{\mathrm{1hz}}_{j,t} + n^{\mathrm{corr}}_{j,t} + n^{\mathrm{w}}_{j,t}.
  \end{split}
  \label{eq:todmodel}
\end{equation}
Here $j$ represents a radiometer label; $t$ indicates a single time
sample; $p$ denotes a single pixel on the sky; and $c$ represents one
single astrophysical signal component. Furthermore, \g\ denotes the
instrumental gain; \P\ denotes the pointing matrix,;$\B^{\mathrm{symm}}$ and $\B^{\mathrm{asymm}}$ denote the symmetric
and asymmetric beam matrix, respectively; \a\ represents the astrophysical
signal amplitudes; $\beta$ shows the corresponding spectral parameters;
$\Dbp$ are the bandpass corrections; $\M_{cj}$ denotes the
bandpass-dependent component mixing matrix; $s^{\mathrm{orb}}$ is the
orbital dipole; $s^{\mathrm{fsl}}$ are the far sidelobe corrections;
$s^{\mathrm{1hz}}$ represents electronic 1\,Hz spike corrections;
$n^{\mathrm{corr}}$ is the correlated noise; and $n^{\mathrm{w}}$ is
the white noise. This simple model has already been
  demonstrated to be an excellent fit to the \Planck\ LFI measurements
  by
  \citet{planck2013-p02,planck2014-a03,planck2014-a12,planck2016-l02},
  although the current work stands as the first time it has been
  formulated in terms of one single equation. It is also important to
  note that the reason this model actually does work for \Planck\ LFI
  is to a large extent, thanks to the fact that the LFI radiometers have
  very well behaved systematics; for other detector types, more
  complex models are very likely needed. For a more detailed
explanation of the current model, we refer to
\citet{bp01} and companion papers.

Data sets may also be included in the form of
preprocessed pixelized sky maps, $\m_{\nu}$, in which case the above
data model is simplified to:
\begin{equation}
  \m_{\nu,p} = g_{\nu}\B^{\mathrm{symm}}_{pp',\nu}\sum_{c}
    \M_{c\nu}(\beta_{p'}, \Dbp^{j})a^c_{p'} + n^{\mathrm{w}}_{\nu,p}.
  \label{eq:mapmodel}
\end{equation}
When producing $\m_{\nu}$, all time-dependent quantities (i.e., the
far sidelobe, orbital dipole, 1\,Hz spike, and correlated noise
contributions) in Eq.~\eqref{eq:todmodel} are explicitly subtracted
from the TOD prior to mapmaking, leaving only sky stationary
contributions in the final pixelized map. However, at this level only
a very limited set of instrumental parameters may be accounted for per
frequency, namely an overall absolute calibration factor, $g_{\nu}$,
an azimuthally symmetric beam, $\B^{\mathrm{symm}}$, and white noise,
$n^{\mathrm{w}}_{\nu,p}$. This latter expression may be written on the
following compact form:
\begin{equation}
  \m_{\nu} = \A_{\nu}(\beta)\a +\n^{\mathrm{w}}_{\nu},
  \label{eq:binned_map}
\end{equation}
where $\A_{\nu}(\beta)$ is an effective mixing matrix that takes into
account both the frequency scaling of each component and beam
convolution.

The goal of the Bayesian approach is to sample from the joint posterior distribution,
\begin{equation}
  P(\g,\n_{\mathrm{corr}},\xi_n,\Dbp,\a,\beta,C_{\ell}\,\mid\,\d).
  \label{eq:joint_posterior_full}
\end{equation}
This is a large and complicated distribution, with many degeneracies.
However, exploiting the Gibbs sampling algorithm \citep{geman:1984} we
may factorize the sampling process into a finite set of simpler
sampling steps. In this algorithm, samples from a multi-dimensional
distribution are generated by sampling from all corresponding
conditional distributions. The \BP\ Gibbs chain may
be written schematically as follows \citep{bp01},
\begin{alignat}{10}
\g &\,\leftarrow P(\g&\,\mid &\,\d,&\, & &\,\xi_n, &\,\Dbp, &\,\a, &\,\beta, &\,C_{\ell})\label{eq:conditional_gain},\\
\n_{\mathrm{corr}} &\,\leftarrow P(\n_{\mathrm{corr}}&\,\mid &\,\d, &\,\g, &\,&\,\xi_n,
&\,\Dbp, &\,\a, &\,\beta, &\,C_{\ell}),\\
\xi_n &\,\leftarrow P(\xi_n&\,\mid &\,\d, &\,\g, &\,\n_{\mathrm{corr}}, &\,
&\,\Dbp, &\,\a, &\,\beta, &\,C_{\ell}),\\
\Dbp &\,\leftarrow P(\Dbp&\,\mid &\,\d, &\,\g, &\,\n_{\mathrm{corr}}, &\,\xi_n,
&\,&\,\a, &\,\beta, &\,C_{\ell}),\\
\beta &\,\leftarrow P(\beta&\,\mid &\,\d, &\,\g, &\,\n_{\mathrm{corr}}, &\,\xi_n,
&\,\Dbp, & &\,&\,C_{\ell})\label{eq:conditional_beta},\\
\a &\,\leftarrow P(\a&\,\mid &\,\d, &\,\g, &\,\n_{\mathrm{corr}}, &\,\xi_n,
&\,\Dbp, &\,&\,\beta, &\,C_{\ell})\label{eq:conditional_amp},\\
C_{\ell} &\,\leftarrow P(C_{\ell}&\,\mid &\,\d, &\,\g, &\,\n_{\mathrm{corr}}, &\,\xi_n,
        &\,\Dbp, &\,\a, &\,\beta&\,\phantom{C_{\ell}})&,\label{eq:conditional_bp}
\end{alignat}
where $\leftarrow$ indicates sampling from the distribution on the right-hand side.

We note that not all of these steps follow the strict Gibbs approach
of conditioning on all other parameters. Most notably for us, this is
the case in Eq.~\eqref{eq:conditional_beta} for the spectral
parameters sampler, $P(\beta\mid\d,\ldots)$, which places conditions on all parameters,
except for \a. Instead, we effectively sample \a\ and $\beta$ jointly by
exploiting the definition of a conditional distribution, as detailed
in Sect.~\ref{subsec:gibbs}. The advantage of a joint sampling step is
a significantly shorter Markov correlation length as compared to
standard Gibbs sampling.

A very convenient property of Gibbs sampling is its modular nature, as
the various parameters are sampled within each conditional
distribution, but joint dependencies are explored through the
iterative scheme. In this paper, we are therefore only concerned with
the sampling of two of the above steps, namely
Eqs.~\eqref{eq:conditional_beta} and \eqref{eq:conditional_amp}. For all other sampling
steps, we refer to \citet{bp01} and references therein.

\begin{table*}[tbh]
\centering
\caption{Summary of main parametric signal models for the temperature analysis.
  The symbol ``$\sim$'' implies that the respective parameter has a prior as given by the
  right-hand side distribution; Uni denotes a uniform distribution within the indicated
  limits; $N(\mu,\sigma^2)$ denotes a (normal) Gaussian distribution with the indicated mean and
  variance; and $a_i$ denotes the component amplitude of component $i$ at the
  given reference frequency $\nu_{0,i}$, and $s_i$ is spectral energy density,
  i.e., the observed signal at a given frequency, $\nu$.  }

\renewcommand{\arraystretch}{1.4}
 \setlength{\tabcolsep}{4pt}
\begin{tabular}{p{2.5cm}lll}
  \hline 
  \hline
  \multicolumn{1}{l}{\textsc{Component}} & \multicolumn{1}{l}{\textsc{Free parameters}} & \multicolumn{1}{l}{\;\;\textsc{Spectral energy density}, $s_\nu [\muKRJ]$ } & \multicolumn{1}{l}{\textsc{Additional information}} \\
  \multicolumn{1}{l}{} & \multicolumn{1}{l}{\textsc{and priors}} & \multicolumn{1}{l}{ } & \multicolumn{1}{l}{} \\
  \hline
  CMB\dotfill& $\Acmb \sim \mathrm{Uni} (-\infty , \infty)$ & \begin{tabular}{@{}r @{\;=\;}l} $x$ & $\frac{h\nu}{\kB\Tcmb}$ \\ $g(\nu)$ & $(\exp(x)-1)^2/(x^2 \exp(x))$ \\ $\scmb$ & $\Acmb /g(\nu)$\end{tabular} & $\Tcmb = 2.7255$ K \vspace{0.5cm} \\
   \begin{tabular}{@{}l} Relativistic CMB \\ quadrupole\dotfill \end{tabular} & $\Aquad = \Tcmb\,\bsun^2\, z^2$ & \begin{tabular}{@{}r @{\;=\;}l} $x$ & $\frac{h\nu}{\kB\Tcmb}$ \\ $g(\nu)$ & $(\exp(x)-1)^2/(x^2 \exp(x))$ \\ $\mathcal{Q}(\nu)$ & $(x/2)(\exp(x)+1)/(\exp(x)-1)$ \\ $\squad$ & $\Aquad\, \mathcal{Q}(\nu)/g(\nu)$\end{tabular} & \begin{tabular}{@{}r @{\;=\;}l} \Tcmb\ & 2.7255 K\\ \bsun & $1.2343\cdot 10^{-3}$ \\ $\vec{\hat{\beta}}_{\mathrm{sun}}$ & $(264.00\deg,48.24\deg)$ \\ $z$ & $\vec{\hat{n}}\cdot\vec{\hat{\beta}}_{\mathrm{sun}}$   \end{tabular} \vspace{0.5cm} \\
  Synchrotron\dotfill &  \begin{tabular}{@{}r @{}l} $\Asynch \sim$ & $\;\mathrm{Uni} (-\infty , \infty)$ \\ $\bsynch \sim$ & $\;N(-3.3\pm 0.1),$ \\ & \;fullsky \end{tabular} & $\ssynch = \Asynch \left(\frac{\nu}{\nuzeros}\right)^{\bsynch+\,C\,\mathrm{ln}\,\nu/\nuzeros}$ & \begin{tabular}{@{}r @{\;=\;}l} $\nuzeros$ & 30 GHz \\ $C$ & 0, low signal-to-noise\end{tabular} \vspace{0.5cm}\\
  Free-free\dotfill & \begin{tabular}{@{}r @{} l}$a_\mathrm{ff}$ $\sim$& $\,\mathrm{Uni} (-\infty , \infty)$\\ \Te\ =& $\;7000\,\mathrm{K}$, \\ & \;fullsky \end{tabular} & \begin{tabular}{@{}r @{\;=\;}l} $\gff$ & log$\,\left\{\exp\left[ 5.960 - \sqrt{3}/\pi \,\mathrm{log}\left(\nu_9 \, T_4^{-3/2}\right)\right] + e  \right\}$ \vspace{0.1cm}\\ $\sff$ & $\Aff \, \left(\frac{\nuzeroff}{\nu}\right)^2 \frac{\gff(\nu)}{\gff(\nuzeroff)}$ \end{tabular} & \begin{tabular}{@{}r @{\;=\;}l} $\nuzeroff$ & 40.0 GHz \\ $T_4$ & $\Te / 10^4$ \\ $\nu_9$ & $\nu/(10^9\;\mathrm{Hz})$ \\ $e$ & Euler's number\end{tabular} \vspace{0.5cm}\\
    \begin{tabular}{@{}l} AME/\\ spinning dust\dotfill \end{tabular} & \begin{tabular}{@{}r @{}l} $\Aame \sim$& $\;\mathrm{Uni} (-\infty , \infty)$\\ $\nup \sim$ & $\; N(22\pm 3\;\mathrm{GHz}),$ \\ & \;fullsky \end{tabular} & $\same = \Aame \,\left(\frac{\nuzeroame}{\nu}\right)^{2} \frac{\fame\left(\nu\cdot\nu_{\rm p0}/\nup\right)}{\fame\left(\nuzeroame\cdot\nu_{\rm p0}/\nup\right)}$ &  \begin{tabular}{@{}r @{\;=\;}l} $\nuzeroame$ & 22.0 GHz \\ $\nu_{\rm p0}$ & 30.0 GHz \\ $\fame(\nu)$ & External template\end{tabular} \vspace{0.5cm}\\
  Thermal dust\dotfill &  \begin{tabular}{@{}r @{}l} $\Adust \sim$ & $\;\mathrm{Uni} (-\infty , \infty)$ \\ $\bdust \sim$ & $\;N(1.56 \pm 0.03),$ \\ & \;fullsky \end{tabular} & \begin{tabular}{@{}r @{\;=\;}l} $\gamma$ & $\frac{h}{\kB\;\Tdust}$ \\ $\sdust$ & $\Adust \,\left(\frac{\nu}{\nuzerod}\right)^{\bdust+1}\;\frac{\exp(\gamma\nuzerod)-1}{\exp(\gamma\nu)-1}$\end{tabular} & \begin{tabular}{@{}r @{\;=\;}l} \Tdust & \npipe\ template \\ $\nuzerod$ & $545\;\mathrm{GHz}$\end{tabular} \vspace{0.5cm}\\
  Radio sources\dotfill &  \begin{tabular}{@{}r @{}l} $\Asrc >$ & \;0 \\ $\alphasrc \sim$ & $\;N(-0.1 \pm 0.3)$ \end{tabular} & $\ssrc = U_{\mathrm{mJy}}(\nuzerosrc)\, \Asrc\, \left(\frac{\nu}{\nuzerosrc}\right)^{\alphasrc-2}$ & \begin{tabular}{@{}r @{\;}l}  $\nuzerosrc =$ & $30\GHz$ \\ $U_{\mathrm{mJy}}(\nuzerosrc) =$ & Unit conversion \\ & factor \end{tabular}  \vspace{0.1cm}\\
  \hline
 \end{tabular}
 \label{tab:components}
 \endPlancktablewide
\end{table*}

\subsection{Astrophysical sky model}
\label{subsec:sky_model}

The dominant astrophysical foreground components in the \Planck\ LFI frequencies are synchrotron, AME, free-free, thermal dust emission, and compact radio sources. The modeling of each of these components is detailed in \citet{bp01}, so we only review the relevant details in this paper. In Table~\ref{tab:components}, we summarize the models for each component in terms of free parameters, priors, and SEDs. In addition, each diffuse component is modeled in terms of an amplitude sky map, $\vec{a}$, at a given reference frequency $\nu_0$ in brightness temperature units. Scaling to arbitrary frequencies is performed through the SEDs, such that the actual observed signal at a given frequency, $\nu,$ may generally be written as:\  
\begin{equation}
  \label{eq:signal}
  s_{\mathrm{RJ}}^i (\nu)=a_i\cdot f_i(\nu,\nu_{0,i},\beta_i),
\end{equation}
where $i$ denotes the specific component, $\nu_{0,i}$ is the reference frequency of the given component, $\beta_i$ is a set of component-specific spectral parameters, and $f_i$ is the SED. For diffuse components, $a_i$ is defined in terms of spherical harmonic space with a maximum multipole, \lmax\, defined for each component, depending on the signal-to-noise ratio (S/N) and angular resolution of the data sets supporting that component. For instance, synchrotron and AME have lower values of \lmax than the thermal dust and CMB. In addition, as discussed in Sect.~\ref{subsec:amplitude_priors}, we regularize the high-$\ell$ multipoles of each component with some smoothing prior, either derived from the known physical behaviour of the respective component \citep[e.g.,][]{planck2014-a12} or by a Gaussian smoothing operator. For compact sources, $a_i$ represents simply the flux density in mJy, with the spectral index $\alpha$ also defined in mJy, and an explicit unit conversion factor, $U_{\mathrm{mJy}}$, converts from flux density to brightness temperature units.

With this notation, the astrophysical sky model used for the current \BP\ analysis may be written as follows:
\begin{align}
  \vec{s}_{\mathrm{RJ}} &= \left(\vec{a}_{\mathrm{CMB}}+\vec{a}_{\mathrm{quad}}(\nu)\right) \frac{x^2 e^x}{(e^x -1)^2}+\label{eq:cmb_astsky},\\
  &+ \vec{a}_{\mathrm{s}} \left(\frac{\nu}{\nuzeros}\right)^{\bsynch} + \label{eq:synch_astsky},\\
  &+ \vec{a}_{\mathrm{ff}} \left(\frac{\nuzeroff}{\nu}\right)^2 \frac{g_{\mathrm{ff}}(\nu;\Te) }{g_{\mathrm{ff}}(\nuzeroff;\Te)} +\label{eq:ff_astsky},\\
  &+ \vec{a}_{\mathrm{ame}} \left(\frac{\nuzeroame}{\nu}\right)^2 \frac{f_{\mathrm{ame}} \left(\nu\cdot \frac{30.0\GHz}{\nup}\right)}{f_{\mathrm{ame}} \left(\nuzeroame\cdot \frac{30.0\GHz}{\nup}\right)}+ \label{eq:ame_astsky},\\
  &+ \vec{a}_{\mathrm{d}} \left(\frac{\nu}{\nuzerod}\right)^{\bdust+1} \frac{e^{h\nuzerod/\kB\Tdust}-1}{e^{h\nu/\kB\Tdust}-1}+ \label{eq:dust_astsky}\\
  &+ U_{\mathrm{mJy}} \sum_{j=1}^{N_{\mathrm{src}}} \vec{a}_{j,\mathrm{src}} \left(\frac{\nu}{\nuzerosrc}\right)^{\alpha_{j,\mathrm{src}}-2}, \label{eq:ptsrc_astsky}
\end{align}
where $\vec{a}_{\mathrm{CMB}}$ and $\vec{a}_{\mathrm{quad}}$ are given
in thermodynamic temperature units ($\mathrm{K_{CMB}}$),
$\vec{a}_{j,\mathrm{src}}$ in flux density units (mJy),
and all other amplitudes $\vec{a}_i$ are given in terms of brightness
temperature ($\mathrm{K_{RJ}}$). The amplitude
of component $i$ is equal to that observed at a monochromatic
frequency, $\nu_{0,i}$. The sum in Eq.~\eqref{eq:ptsrc_astsky} runs over
all compact sources brighter than some flux threshold as defined by an
external source catalogue. In particular, we adopt the same catalogue
as \citet{planck2016-l04}, which is a hybrid of the AT20G
\citep{murphy2010}, GB6 \citep{gregory1996}, NVSS \citep{condon1998}
and PCCS2 \citep{planck2014-a35} catalogs, comprising a total of
12\,192 individual sources.

When comparing the results from the above model with previous work, it
is important to note that we fit a straight power-law for the
synchrotron SED. This means that the effect of any potential negative
curvature between 408\,MHz and 30\,GHz, as, for instance, assumed by
\citet{planck2014-a12}, will instead by interpreted as a slightly
steeper spectral index in the current analysis. This is more
explicitly demonstrated in Sect.~\ref{sec:results}.

For further information regarding this model and a brief discussion
of each individual component, we refer to
\citet{bp01} and references therein. The main goal of the present
paper is to establish efficient sampling algorithms for the amplitudes
and spectral parameters in Eq.~\eqref{eq:cmb_astsky}--\eqref{eq:ptsrc_astsky}. 

\section{Data Selection}
\label{sec:data}

\subsection{\BP\ Data Selection}
\label{subsec:bp_data_selection}

As discussed in Sect.~2 of \cite{bp01}, the only data set which is considered at the time-ordered level is the \Planck\ LFI data. With the minimal sky model, discussed in Sec.~\ref{subsec:sky_model}, a total of four unpolarized astrophysical sky components are considered. Seeing as LFI only contains three frequency channels, it is clear that the LFI data itself is unable to properly constrain this sky model. A set of selected external data is therefore included in the foreground analysis in order to constrain the sky model presented within \BP.

Table~\ref{tab:data_survey_char} provides an overview of all frequency
maps included in the intensity component separation procedure. We note again that the main motivation underlying the \BP\
analysis is not to derive a
novel state-of-the-art intensity sky model, but rather to develop and
demonstrate the Bayesian end-to-end analysis framework using
\Planck\ LFI as a worked case. Accordingly, to ensure that the main
results are dominated by \Planck\ LFI, all CMB-dominated \Planck\ HFI
bands and the \WMAP\ \textit{K}-band channel are excluded from the analysis;
the data summarized in Table~\ref{tab:data_survey_char} represent a
minimum set that is able to algebraically resolve all main foreground
components relevant for \Planck\ LFI. All sky maps (including LFI and
others) are discretized using the
\healpix\footnote{\url{http://healpix.jpl.nasa.gov}} \citep{gorski2005}
pixelization. 

For all non-LFI bands, we adopt nominal bandpass profiles as
recommended by the respective references. However, as an exception, we
adopt the simplified (and commonly used) delta function approximation
for the Haslam 408\,MHz channel. For LFI 30\,GHz, we allow for both an
absolute bandpass shift for the full frequency band and relative
differences between individual detectors; whereas for the 44 and 70\,GHz
channels, we only allow relative detector shifts, but with no overall
absolute shifts; see \citet{bp09} for further details.

The noise is assumed to be uncorrelated and Gaussian for all channels
except \Planck\ LFI, with a spatially varying root mean square (RMS) as defined by the
number of hits per pixel. Again, the only exception is Haslam
408\,MHz, which is nominally strongly signal-dominated per pixel and dominated
by systematic uncertainties, not statistical. In this case, we instead
adopted a noise rms model that is the sum of an isotropic 0.8\,K term
(representing statistical uncertainties) and 1\,\% of the actual map
itself, representing multiplicative uncertainties; this is the same
approach as taken by \citet{planck2014-a12}. For LFI, correlated noise
is accounted for on all angular scales through explicit time-domain
sampling, as discussed by \citet{bp06}.

In the data model given in Eq.~\eqref{eq:todmodel}, only the orbital CMB
dipole and the far sidelobes are modeled with the full asymmetric beams. 
For astrophysical component modeling,
all beams are assumed to be azimuthally symmetric, with window
functions, $b_{\ell}$, provided individually by each experiment. No
uncertainties on these are propagated in the current analysis, but
support for this will be added in future work. The Haslam 408\,MHz and
\Planck\ DR4 (\npipe) 857\,GHz maps are smoothed from their native
resolutions to $60\arcm$ and $10\arcm$, respectively. The latter is
additionally re-pixelized to a \healpix\ resolution of $\nside = 1024$
to reduce CPU and memory requirements; note that this channel still
has a higher angular resolution than the 70\,GHz LFI channel, which is
the highest resolution channel of the main \BP\ analysis. 

An important and novel aspect of the current analysis is
component-based monopole (or ``zero-levels'' or ``offsets'')
determination, as discussed in Sect.~\ref{subsec:monopole_sampler}. Rather than
attempting to set the monopoles for each channel before component
separation, we impose physical priors on the monopole for each
astrophysical component. This astrophysical model is then used to
determine deterministically the zero-level for each frequency
map. Reasonable priors may be defined for all components except
synchrotron emission, and for this component we instead adopt explicit
literature values. Specifically, we adopted a monopole value of
$8.9\pm1.3\,\KCMB$ for synchrotron emission at 408\,MHz, as estimated
by \citet{wehus2014} and we thereby neglect possible contributions
from free-free emission to Haslam 408\,MHz outside the very
conservative Galactic mask employed by \citet{wehus2014}. We also
applied the dipole corrections to the Haslam map derived by the same
analysis.

Any additional pre-processing applied to the various maps was kept at a
minimal level. Specifically, for \WMAP\ we added the \WMAP\ solar CMB
dipole of $(d, l, b) = (3355\,\mu\mathrm{K},\,263.99\deg,\,48.26\deg)$ to each map \citep{Hinshaw_2009};
while for \Planck\ 857\,GHz, we apply a zodiacal light correction,
following \citet{npipe}. 

No calibration corrections are applied to any non-LFI data sets, and
we thus rely on the calibration of the original analyses for these
channels. This is particularly important with respect to the
\WMAP\ channels, which have a non-negligible impact on the solar CMB
dipole; consequently, the final \BP\ solar dipole estimate represents
a noise-weighted average between the \WMAP\ and \BP-based
LFI estimates.

\subsection{Simulation data}
\label{subsec:sim_data}

Proper testing of the algorithms presented in the current paper constitutes a vital component in the verification of the feasibility of these algorithms within the \commander\ framework. As such, a suite of simulated data (with controlled noise) and total offsets was constructed. These simulated data are created to represent the full \BP\ sky model, with frequency channels equivalent to the conditions placed on the data selection (discussed in Sect.~\ref{subsec:bp_data_selection} and presented in Table~\ref{tab:data_survey_char}).

The simulations are entirely created within the map-space domain. Using a sample from the \BP\ sky model ensemble, mock frequency maps were created. Using  \commander\ we output each of the foreground components, as determined by the \BP\ sky model, at each of the input frequency channels listed in Table~\ref{tab:data_survey_char}. As a result, we were able to create mock frequency maps by co-adding each of these sky components at the corresponding frequencies. Thanks to this simple method, the full sky model, including the spectral parameters, are encapsulated in the set of simulated frequency maps.

Noise was then added to each of the simulated frequency maps by taking a random realization of the noise rms sky maps, assuming that the noise is white. For the noise realizations, we utilize the actual noise rms data as used and produced within the full \BP\ results.

\section{\commander\ extensions for efficient intensity-based component separation}
\label{sec:algorithms}

The specific \BP\ computer code implementation is called
\commanderthree\ \citep{bp03}, and this is a direct generalization of the
code first introduced for CMB power spectrum estimation purposes by
\citet{eriksen:2004} and later generalized to also account for
astrophysical component separation by
\citet{eriksen2008,seljebotn:2013,seljebotn:2019}. It was one of four
main component separation algorithms adopted by the
\Planck\ collaboration \citep{planck2013-p06, planck2014-a12,
  planck2016-l04, npipe}. Unless it is useful for context, we do not
distinguish between the different code versions and we simply refer to all versions as \commander.
In this section, we describe the four algorithmic improvements we have made to \commander,
as introduced in Sect.~\ref{sec:introduction}.

\begin{figure*}
  \center
  \includegraphics[width=0.46\linewidth]{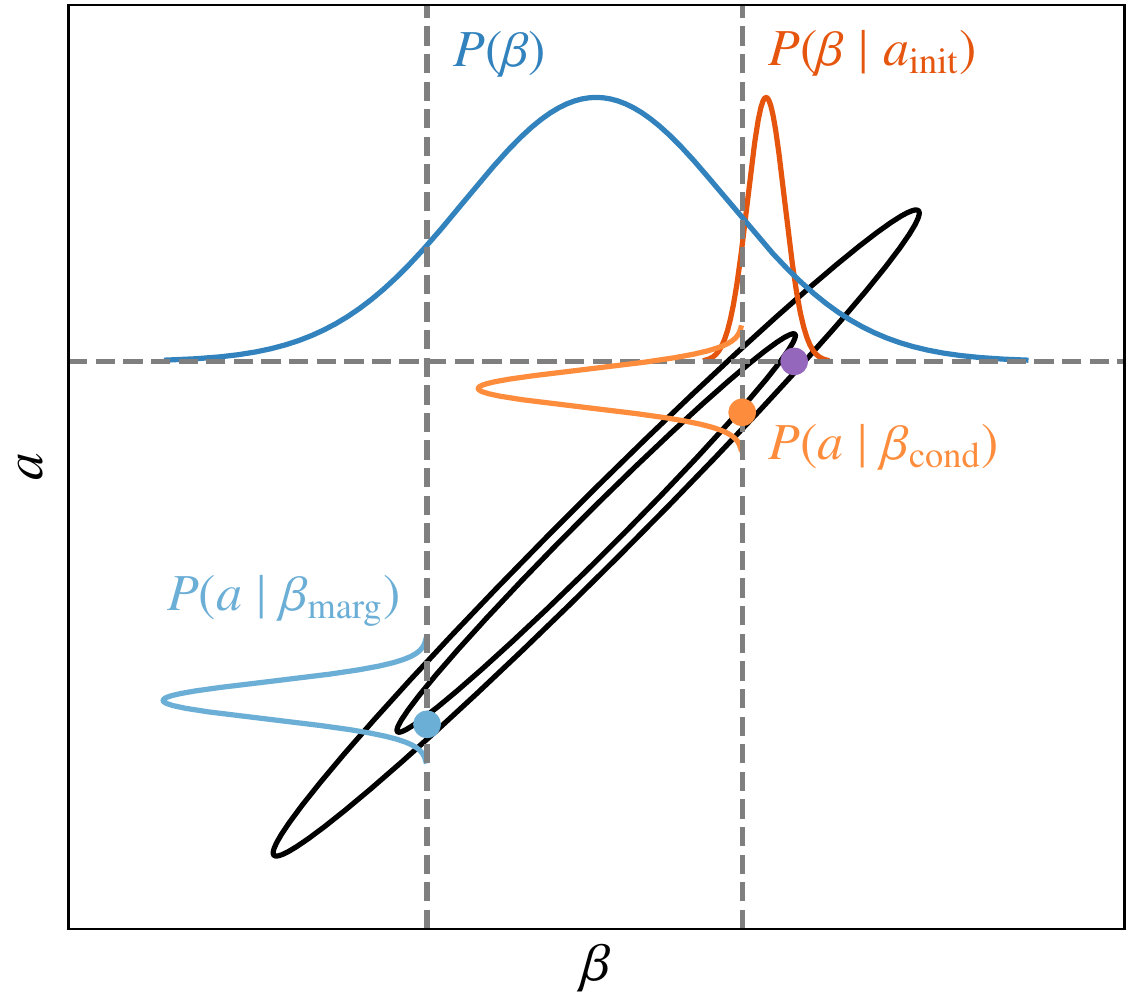}\hspace*{1cm}
  \includegraphics[width=0.46\linewidth]{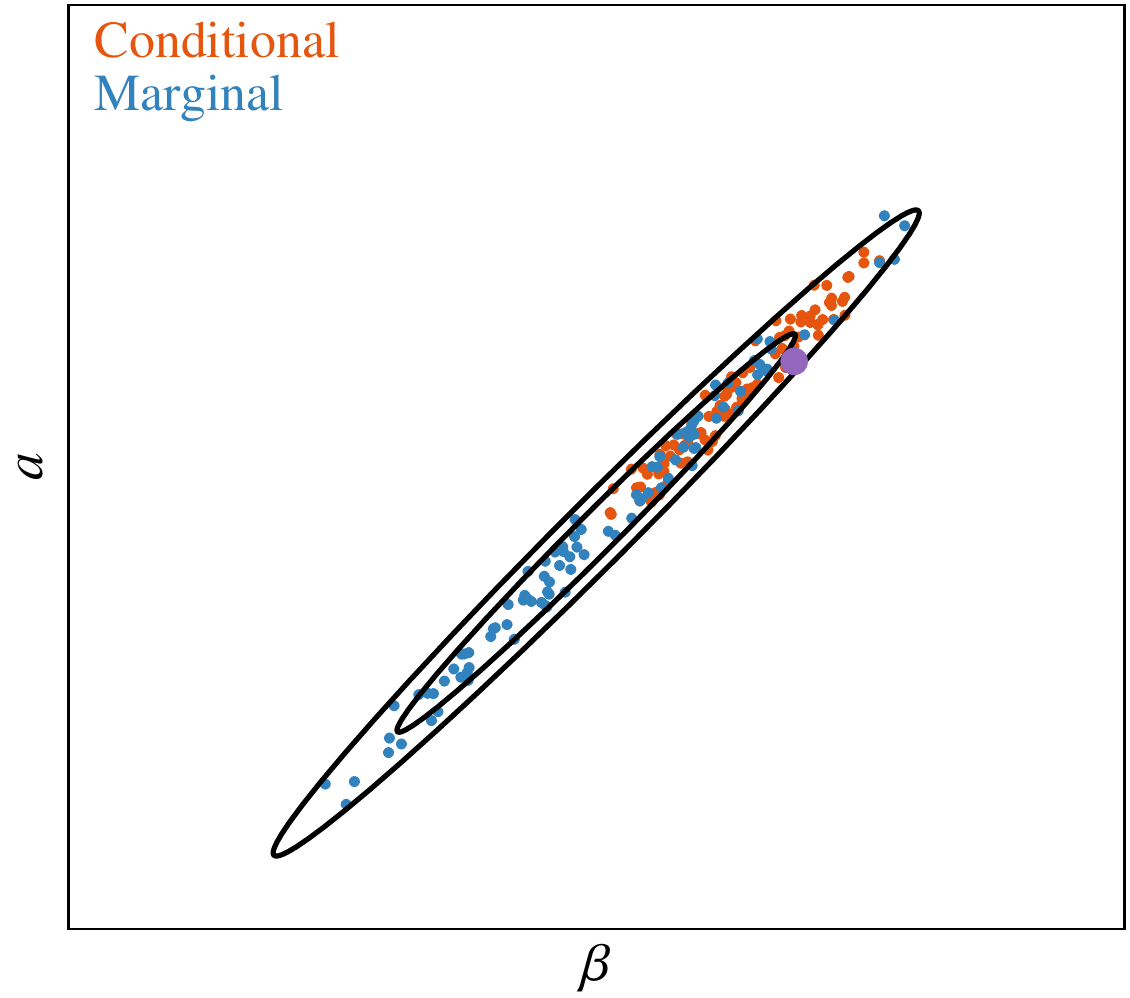}
  \caption{Illustration of conditional (pure Gibbs; orange) and marginal
    sampling (blue) algorithms for a highly correlated (Pearson's
    correlation coefficient of $\rho=0.99$) two-dimensional Gaussian
    distribution (black contours). The initial position $(\beta_{\mathrm{init}},a_{\mathrm{init}})$ is
    indicated by a purple dot. Left:~Comparison of un-normalized
    conditional $P(\beta\mid a_{\mathrm{init}})$ distribution evaluated at the initial position
    and the corresponding marginal $P(\beta)$ distribution; note that the
    latter is much wider than the former. Assuming $\beta$ samples drawn
    at the 10th percentile, the graphs along the vertical lines
    represent un-normalized conditial distributions $P(a\mid\beta)$ evaluated
    at the $\beta$ values drawn with the
    conditional (orange) and marginal (blue) distributions of $\beta$;
    note that the marginal sampling case results in a much longer step
    length between the initial and final sample values.
    Right:~Samples of a
    standard Gibbs sampling chain (orange) using conditional sampling for
    both $a$ and $\beta$, and a sampling chain (blue) using marginal
    sampling for $\beta$ and conditional sampling for $a$. Both cases
    show the first 100 samples initialized from the purple point.}
  \label{fig:marginal_conditional_plot}
\end{figure*}

\subsection{Joint amplitude and spectral parameter sampling}
\label{subsec:gibbs}

As summarized in
Eqs.~\eqref{eq:conditional_gain}--\eqref{eq:conditional_bp}, the
\BP\ pipeline implements a Gibbs sampling chain iterating over all
free parameters in the data model. While Gibbs sampling in general is
a very powerful method for exploring complicated distributions, its
main weakness is the inability to probe degenerate distributions. This
problem is illustrated for a toy example in the left panel of
Fig.~\ref{fig:marginal_conditional_plot}: the black contours
represents the 68 and 95\,\% confidence limits of a two-dimensional
Gaussian distribution with a Pearson's correlation coefficient of
$\rho=0.99$. The purple point indicates the starting position of a
Markov chain, while the dashed, grey horizontal line indicates the baseline
of the corresponding conditional distribution
$P(\beta\mid a_{\mathrm{init}})$, which itself is shown as an orange
curve. Since the Gibbs sampling algorithm works by moving according to
conditional distributions alone, the allowed step size in each
iteration is very narrow compared to the full marginal distribution,
shown as a blue distribution.

To illustrate the step size effect in the Gibbs sampler, we consider a Gibbs move in Fig.~\ref{fig:marginal_conditional_plot},
starting with $\beta$ from the purple point. The
relevant conditional distribution for $\beta$ is shown as an orange curve
along the horizontal dashed gray line passing through the purple
point and we can thus draw a random value from this distribution. This
could for instance be the value indicated by the right-most vertical
dashed gray line. According to the Gibbs sampling algorithm, we are required to draw a sample from the corresponding conditional
distribution for $a$, which is indicated by the orange distribution
aligned with the vertical dashed line. One possible outcome of this,
after completing one full Gibbs iteration, is the orange point. Now,
because each conditional distribution is much narrower than the
corresponding marginal distributions, the relative Gibbs step size is
very short, and it takes a very long time to move from one side of the
joint distribution to the opposite. The result is poor Markov chain
mixing and a very long correlation length. As a real-world
illustration of this, the orange points in the right panel of
Fig.~\ref{fig:marginal_conditional_plot} show the 100 first steps of
an actual Gibbs chain with this precise target distribution. We see
that less than half of the distribution is actually explored, and many
thousands of samples will be required in order to probe the full
distribution with this algorithm.

This problem is directly relevant for modern Bayesian intensity-based
CMB component separation. For experiments such as \Planck\ and \WMAP,
characterized by very high S/N, there are strong
degeneracies between the foreground amplitudes, the foreground
spectral parameters, and map-level monopoles. Explicitly, if one
assumes that all spectral parameters and monopoles are known, then the
conditional amplitude uncertainty is very small. Conversely, if we
assume the amplitude and monopoles to be known, then the conditional
spectral parameter uncertainties are small. However, when all
parameters are unknown, the full uncertainties are significant.

For this reason, Gibbs sampling should usually be considered a last
resort to handle an otherwise intractable distribution. If direct
joint sampling methods are available, then those are usually more
efficient. Fortunately, the Gibbs sampling method can be interleaved
by any combination of conditional and joint steps while still
maintaining the requirement of detailed balance \citep{geman:1984}; also, the more steps that can be handled jointly, the more efficient the
overall Gibbs chain will be. For the purposes of intensity-based CMB
component separation, we therefore introduce a new special-purpose
joint amplitude--spectral parameter step by exploiting the definition
of a conditional distribution as follows, 
\begin{equation}
  P(\a,\beta\mid\m_\nu) = P(\beta\mid\m_\nu)\, P(\a\mid\m_\nu,\beta).
  \label{eq:joint_P_to_marg_cond}
\end{equation}
The first distribution on the right hand side is the marginal
distribution of $\beta$ with respect to the data $\m_\nu$, and the second
distribution is the conditional distribution of $\a$ with respect to
$\beta$. This equation therefore implies that we may generate a joint
sample by first drawing $\beta$ from its \emph{}
"marginal"\ distribution, and then sample $\a$ from the corresponding
"conditional"\ distribution:
\begin{align*}
 \beta \leftarrow & P(\beta\mid\m_{\nu}),\\\label{eq:beta_marg} 
 \a\leftarrow & P(\a\mid\m_{\nu},\beta).
\end{align*}
We note the absence of $\a$ in the first distribution.
We then need to derive sampling procedures for each of these two
distributions. According to Bayes' theorem:
\begin{equation}
 P(\a,\beta\mid\m_{\nu}) = \frac{P(\m_{\nu}\mid\a,\beta) P(\a,\beta)}{P(\m_{\nu})}, \label{eq:joint_P_Bayes}
\end{equation}
where $P(\m_{\nu})$ is just a normalization factor (often called the
``evidence''), $P(\a,\beta)$ denotes optional priors, and the final
factor is the likelihood function, $P(\m_{\nu}\mid\a,\beta) \equiv
\mathcal{L}(\a,\beta)$. From the compact data model in
Eq.~\eqref{eq:binned_map}, we note that
\begin{equation}
  \m_{\nu} - \A_{\nu}(\beta)\a = \n^{\mathrm{w}}_{\nu},
  \label{eq:map_residual}
\end{equation}
and since $\n^{\mathrm{w}}_{\nu}$ is assumed to be zero-mean and
Gaussian distributed with known variance, we can immediately use the following expression for the likelihood function:
\begin{equation}
  \ln\mathcal{L}(\a,\beta) \propto -\frac{1}{2}\sum_{\nu}(\m_\nu-\A_\nu(\beta) \a)^T\, \iN_\nu\, (\m_\nu-\A_\nu(\beta) \a),
  \label{eq:likelihood_full}  
\end{equation}
where $\N_\nu$ is the noise covariance matrix of band $\nu$, 
which is diagonal in the case of pure white noise.
The priors are less well-defined, and are left to the user to determine. In
the following, we adopt Gaussian priors for spectral parameters
and for notational convenience, we assume no spatial amplitude priors, $\S^{-1}=\tens 0$.

We first consider the marginal spectral parameter distribution,
$P(\beta\mid\m_{\nu})$. This is derived by integrating
Eq.~\eqref{eq:likelihood_full} with respect to $\a$, and this was done
by \citet{2009MNRAS.392..216S} and \citet{stivoli:2010} as part of developing the
\textsc{Miramare} component separation code. The result is expressed as:
 \begin{align}
   \mathrm{ln}\,\mathcal{L}_{\mathrm{marg}}(\beta) &= \mathrm{ln} \int d\a \; \mathrm{exp} \left[-\frac{1}{2}(\m-\A\a)^T\, \iN\, (\m-\A\a)\right] \nonumber \\
   &= \mathrm{const} -\frac{1}{2} (\A^T\,\iN\, \m)^T\, (\A^T\,\iN\, \A)^{-1}\,(\A^T\,\iN\, \m) \nonumber \\
   &\quad\quad+ \frac{1}{2}\mathrm{ln} \left| (\A^T\,\iN\, \A)^{-1}\right|,  \label{eq:likelihood_beta_marg}
\end{align}
where all terms should be interpreted as sums over frequencies. The
same authors also introduced a so-called ``ridge likelihood,'' in which
one does not marginalize over $\a$, but rather sets $\a$ equal to its
maximum likelihood value for a given value of $\beta$. This may also
be analytically evaluated, and is in fact identical to the above
expression, with the exception of  the last determinant term being excluded. We implemented
support for both options in our codes. We note that this expression
requires all data to be defined at the same angular resolution and so,
all data must be smoothed to a common resolution before evaluating
Eq.~\eqref{eq:likelihood_beta_marg}.

The second required distribution is $P(\a\mid\m_{\nu},\beta)$. This is a
simple multi-variate Gaussian distribution in $\a$, for which there
are efficient samplers readily available (see, e.g., Appendix~A of
\citealp{bp01} for details). One particularly efficient sampling equation
is as follows:
\begin{align}
  \biggl(\sum_{\nu}\A_{\nu}^t\N_{\nu}^{-1}\A_{\nu}\biggr)\a = \sum_{\nu}\A_{\nu}^t\N_{\nu}^{-1}\m_{\nu} + \sum_{\nu}\A_{\nu}^t\N_{\nu}^{-1/2}\eta,
\label{eq:ampl_samp_wiener}
\end{align}
where $\eta$ is a vector of random Gaussian $N(0,1)$
variates. This equation may be solved efficiently using preconditioned
Conjugate Gradient methods \citep{shewchuk:1994}, as discussed by
\citet{seljebotn:2019}. 

These equations are integrated into the main \BP\ Gibbs sampling loop
according to the following steps. First, we run a short (typically a few hundred steps)
standard Metropolis sampler (see Appendix~A of \citealp{bp01}) for
each spectral parameter, using the product of
Eq.~\eqref{eq:likelihood_beta_marg} and any desired priors to define
the accept rate, that is, the relative number of Metropolis proposals
being accepted (which should preferably stay between 0.3 and 0.7
for an efficient sampler).
All data are smoothed to a common angular and pixel
resolution before evaluating the expression. Immediately following the
last Metropolis step, we draw one sample from $P(\a\mid\m_{\nu},\beta)$
using Eq.~\eqref{eq:ampl_samp_wiener}; it is critically important that
no other parameters are updated between $\beta$ and $\a$, as the
previous value of $\a$ is completely inconsistent with the new $\beta$
value, which is drawn marginally with respect to $\a$.

Returning to Fig.~\ref{fig:marginal_conditional_plot}, the improvement
achieved by this joint two-step sampler is illustrated as blue
distributions and points. Starting with the left panel, the
fundamental difference between the joint and Gibbs samplers is that
the first step in the $\beta$-direction is drawn from the full
marginal distribution (horizontal blue distribution) instead of the
conditional distribution (horizontal orange distribution). This is
much wider, and covers by construction the full width of the
underlying target distribution. One single proposal may therefore move
from one side of the distribution to the other, and there is no memory
of the previous parameter state. However, to obtain a valid sample, it is critically important to draw a corresponding sample from the
appropriate conditional amplitude distribution (vertical blue
distribution) immediately after the marginal move. Correspondingly,
the blue points in the right panel shows 100 samples drawn with the
joint sampler. In this case, they cover the full distribution very
efficiently.

\begin{figure}[t]
  \center
  \includegraphics[width=\linewidth]{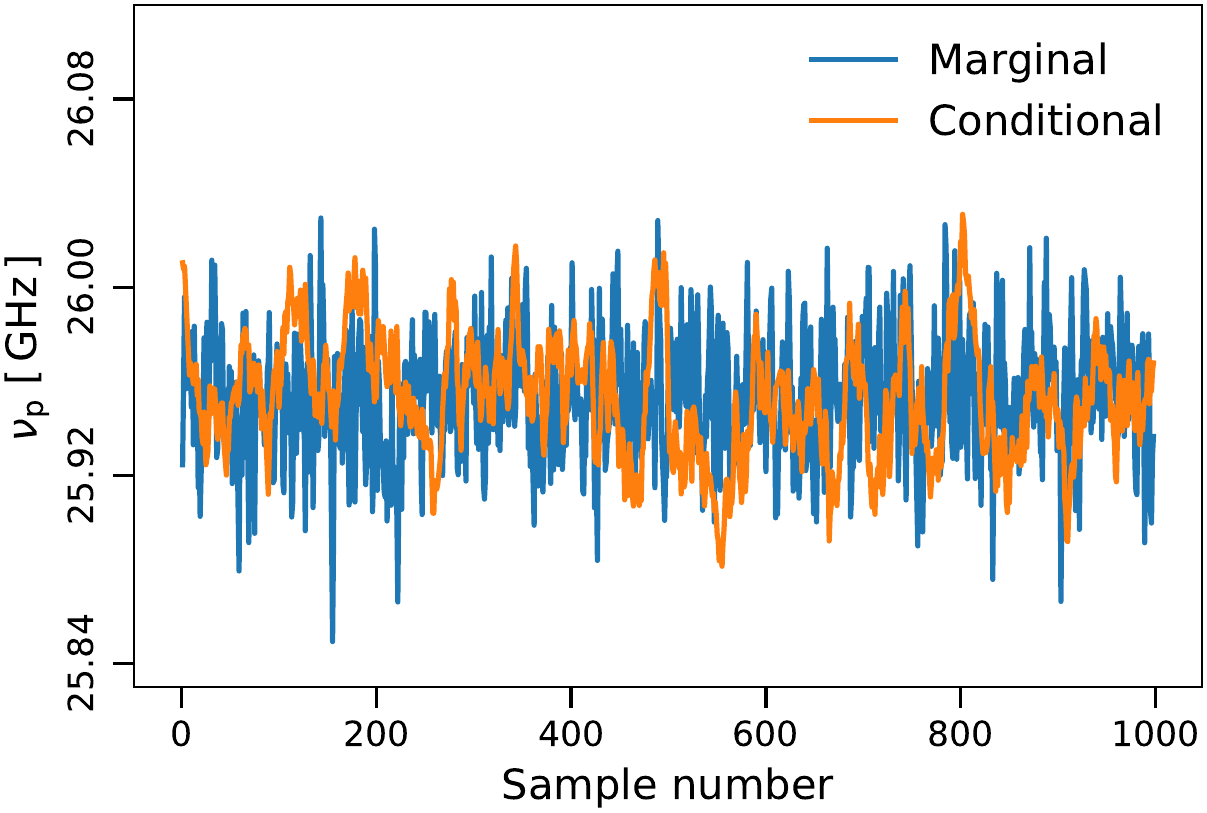}
  \caption{Comparison of AME $\nup$ chains derived using the conditional (orange; Eq.~\ref{eq:likelihood_full}) and marginal (blue; Eq.~\ref{eq:likelihood_beta_marg}) samplers discussed in the text. For the purposes of this illustration, all other parameters than $\a_{\mathrm{AME}}$ and $\nup$ are fixed.}
  \label{fig:combined_sampler_chains}
\end{figure}

Figure~\ref{fig:combined_sampler_chains} shows a similar comparison
for AME $\nup$ for a test chain that only explores the AME parameters
in the \BP\ data model with both methods. Also in this case, we see that the marginal
sampler explores the full range much more efficiently than the
conditional sampler. 

\subsection{Component-based monopole determination} 
\label{subsec:monopole_sampler}

Next, we considered the problem of monopole determination for CMB
experiments, which has long been one of the main challenges for
parametric component separation methods \citep[see, e.g.,][and
  references therein]{planck2014-a12, wehus2014}. The problem stems
from the following challenge: For traditional CMB experiments and
maximum likelihood mapmaking methods, there are no data-driven
constraints on the monopoles in the derived frequency sky maps. For
example, \WMAP\ is explicitly differential in nature, measuring only
differences between pairs of points, and therefore cannot by
construction constrain the zero-level. For \Planck, the high level of
$1/f$ noise prohibits any useful constraints on the zero-levels. An
important exception to this is \COBE/FIRAS \citep{mather:1994}, which
is absolutely calibrated; but its mK-level uncertainties are still
orders of magnitude too large to be useful for modern CMB component
separation purposes.

For this reason, several indirect methods have been established to
determine the frequency map monopoles based on the morphology of the
maps themselves. Four examples are mean subtraction in a small region
\citep{planck2013-p02b}; fitting a plane-parallel co-secant model
\citep{bennett2003b,planck2014-a03}; imposing foreground SED
consistency between neighboring frequencies \citep{wehus2014}; and
cross-correlation with external data sets with known zero-levels
\citep{planck2014-a09}. However, all of these methods have in common
the fact that they operate on the basis of frequency maps and are aimed to
determine the zero-level at a given frequency channel, before feeding
these into traditional component separation algorithms. In this study,
we have made the observation that it is, in fact, much simpler
to determine the monopoles of the component amplitude maps and
to then use these to deterministically set the frequency map monopoles
through the resulting sky model. The frequency map zero-levels have
thus no independent impact on any higher order analyses (most notably,
the spectral parameters), but simply adjust to whatever the
model dictates at any given moment.

\begin{figure}
  \center       
  \includegraphics[width=\linewidth]{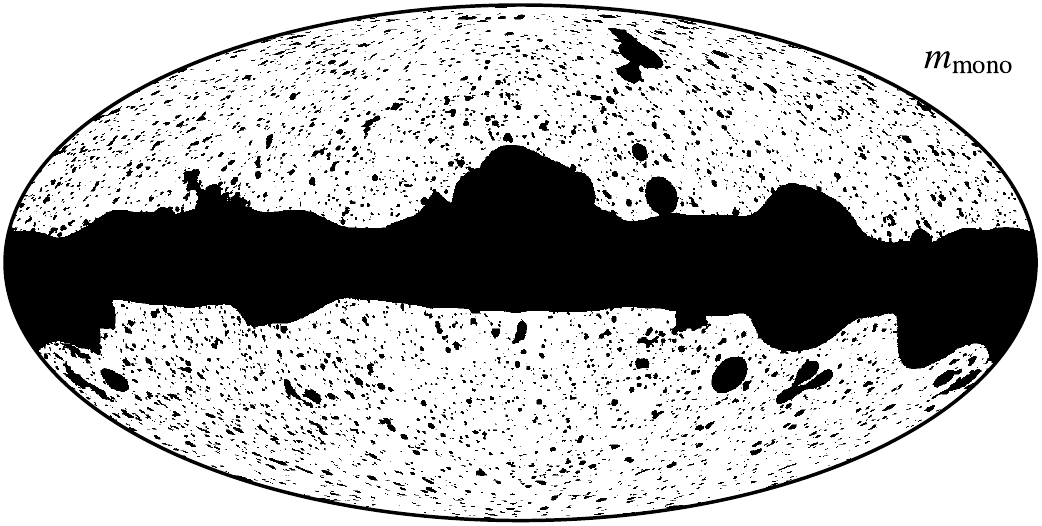}\\
  \includegraphics[width=\linewidth]{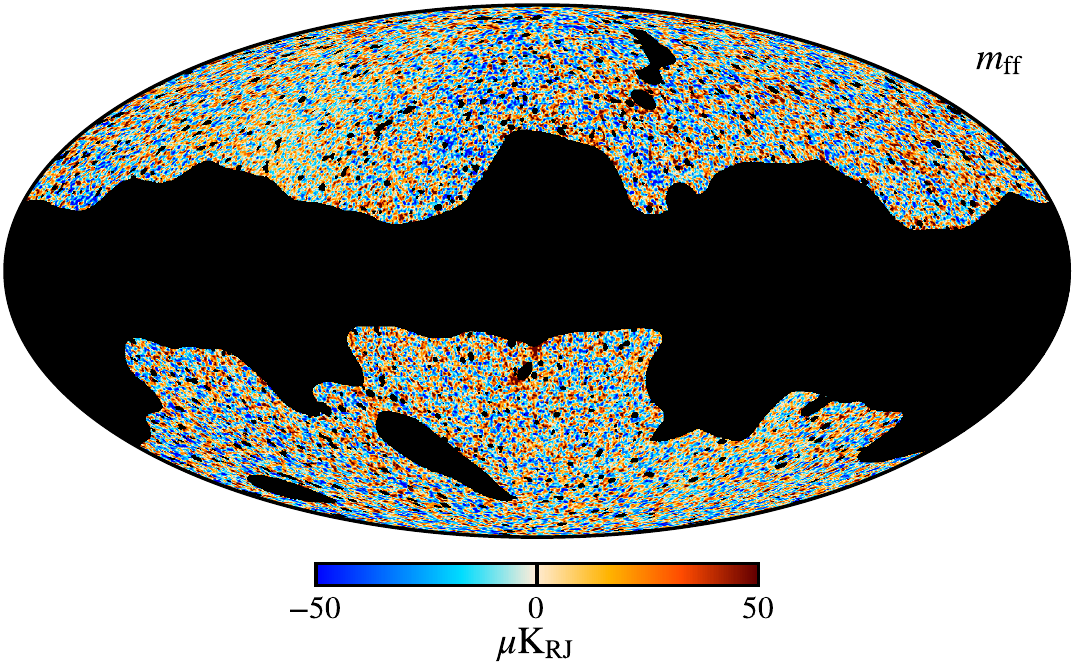}\\
  \caption{Masks used for signal amplitude zero-level and band monopole sampling:
    Frequency-band monopole and CMB amplitude zero-level (top) and
free-free amplitude zero-level (bottom), with a free-free amplitude
    sample at an angular resolution of 30\arcm\ FWHM plotted underneath.
  }
  \label{fig:monopole_masks}
\end{figure}

As a result, the\ question that immediately rises considers how we may, in fact, determine
the component monopoles. This must be done on a case-by-case basis,
applying the most natural prior for each component.
We note that any true monopole signal in the components that
do not agree with the chosen priors will end up in the frequency
map monopoles.
Starting with the CMB case, this can either be set to zero or
$2.7255\pm0.0006\,\mathrm{K}$ \citep{fixsen2009}, depending on whether
we want sky maps without or with the CMB monopole. In practice, we
additionally account for sub-optimal foreground modeling by applying a
mask. For the current analysis, we derived the CMB monopole mask from a
set of smoothed component amplitude maps, namely, by thresholding the
sum of synchrotron, AME, free-free and thermal dust emission, all
smoothed to 10\deg\ FWHM. In addition, we masked out radio sources and
any pixel with a reduced normalized \chisq\ higher than
$5\,\sigma$. The resulting mask is shown in
Fig.~\ref{fig:monopole_masks}, and has an accepted sky fraction of
$f_{\mathrm{sky}}=0.64$. The monopole of the CMB component map is set
to zero (or $2.7255\,\mathrm{K}$) outside this mask,
while simultaneously fitting for (but not
modifying) the dipole component.

Regarding the free-free component, \citet{planck2014-a12} found that
the measured emission is strongly noise-dominated over large areas of
the sky, with no detectable amplitude. Also, in this case, we therefore
set the monopole to have zero mean outside a conservative mask.  In
this case, the mask is derived from the free-free amplitude map
itself, evaluated at 30\GHz\ and smoothed to 10\deg\ FWHM, and
truncated at 5\muKRJ. In addition, we excluded areas where the other
foreground signals are high to account for signal-leakage into the
free-free component, similarly to the CMB mask, but while thresholding the sum
of all other components evaluated at 44\GHz. The resulting mask is
shown in the bottom panel of Fig.~\ref{fig:monopole_masks},
accepting $f_{\mathrm{sky}}=0.50$. We do note that any true unmasked
free-free signal is by definition positive and this can bias the
monopole also in the noise-dominated regime. Future works should aim to
correct for this bias by directly estimating the residual free-free
monopole in the unmasked region. We do note, however, that this
bias will decrease as more high sensitivity data become available, as
more and more of the free-free signal may be masked directly. 

In contrast, synchrotron emission as observed by the Haslam 408\,MHz
map is highly diffuse on the sky and there are no regions on the sky
that can be assumed to be approximately clean of synchrotron emission,
namely, exhibiting no synchrotron signal.
In this case, the best estimates of the synchrotron amplitude are those
already estimated for the Haslam 408\,MHz map itself. In this paper,
we adopt the zero-level correction of $8.9\pm 1.3\KCMB$ derived by
\citet{wehus2014}. We marginalized over the uncertainty by drawing a
random offset correction in every Gibbs iteration, as defined by
a Gaussian distribution with the quoted mean and standard deviation.

\begin{figure}
  \center       
  \includegraphics[width=\linewidth]{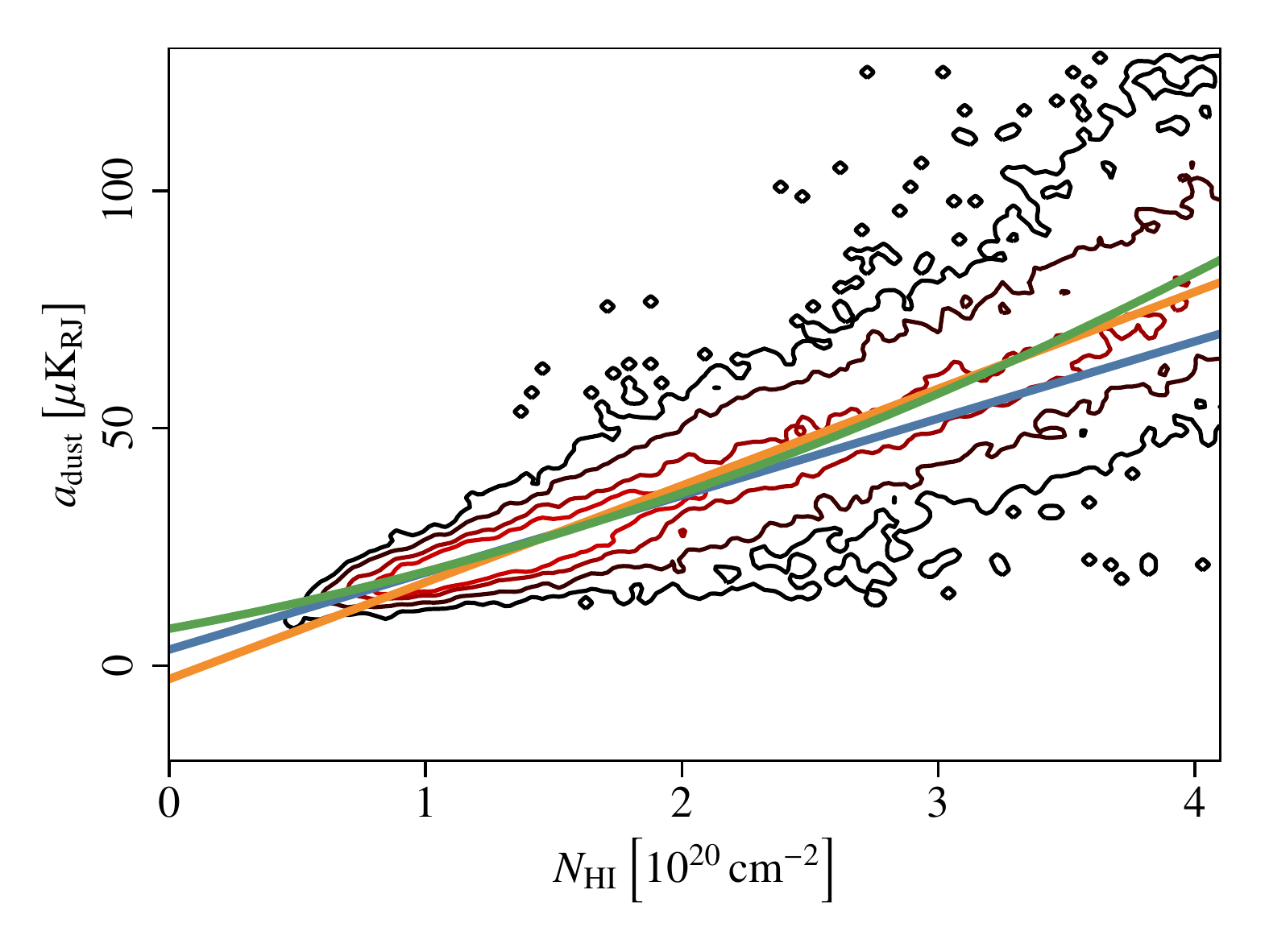}\\
  \includegraphics[width=\linewidth]{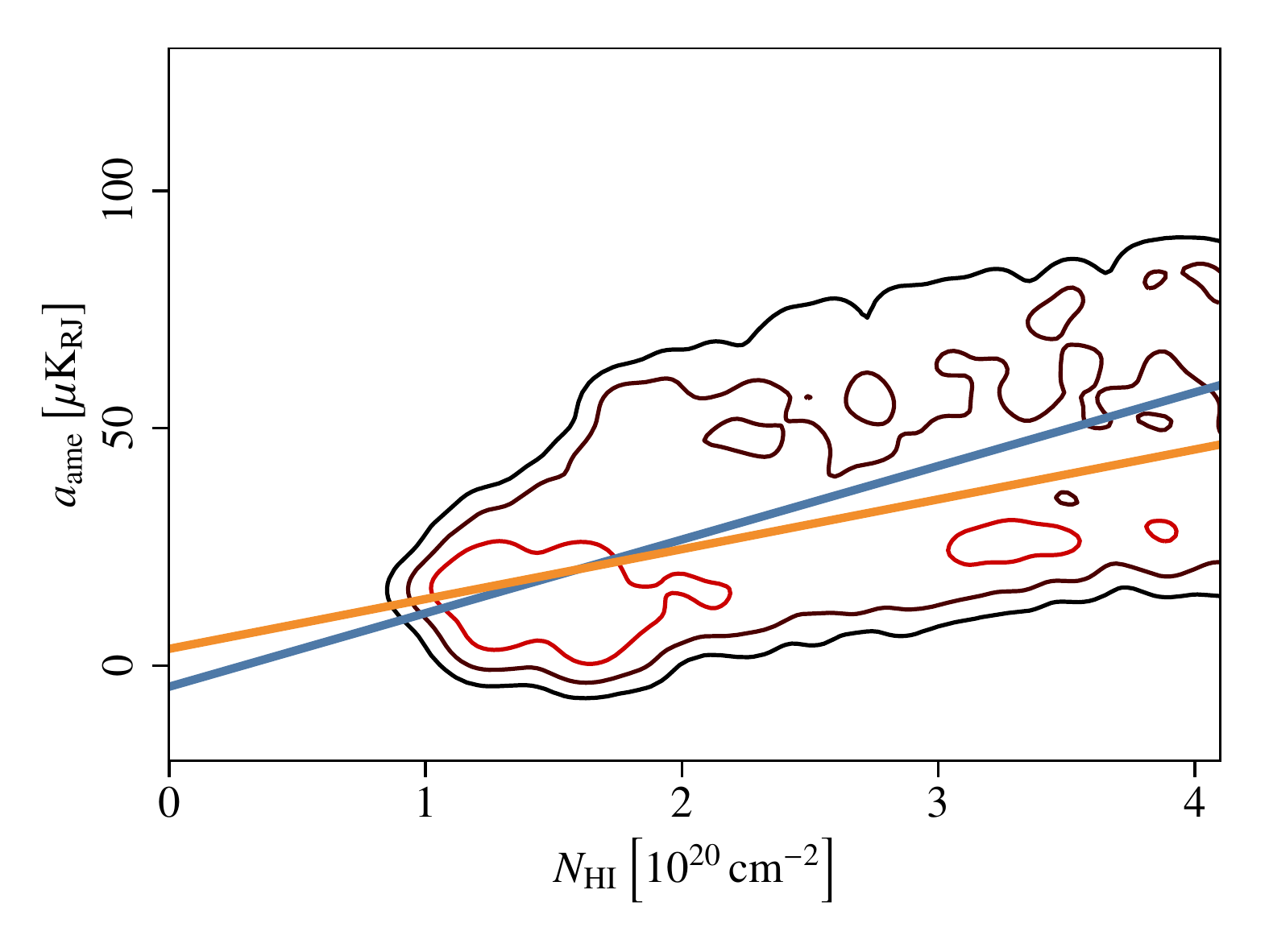}
  \caption{Cross-correlations between the H\,\textsc i column density
    $N_{\mathrm{H\,\textsc i}}$ and (top) the thermal dust amplitude and
    (bottom) the AME amplitude. The thermal dust
    cross-correlation is evaluated at \healpix\ resolution $\nside
    = 64$ and a common angular resolution of 60\arcm\ FWHM, while the
    AME cross-correlation is evaluated at \healpix\ resolution
    $\nside = 16$ and a common angular resolution of 10\deg\ FWHM. The
    lines represent best-fit lines for pixels with H\,\textsc i column densities
    less than $2\cdot 10^{20}\cmisq$ (blue) or $4\cdot
    10^{20}\cmisq$ (orange). The green curve is the best-fit
    second degree polynomial to pixels with H\,\textsc i column densities less
    than $4\cdot 10^{20}\cmisq$. 
    The contour lines are plotted at 0.001, 0.01, 0.05, and 0.1 $N_{\mathrm{pix}}/(\,\mu\mathrm{K_{RJ}}\; 10^{20}\,\mathrm{cm}^{-2}\; \nside^2\,)$; where only the lower three contour line values are plotted for AME. The contours have been smoothed for visualization.
     }
  \label{fig:nhi_corrplot}
\end{figure}

For the thermal dust emission, we adopted essentially the same approach as
the \Planck\ HFI DPC \citep[e.g., see][]{planck2016-l03}, setting the
zero-level through cross-correlation with H\,\textsc i column density
observations \citep[e.g., see][]{Lenz_et_al:2017}, although with a few
minor variations. First, we applied this method to the thermal dust
component map, as opposed to individual frequency maps. Second, for
\Planck\ DR4 \citep{npipe}, the HFI 545\,GHz zero-level was set through
a linear fit for pixels with $\NHI <
4\cdot10^{20}\,\mathrm{cm}^{-2}$. However, as the 545--\NHI\ scatter
plot appeared to be non-linear around values of $\NHI =
1.5\cdot10^{20}\,\mathrm{cm}^{-2}$, they also performed a second
degree fit. We show in Fig.~\ref{fig:nhi_corrplot} a similar scatter
plot between \NHI\ and the thermal dust amplitude from one of our
Gibbs samples, where we performed both a first and second degree
polynomial fit to the plot at $\NHI <
4\cdot10^{20}\,\mathrm{cm}^{-2}$. Furthermore, we also performed a
linear fit at $\NHI < 2\cdot 10^{20}\,\mathrm{cm}^{-2}$ and we see
that the intersection of the linear fit with a lower threshold is
close to the intersection of the second degree fit. This raises the
question of the uncertainty of the threshold value for the linear fit,
as \citet{planck2011-7.12} found a good correlation up to at least
$\NHI = 2\cdot 10^{20}\,\mathrm{cm}^{-2}$. We therefore implement this
cross-correlation method as a prior on the thermal dust zero-level
using a range of thresholds, and for each Gibbs iteration we perform
linear fits of \NHI\ and the thermal dust amplitude with
\NHI\ threshold values ranging from 1.5 to 4
$[10^{20}\,\mathrm{cm}^{-2}]$ with increments of 0.5. Then we draw the
intersection value from a Gaussian distribution given the mean and
variance of the linear fits, which is subtracted from the dust
amplitude. This way, the uncertainty of the thermal dust amplitude
zero-level also propagates through the pipeline.

The zero-level of the AME component is determined using the same
procedure, noting that \citet{planck2014-a12} demonstrated a very
tight spatial correlation between AME and thermal dust emission on
large angular scales. The only difference with respect to the thermal
dust procedure is that we smooth all maps to a common angular
resolution of 10\deg\ FWHM and a \healpix\ resolution of $\nside=16$,
and adopt thresholds of 2 to 4 $[10^{20}\,\mathrm{cm}^{-2}]$, with
increments of 0.5; both the smoothing and the lower resolution
are imposed to reduce the impact of instrumental noise.
A scatter plot between AME and \NHI\ is shown in
Fig.~\ref{fig:nhi_corrplot} for one arbitrary Gibbs sample.

With the introduction of component-based monopole priors, all
frequency-band monopoles become free parameters and can be
deterministically fitted. Explicitly, for each frequency channel we
first subtract the predicted sky model as defined by
Eq.~\eqref{eq:mapmodel} and then fit and subtract the residual monopole
outside some mask. The mask should be defined such that it excludes
areas on the sky prone to foreground mismodeling, hence we adopt the
same mask as we do for the CMB monopole prior, shown in the top panel
of Fig.~\ref{fig:monopole_masks}. We note that this monopole
adjustment needs to be done immediately after any change in any of
the component maps, in order to not break the Gibbs chain, fully
analogously to the immediate amplitude update that must follow any
marginal spectral parameter move discussed in the previous section.

\subsection{Joint spectral parameter and frequency-band monopole sampling} 
\label{subsec:monopole_index}
Returning to the AME--H\,\textsc i cross-correlation plot in
Fig.~\ref{fig:nhi_corrplot}, we notice that the zero-level is
associated with a large statistical uncertainty. When sampling the AME
peak frequency, $\nup$, this uncertain monopole is also directly
affected by the resulting SED changes, and corresponding monopole
offsets are induced at all frequencies. If $\nup$ is sampled
conditionally with respect to the band monopoles, these will therefore
tend to pull \nup\ towards the old value and thereby increasing the
overall Markov chain correlation length.

This inefficiency may be alleviated by exploiting the new component-based
monopole sampler described in
Sect.~\ref{subsec:monopole_sampler}. Since all frequency-band
monopoles are now deterministically defined by the sky model, these
can be adjusted jointly whenever that is modified. We therefore
implement internal estimation of frequency band monopoles during the
spectral parameter sampling algorithm, such that for each proposed
spectral parameter value, we estimate a new band monopole value
conditioned on the input amplitude and the proposed parameter
value. The frequency band monopoles are updated together with the
final parameter at the end of the sampling.

\begin{figure}[t]
  \center
  \includegraphics[width=0.9\linewidth]{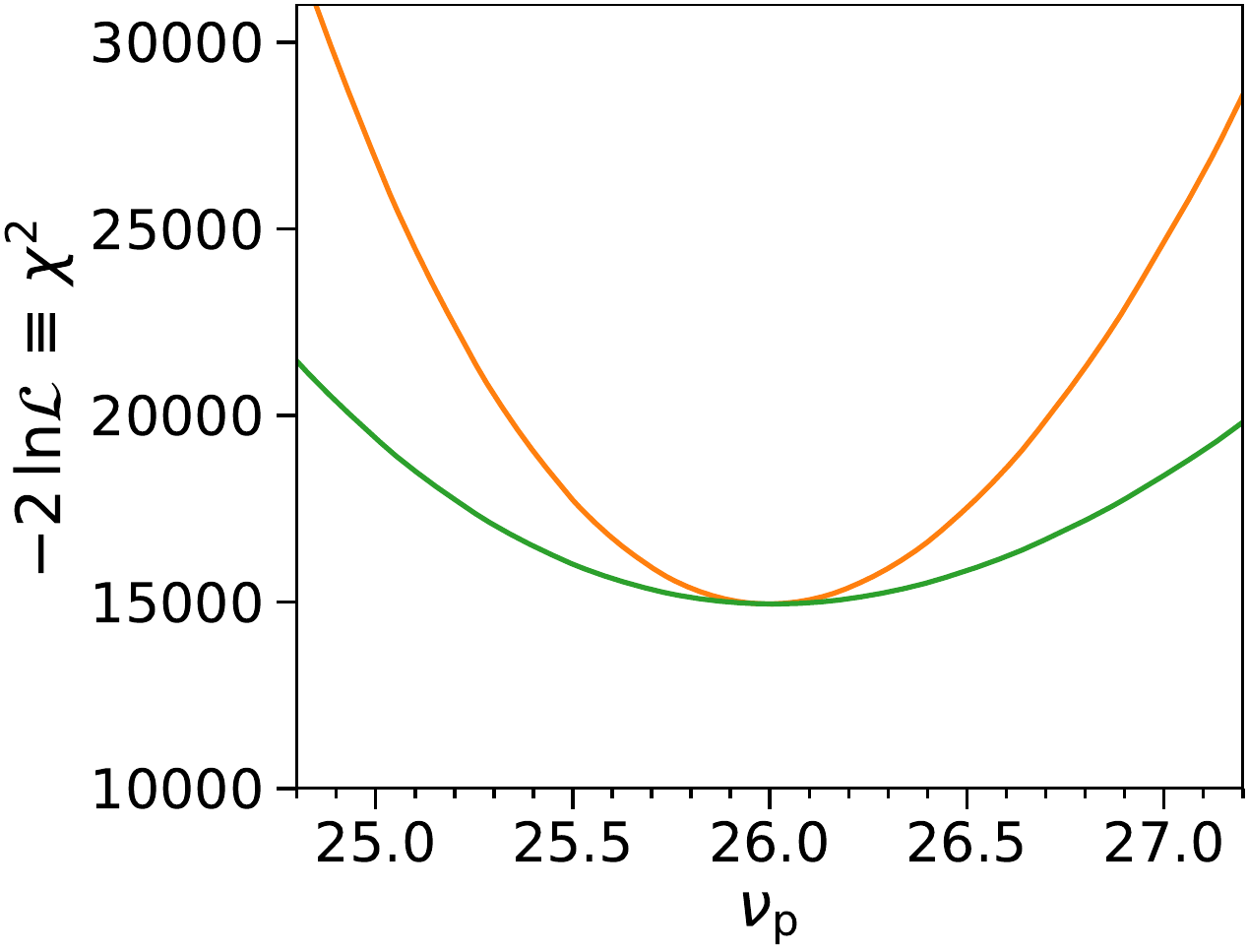}
  \caption{Comparison of AME $\nup$ $\chi^2$ distributions with
    (green) and without (orange) monopole marginalization. These
    distributions are evaluated using the marginal spectral parameter
    likelihood given in Eq.~\eqref{eq:likelihood_beta_marg}, but the
    same qualitative behaviour holds irrespective of which spectral
    parameter distribution is used (conditional, ridge, or marginal):
    the distribution becomes significantly wider when marginalizing
    over band monopoles.  }
  \label{fig:combined_sampler}
\end{figure}

To illustrate the usefulness of this combined sampling step, we
generate an idealized simulation that includes only AME signal and
noise at each frequency channel, with an angular resolution of
10\deg\ FWHM and a \healpix\ resolution of $\nside=16$. We choose an
input peak frequency of $\nup=26$\,GHz, and adopt a zero-level from
H\,\textsc i column density cross-correlation to
2\muKRJ. Figure~\ref{fig:combined_sampler} shows the resulting
log-likelihood (or $\chi^2$) distributions as evaluated from the
marginal definition in Eq.~\eqref{eq:likelihood_beta_marg}, for cases
both with (green curve) and without (orange curve) marginalizing over
the frequency band monopoles. We see that by marginalizing over the
band monopoles the log-likelihood function widens by a factor of
1.5--2. This translates into correspondingly longer Metropolis step
sizes in the spectral index sampling steps, and thereby faster
exploration of the full posterior distribution. The higher conditional
S/N a given component has, the more important this
effect will be.

\subsection{Breaking small-scale degeneracies through spatial priors} 
\label{subsec:amplitude_priors}

The final algorithmic improvement presented in this paper is the
introduction of informative spatial priors for foreground components,
either in the form of purely algorithmic smoothing power spectrum
priors or as actual informative Gaussian priors with a non-zero
mean. The first of these has already been used in the latest
\Planck\ analyses \citep{planck2016-l04,npipe}, but in the following,
we generalize the approach to non-zero cases and we 
systematically show how different choices affect the final results.

Mathematically speaking, the only difference between an informative
prior and a smoothing prior is whether a pre-existing mean map is
assumed for the astrophysical component in question (in which case the
prior is called ``informative'') or whether the prior mean is assumed
to be zero. Practically speaking, however, there is also an important
difference between the prior variances in the two cases, since for
informative priors the variance quantifies the allowed level of
fluctuations around the mean map; while for smoothing priors, it
quantifies the allowed level of fluctuations around zero. Thus,
for informative priors a prior variance of zero is fully acceptable,
in which case the output component map will be identical to the prior
mean; while for a smoothing prior the variance should be larger than
the actual component fluctuations in order to avoid
oversmoothing. 

We start by revisiting the sampling equation for the component
amplitude maps, as defined by Eq.~\eqref{eq:ampl_samp_wiener}. This
equation provides a sample from a posterior defined only by the
likelihood itself. If we additionally want to impose a Gaussian prior
on the amplitudes, as defined by a multi-variate Gaussian
distribution, $N(\mu,\S)$, then this is generalized to: 
\begin{equation}
  (\S^{-1} + \A^t\N^{-1}\A)\a = \A^t\N^{-1}\m_{\nu} + \S^{-1}\mu + \A^t\N^{-1/2}\vec{\eta}_1 +
    \S^{-1/2}\vec{\eta}_2.
  \label{eq:gauss_highdim_prior}
\end{equation}
We refer to Appendix~A in \citet{bp01} for an explicit derivation. In this
expression, $\mu$ has the same dimension as $\a$, and represents the
prior mean for $\a$, while $\S$ is an associated prior covariance
matrix that defines the ``strength'' of the prior, fully analogous to
the usual standard deviation, $\sigma$, of a Gaussian uni-variate
prior. Thus, if $\S = \tens 0$, the final solution for $\a$ will be
identical to $\mu$, while if $\S\rightarrow\infty$ (or, equivalently,
$\S^{-1} = \tens 0$), then the prior term vanishes, and one is left with the
original likelihood expression in Eq.~\eqref{eq:ampl_samp_wiener}. For
reference, we note that the previous \Planck\ analyses
\citep{planck2016-l04,npipe} set $\mu=0$ in this equation and only
used $\S$ to impose smoothness on $\a$.

Computationally speaking, introducing informative priors with non-zero
means in Eq.~\eqref{eq:gauss_highdim_prior} represents no additional
algorithmic complications compared to the prior-free case: the
equation is in both cases solved using the same preconditioned
conjugate gradient implementation. If anything, the equations are
actually a bit easier to solve with informative priors, as they reduce
degeneracies between different parameters, and thereby reduce the
condition number of the coefficient matrix on the left-hand side of
the equation. As pointed out by \citet{seljebotn:2019}, from an
algorithmic point of view an informative prior defined in terms
  of a mean map with a specified covariance may simply be considered
to be a new independent data set with sensitivity only to the
component in question and it therefore provides orthogonal
information with respect to the likelihood term contributions. Rather,
the main challenge regarding priors is how to define them in a useful
and controlled manner that does not significantly bias or contaminate
the final posterior distribution and this must be assessed on a
case-by-case basis.

In the current \BP\ analysis, we followed \citet{planck2016-l04} and
define $\S$ in harmonic space, giving different prior weights to
different angular scales. Explicitly, each component map is defined in
terms spherical harmonics:
\begin{equation}
  \a = \sum_{\ell,m} a_{\ell m} Y_{\ell m},
\end{equation}
and we define the prior covariance matrix as:
\begin{equation}
  \S = \S_{\ell m,\ell'm'} \equiv \left<a_{\ell m}a^*_{\ell' m'}\right> = P_{\ell}\delta_{\ell\ell'}\delta_{mm'},
\end{equation}
where $P_{\ell}$ is an angular prior power spectrum for $\a$, which,
again, is fully analogous to the standard deviation of a Gaussian
prior, but now defined per angular multipole.

\subsubsection{Algorithmic smoothing priors}

Starting with the algorithmic smoothing priors adopted by
\Planck\ 2018, we note that if we define $P_{\ell}^{\a}$ to be an
estimate of the true angular power spectrum of $\a$, then the
following power spectrum prior, 
\begin{equation}
  P_{\ell} = P^{\a}_{\ell}e^{-\ell(\ell+1)\sigma^2},
  \label{eq:gauss_smooth_prior}
\end{equation}
simply represents a Gaussian smoothing prior of $\a$ with a smoothing
kernel width equal to $\sigma$, which often is defined in terms of
$\theta_\mathrm{FWHM}=\sqrt{8\ln 2}\sigma$. We call this an
algorithmic smoothing prior, as it explicitly pushes the solution to
be smooth on small angular scales. We also note that $P_{\ell}^{\a}$
does not have to be an accurate estimate of the true component power
spectrum, but it should in general be greater than the true
spectrum, in order to prevent the prior from being overly
constraining. This is again fully analogous to choosing a prior width
that is wider than the expected target distribution in standard
univariate analysis problems. Thus, simply setting $P_{\ell}^{\a}$ to
a constant that is a few times larger than the expected spectrum is
usually a perfectly good choice.

\begin{figure}
  \center
  \includegraphics[width=0.49\linewidth]{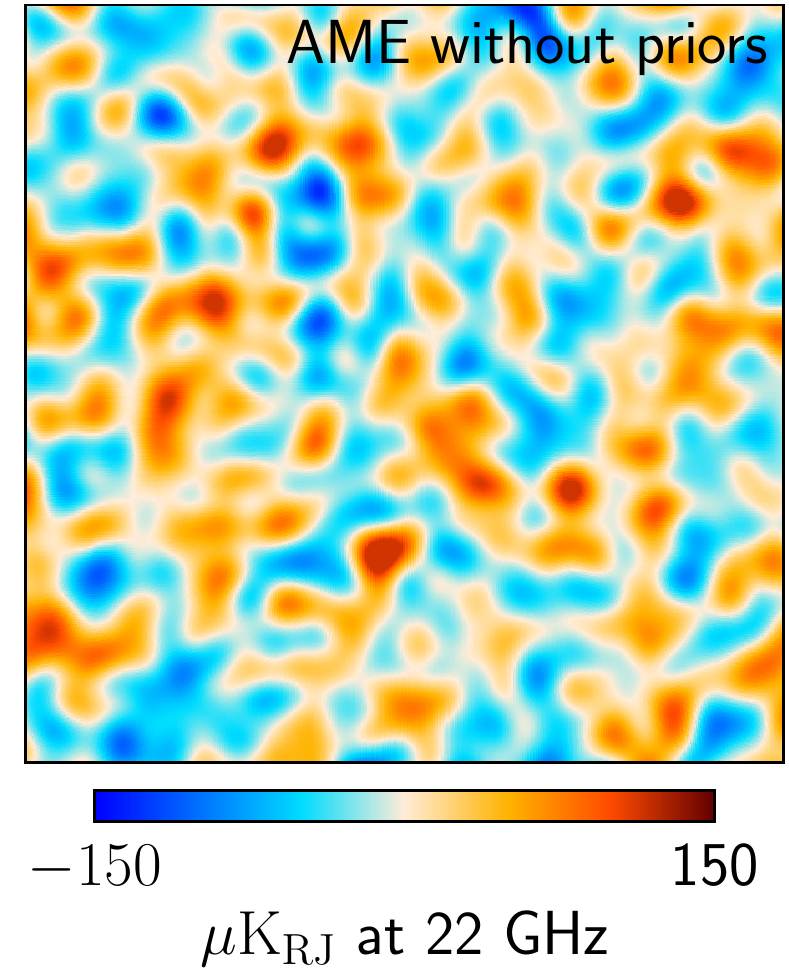}
  \includegraphics[width=0.49\linewidth]{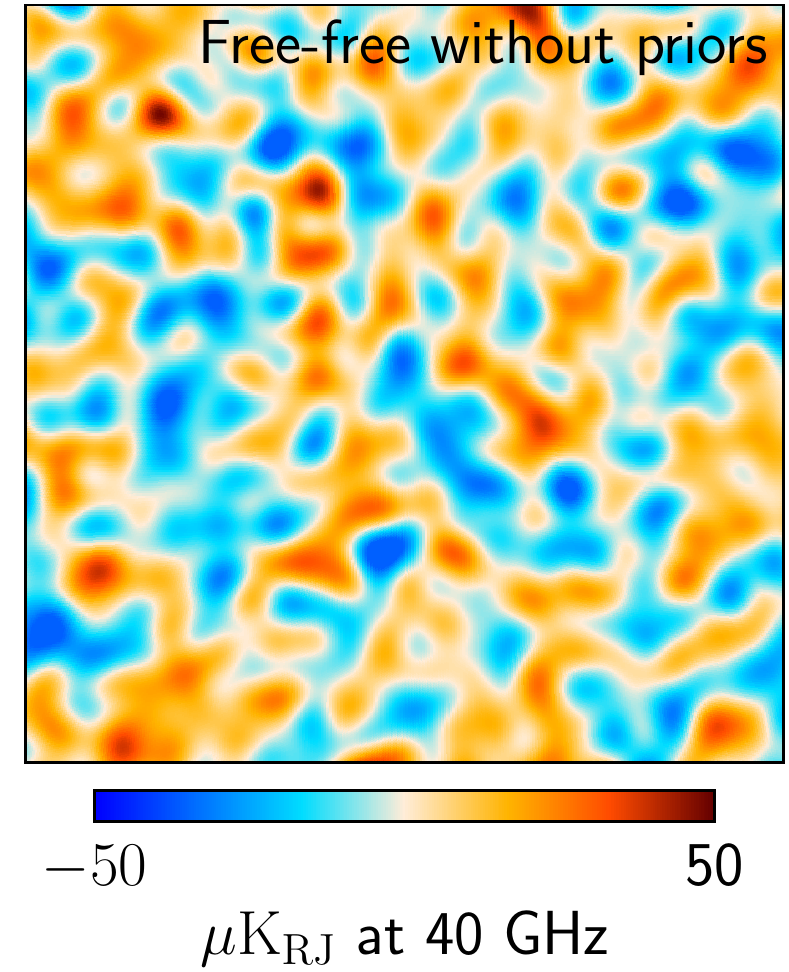}
  \caption{Comparison of AME (left) and free-free
    (right) amplitude maps derived without any spatial priors
    in a $20\times20\deg$ field centered on the Galactic South Pole,
    $(l,b)=(0\deg,-90\deg)$ at an angular resolution of
    60\arcm\ FWHM. We note the striking anti-correlation between the two maps.}
  \label{fig:amp_priors_gnom_60arcmin}
\end{figure}

\begin{figure}
  \center
  \includegraphics[width=\linewidth]{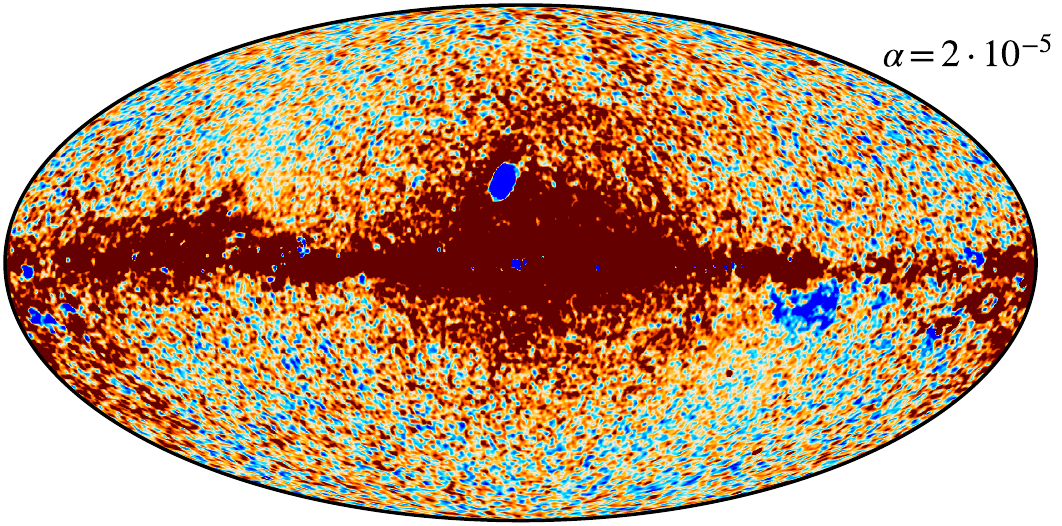}
  \includegraphics[width=\linewidth]{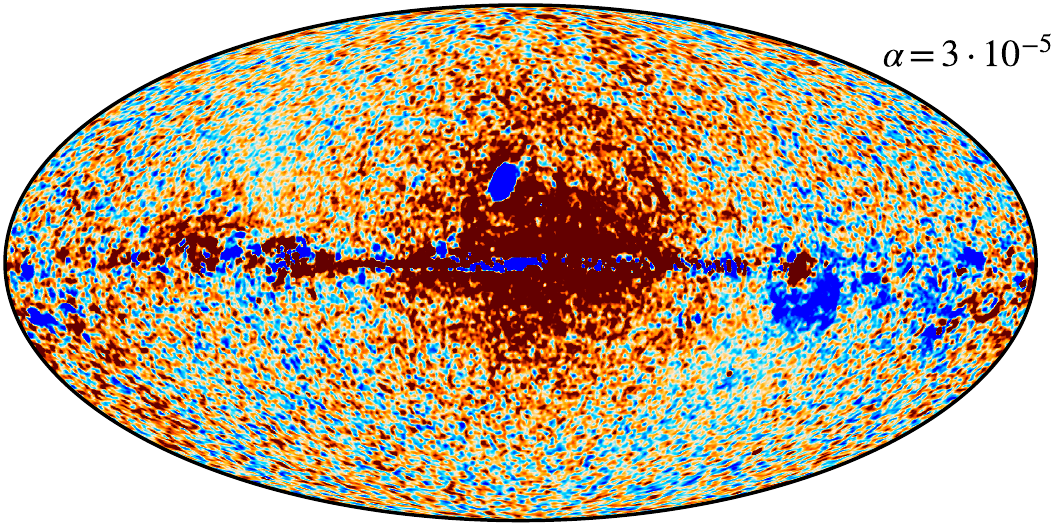}
  \includegraphics[width=\linewidth]{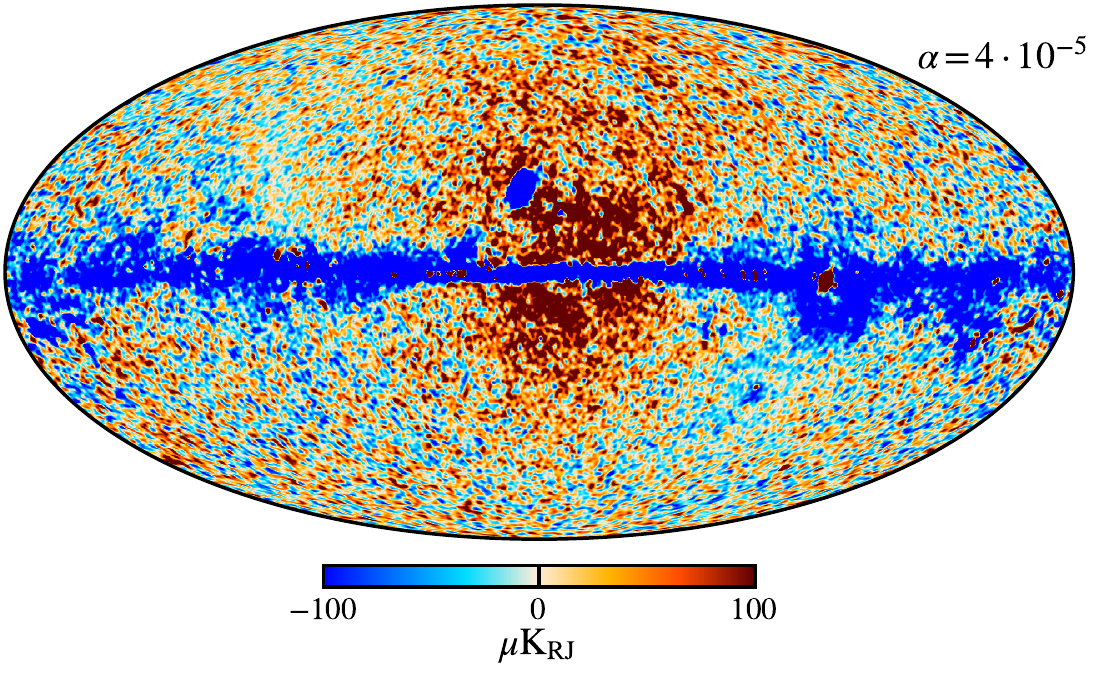}
  \caption{Difference maps between the derived amplitude and prior
    maps for AME for three different 857\,GHz scaling factors. From
    top to bottom, the three panels show scaling factors of
    $\alpha=2\cdot10^{-5}$, $\alpha=3\cdot10^{-5}$, and
    $\alpha=4\cdot10^{-5}$.}
  \label{fig:AME_prior_diff}
\end{figure}

\begin{figure*}
  \center
  \includegraphics[width=0.60\linewidth]{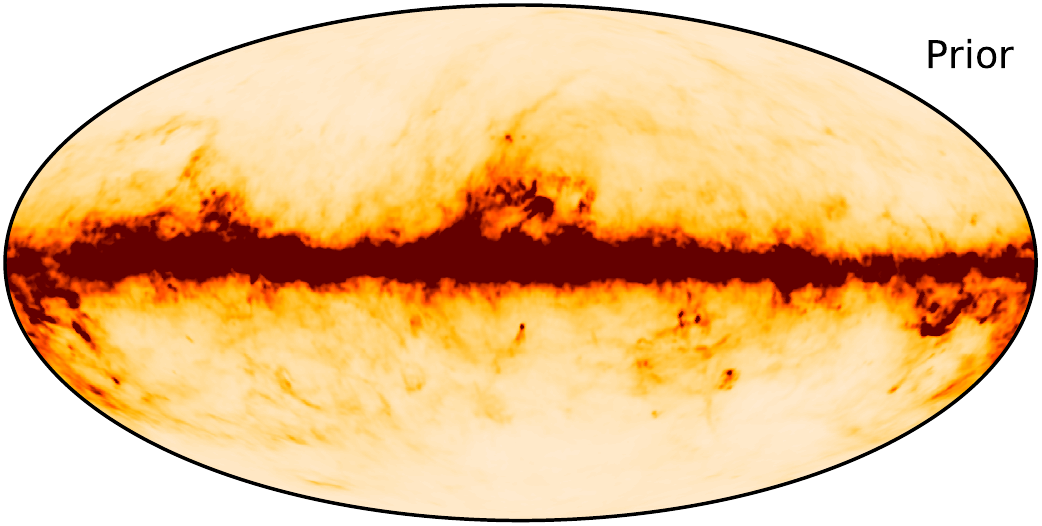}\\ 
  \includegraphics[width=0.49\linewidth]{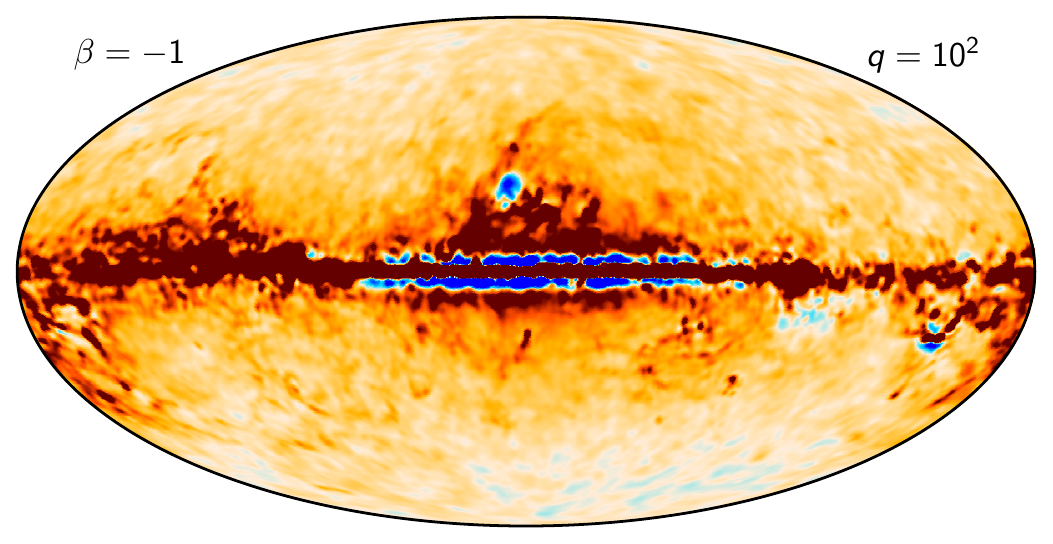}
  \includegraphics[width=0.49\linewidth]{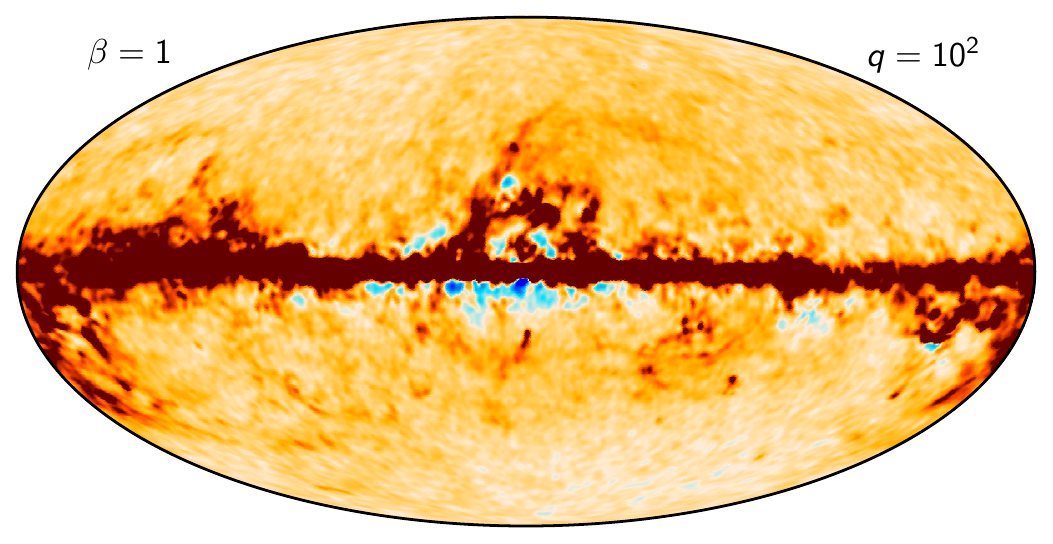}\\
  \includegraphics[width=0.49\linewidth]{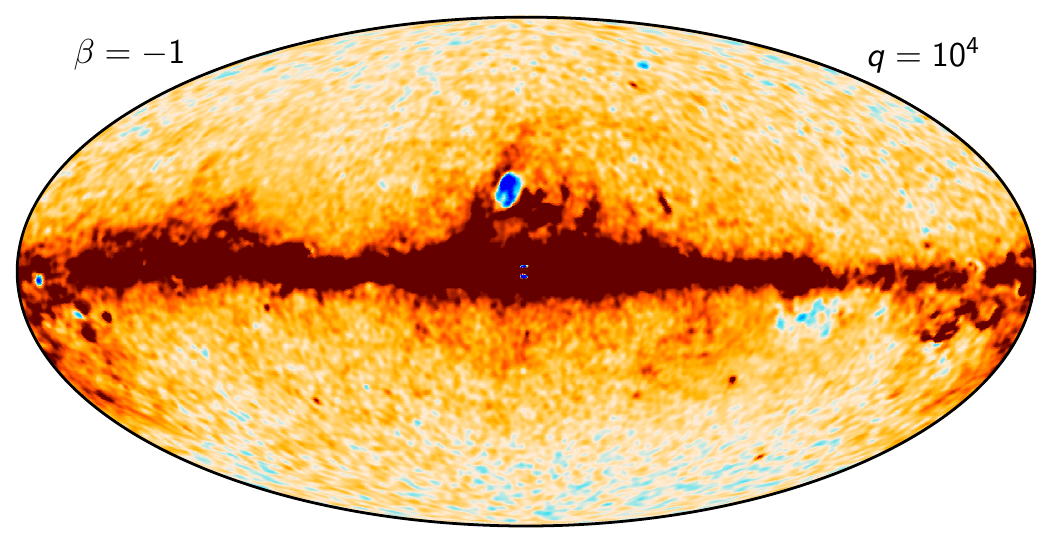}
  \includegraphics[width=0.49\linewidth]{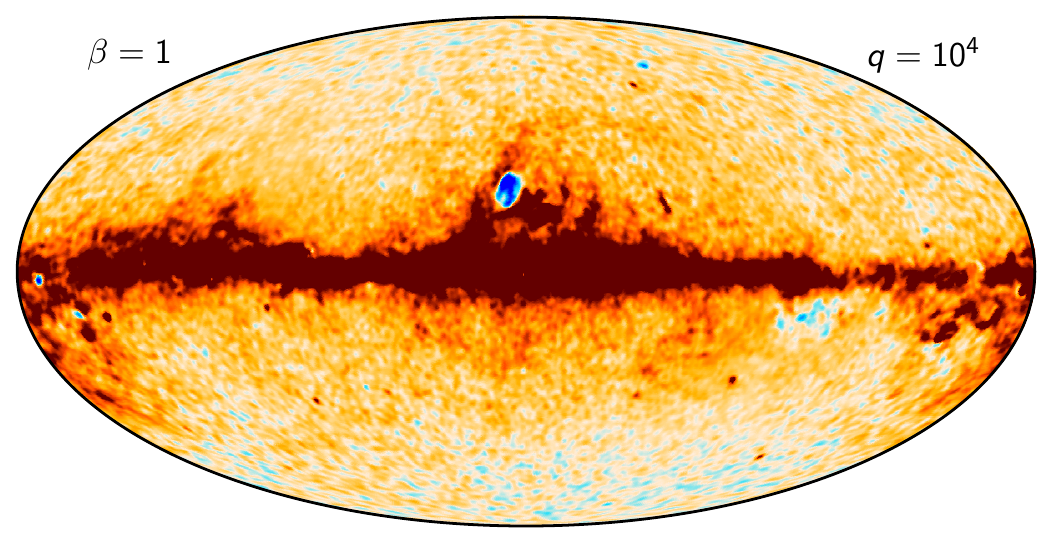}\\
  \includegraphics[width=0.4\linewidth]{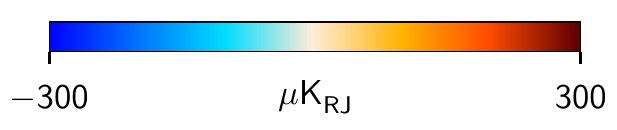}  
  \caption{Effects of the spatial prior on the sampled AME amplitude. AME amplitude prior map, derived by scaling
    the \Planck\ DR4 857\,GHz by $\alpha=3\cdot10^{-5}$ and smoothing
    to $10\arcm$ FWHM (top). Derived AME amplitude maps for four different spatial prior combinations (bottom), $\hat{D}_{\mathrm{AME}}(\ell) = q\,(\ell/\ell_{0})^{\beta}$. Rows show results for $q=10^2\muK_{\mathrm{RJ}}^2$ and $10^4\muK_{\mathrm{RJ}}^2$, respectively, while columns show results for $\beta=-1$ and 1.  }
  \label{fig:ame_amp_priors}
\end{figure*}

\begin{figure}
  \center       
  \includegraphics[width=\linewidth]{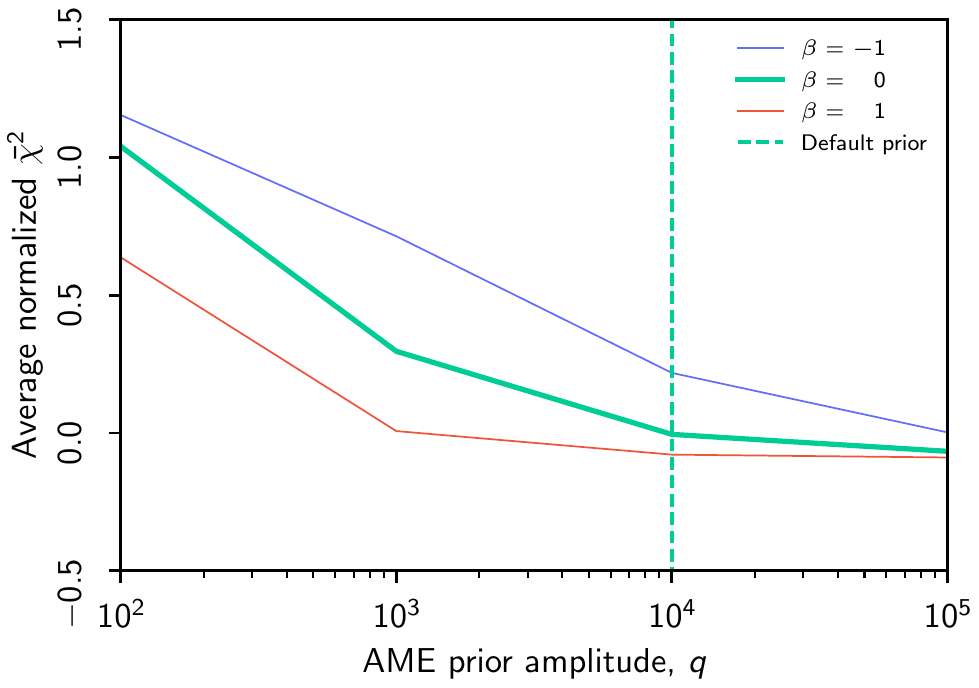}
  \includegraphics[width=\linewidth]{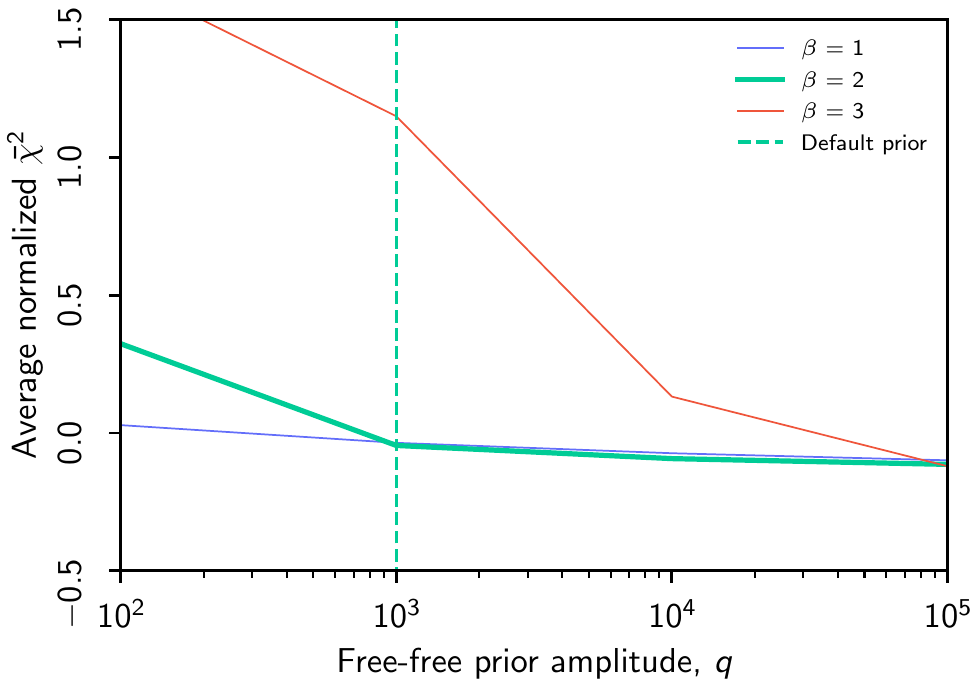}\\
  \caption{Average $\bar{\chi}^2$ per $N_{\mathrm{side}}=16$ pixel as a function of AME (top) and free-free (bottom panel) prior amplitude, $q$, where $\bar{\chi}^2\equiv (\chi^2-n_{\mathrm{dof}})/\sqrt{2n_\mathrm{dof}}$ and $n_{\mathrm{dof}}=15\,400$. Colored solid lines show results for different tilt parameters, $\beta$, and the intersection between the thick green curve and dashed line indicates the prior combination adopted for the main \BP\ analysis.}
  \label{fig:chisq_prior}
\end{figure}

It is important to note that Eq.~\eqref{eq:gauss_smooth_prior} is
indeed just a prior, and not a deterministic postprocessing smoothing
operator. This has both advantages and drawbacks that are important to
be aware of when using the products from the analysis. The main
advantage is that signal-dominated localized objects (for instance
point sources) are not excessively smoothed when applying the
smoothing as a prior. The main disadvantage is that the effective
angular resolution of the amplitude map becomes spatially varying and
depends on the local S/N in a given pixel; if the
S/N is high, the angular resolution will be
determined by the resolution of the data, while if the S/N is low, it is given by $\theta_{\mathrm{FWHM}}$. This is similar
to the GNILC method \citep{Remazeilles2011b}, which also implements
S/N dependent angular resolution. The main difference
between the two methods is that while GNILC requires regions of
different resolutions to be pre-defined, the current approach
automatically and dynamically adopts the resolution while solving
Eq.~\eqref{eq:gauss_highdim_prior}.

In the current \BP\ analysis, where the main scientific target is
CMB power spectrum and cosmological parameter estimation, we do not
impose any priors on the CMB component amplitudes, but we do impose a
Gaussian smoothing prior for thermal dust emission in intensity with
$\theta=5\arcm$ FWHM and $P_{\ell}^{\a} = 10^7\muK_{\mathrm{RJ}}^2$ at
545\,GHz. For studies that are primarily interested in the angular power
spectrum of thermal dust emission, it would be more useful to instead
impose a Lambda-Cold-Dark-Matter ($\Lambda$CDM) spectrum on
the CMB amplitude map, and no priors
on the thermal dust spectrum. Then the resulting thermal dust power
spectrum would be an unbiased estimator; with the current analysis,
the thermal dust spectrum will be biased low on small angular scales
due to the smoothing prior. We also impose a Gaussian smoothing prior
on synchrotron emission in intensity, with $\theta=60\arcm$ FWHM and
$P_{\ell}^{\a} = 3\cdot 10^{14}\muK_{\mathrm{RJ}}^2$ at 408\,MHz.

\subsubsection{Informative Gaussian spatial priors}

For AME and free-free emission, we adopt informative priors with $\mu
\ne 0$. The reason for this is simply that the limited data
combination considered in the current \BP\ analysis (see
Table~\ref{tab:data_survey_char}) is inadequate for constraining all
of AME, free-free, and CMB separately without additional information;
when future observations from, for instance, C-BASS \citep{king2010} and
QUIJOTE \citep{QUIJOTE_I_2015} become publicly available and
integrated in the analysis, this will hopefully no longer be
necessary. As an illustration of the problem,
Fig.~\ref{fig:amp_priors_gnom_60arcmin} compares the AME and free-free
amplitude maps derived without any priors near the Galactic South
Pole; even at a visual level, these two maps are nearly perfectly
anti-correlated, with no true constraining power on their own. This
also makes other components (most importantly the CMB) susceptible to
small systematic residual mismatches between the AME and free-free
components and it significantly increases the CMB noise.

Starting with the AME case, we first note that the prior-free
\Planck\ 2015 analysis \citep{planck2014-a12} found a very strong
spatial correlation between their AME and thermal dust component maps at an
angular resolution of $1^\circ$ FWHM. In general, a high degree of
correlation between these components is expected from current
theoretical AME models
\citep[e.g.,][]{erickson:1957,draine:1998,ali-haimoud:2010,
  silsbee:2011,Hensley2020}, although the specific correlation
coefficient depends on model details. Based on these observations, we
adopted the \Planck\ DR4 857\,GHz map\footnote{The \Planck\ DR4 857\,GHz
  map is corrected for both zodiacal light emission and a zero-level
  of $-0.657\,K_{\mathrm{CMB}}$, and smoothed to $10'$ FWHM, before
  adopted as an AME prior.} \citep{npipe} as a spatial mean template
for AME, that is, $\mu$ in Eq.~\eqref{eq:gauss_highdim_prior}.
However, before it can be inserted into
Eq.~\eqref{eq:gauss_smooth_prior}, it must be adjusted in amplitude to
account for the mean SED difference between thermal dust emission at
857\,GHz and AME at 22\,GHz. To do this, we solved
Eq.~\eqref{eq:gauss_smooth_prior} using the 857\,GHz map scaled by a
range of values between $\alpha=2\cdot10^{-5}$ and $4\cdot10^{-5}$ as
the AME prior. We then took the difference between the derived
amplitude map and the input prior and adopted the scaling factor for
which the difference is smallest as our default prior. Example
difference maps are shown in Fig.~\ref{fig:AME_prior_diff} and we
adopted $\alpha = 3\cdot10^{-5}$ as our default prior.

The final step is to define the strength of this prior, as given by
$P_{\ell}$. Ideally, we want the prior to be stronger (i.e., $\S$ to
be smaller) for the noise-dominated small angular scales and looser
for the signal-dominated large angular scales. To quantify these
considerations, we must define the following a power-law prior power spectrum for AME:
\begin{equation}
  P_{\ell}^{\mathrm{AME}} = q\left(\frac{\ell}{\ell_0}\right)^\beta,
  \label{eq:prior_AME}
\end{equation}
where $q$ is an overall amplitude at a pivot multipole, $\ell_0$, and
$\beta$ is a tilt parameter. A negative (positive) $\beta$ results in
a stronger (weaker) prior at high multipoles and a smaller (higher)
amplitude, $q,$ gives a stronger (weaker) prior on all angular
scales.

Figure~\ref{fig:ame_amp_priors} compares the resulting AME amplitude
maps for various choices of $q$ and $\beta$ (bottom panels) with the
prior mean map (top panel). Here we see that a high value of $q=10^4\muK_{\mathrm{RJ}}^2$
leads to very similar solutions for $\beta=-1$ and $\beta=1$,
indicating that the prior is largely irrelevant, and the solution is
data dominated. We also note that there is substantial instrumental
noise at high latitudes. For $q=10^2\muK_{\mathrm{RJ}}^2$, the maps are notably smoother
at high latitudes, but we also see clear smoothing artefacts near the
Galactic plane, corresponding to harmonic space ringing from the
Galactic plane. A value of $\beta=-1$ leads to sharper edges than
$\beta=1$. 

\begin{figure*}
  \center
  \includegraphics[width=0.60\linewidth]{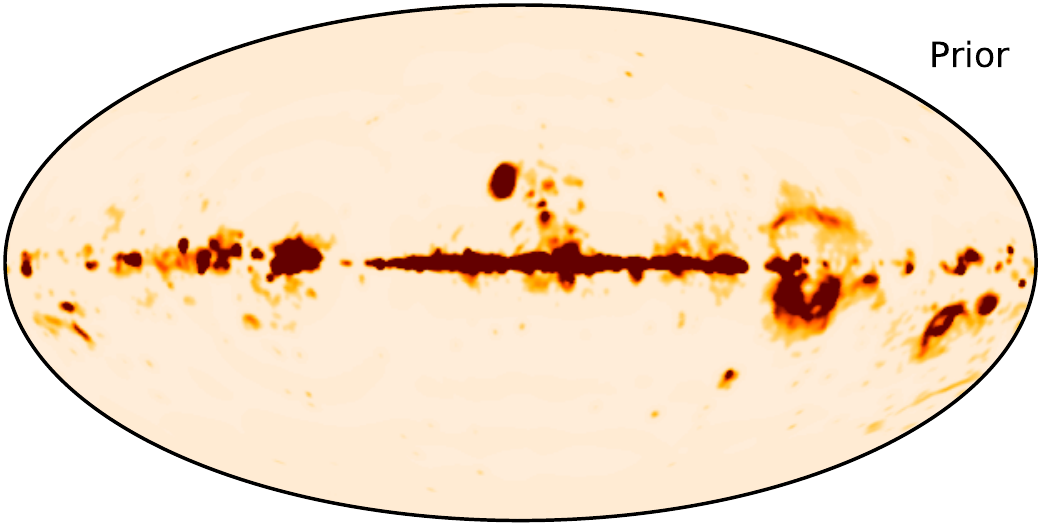}\\    
  \includegraphics[width=0.49\linewidth]{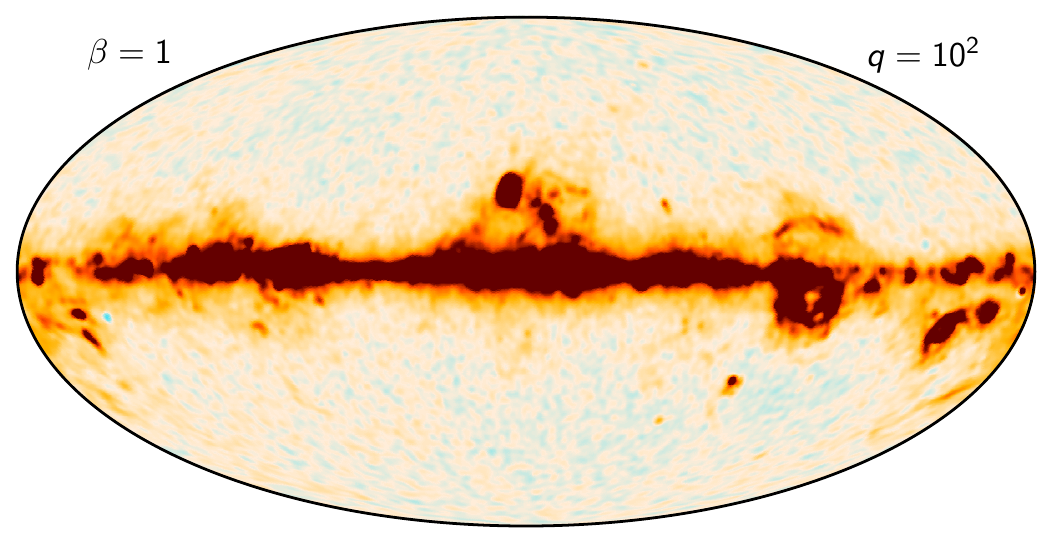}
  \includegraphics[width=0.49\linewidth]{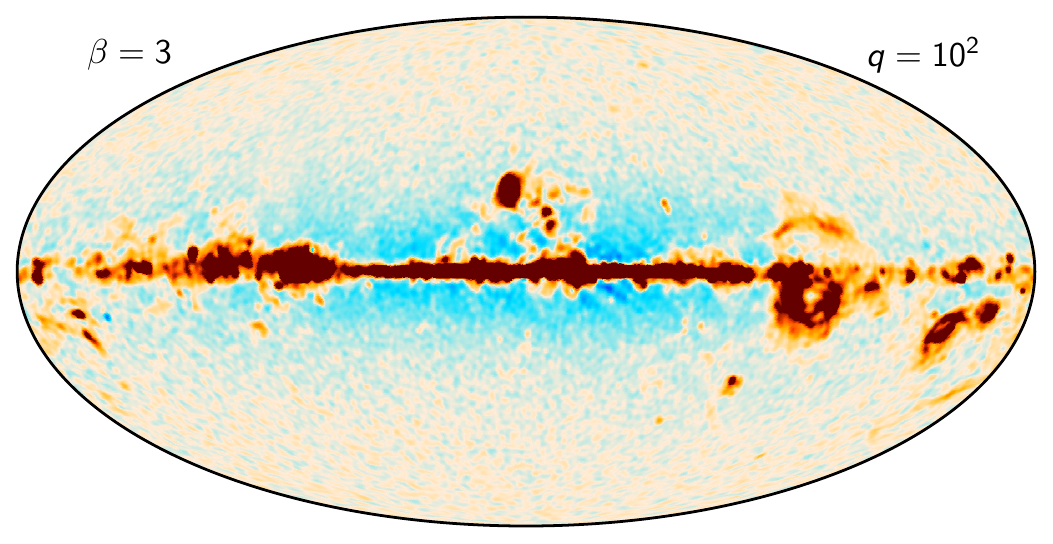}\\
  \includegraphics[width=0.49\linewidth]{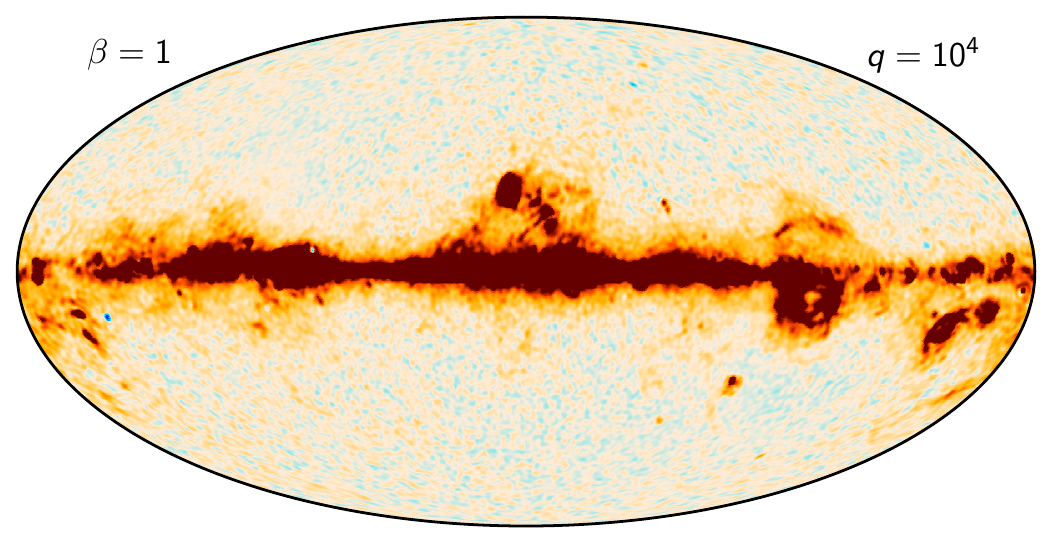}
  \includegraphics[width=0.49\linewidth]{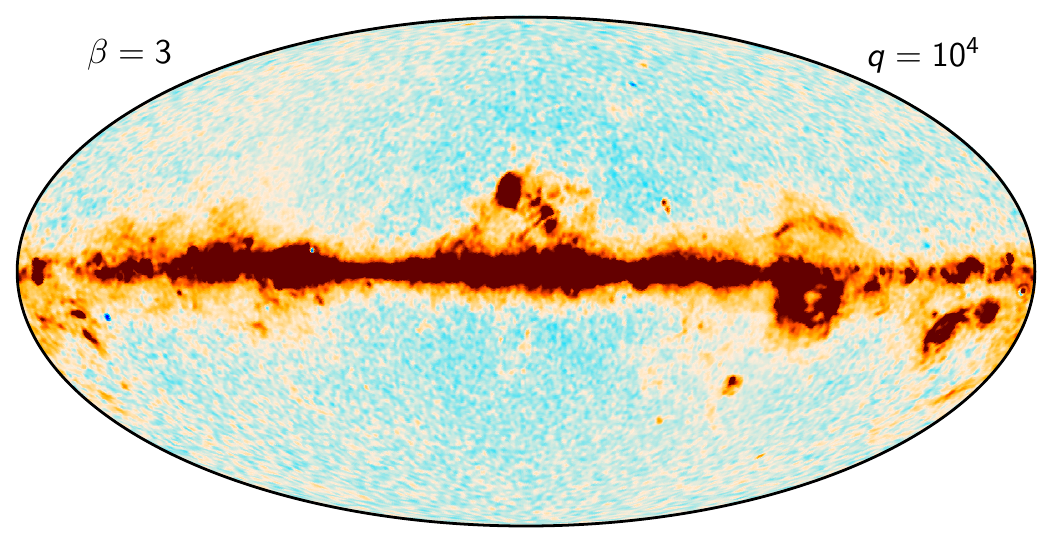}\\
  \includegraphics[width=0.4\linewidth]{figs/colourbar_300uK_RJ.pdf}
  \caption{Free-free amplitude prior map, adopted from
    the \Planck\ 2015 analysis which includes HFI observations (top). Derived free-free amplitude map for four different spatial prior combinations (bottom), $\hat{D}_{\mathrm{ff}}(\ell) = q\,(\ell/\ell_{0})^{\beta}$. Rows show results for $q=10^2\muK_{\mathrm{RJ}}^2$ and $10^4\muK_{\mathrm{RJ}}^2$, respectively, while columns show results for $\beta=1$ and 3.  }
  \label{fig:ff_amp_priors}
\end{figure*}

\begin{figure*}
  \center       
  \includegraphics[width=0.49\textwidth]{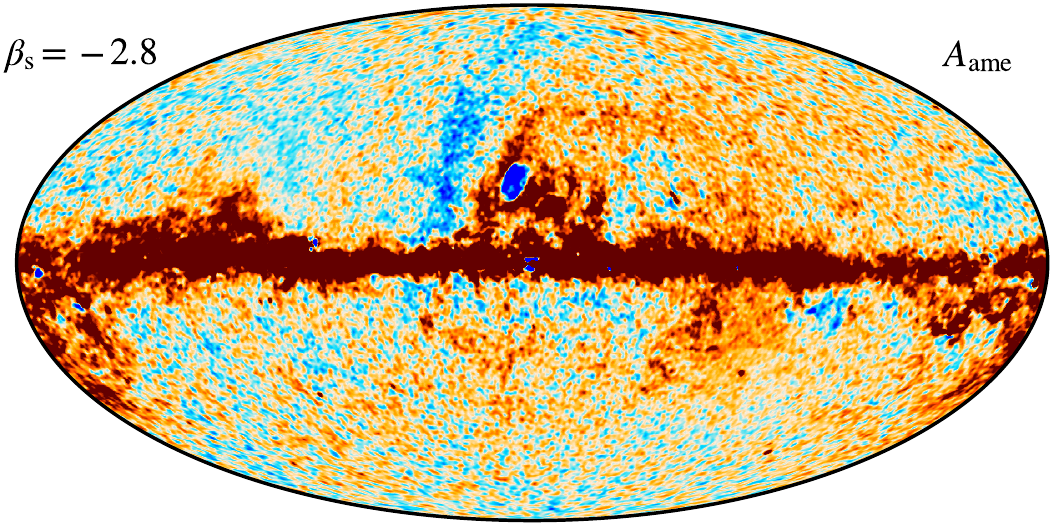}
  \includegraphics[width=0.49\textwidth]{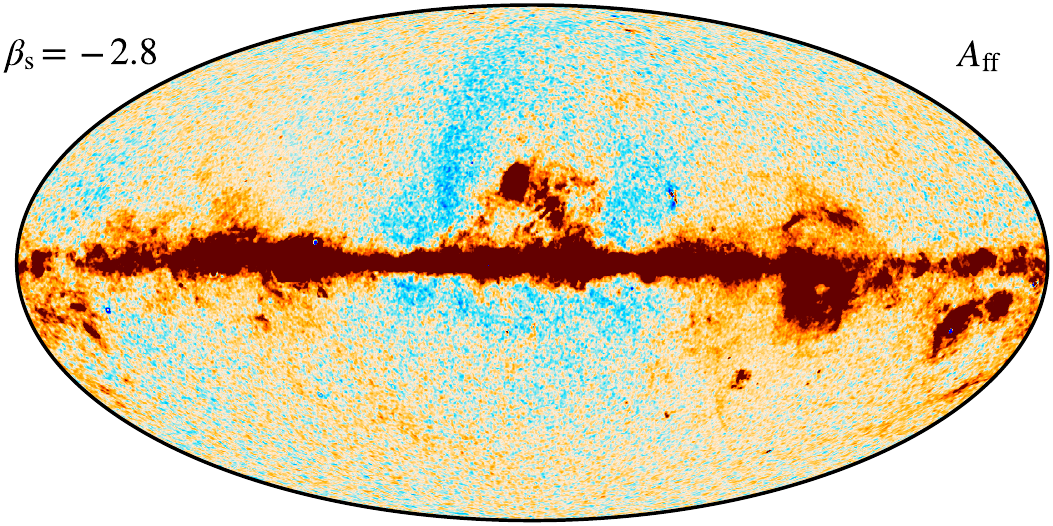}\\
  \includegraphics[width=0.49\textwidth]{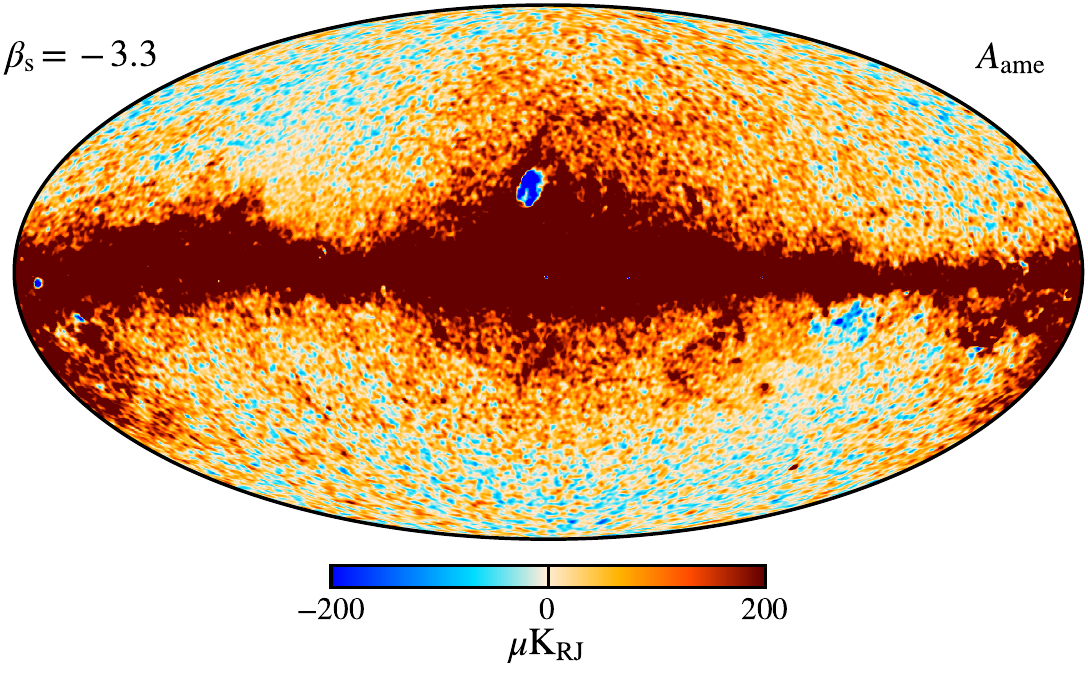}
  \includegraphics[width=0.49\textwidth]{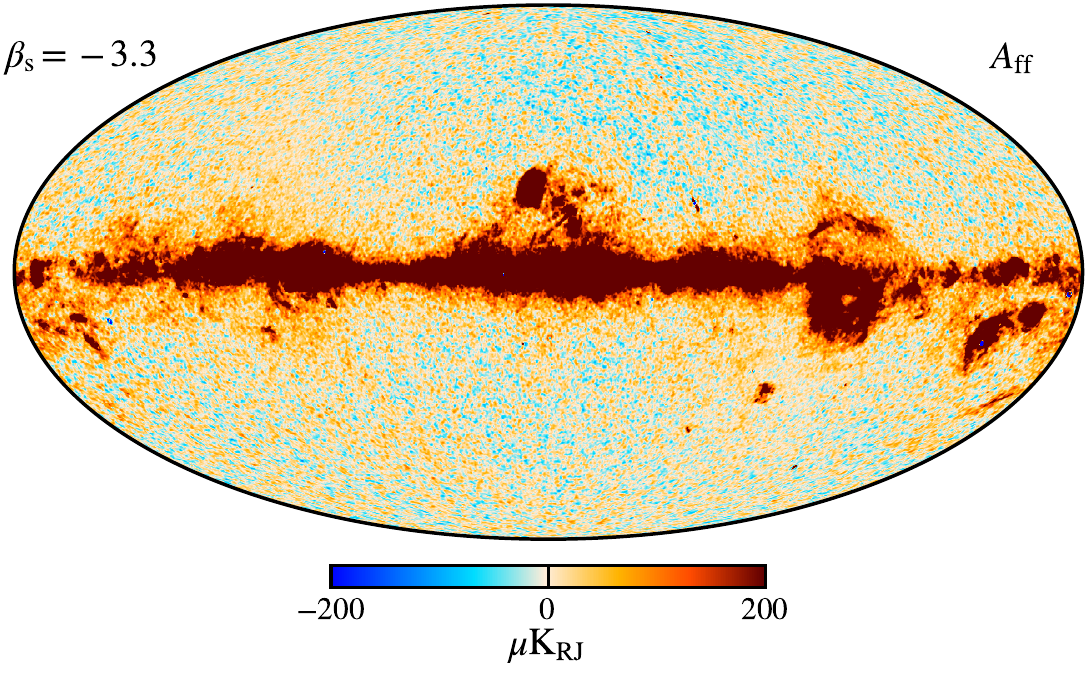}
  \caption{AME (left) and free-free (right) amplitude
    maps derived for two different synchrotron spectral indices. The
    top row shows results for $\beta_{\mathrm{s}}=-2.8$, while
    the bottom row shows results for $\beta_{\mathrm{s}}=-3.3$.
    In the top row, we see a negative synchrotron-like
    (see Fig.~\ref{fig:diff_mean}) imprint, which is not present in
    the bottom row.
  }
  \label{fig:AME_vs_synch_beta}
\end{figure*}

To actually determine the parameters used for the final prior, we
solve Eq.~\eqref{eq:gauss_smooth_prior} over a grid in $q$ and
$\beta$ and evaluate, as follows,
\begin{equation}
  \chi^2(q,\beta) \equiv \sum_{\nu,p}
  \left(\frac{m_{\nu,p}-s_{\nu,p}(q,\beta)}{\sigma_{\nu,p}}\right)^2
\end{equation}
for each configuration, where $s$ is the derived sky model in each
case. The results from this evaluation are shown in the top panel of
Fig.~\ref{fig:chisq_prior} in terms of the normalized reduced
$\bar{\chi}^2 \equiv (\chi^2-n_{\mathrm{dof}})/(2n_{\mathrm{dof}})$
where $n_{\mathrm{dof}}$ is the number of degrees of freedom; for a
perfect model fit and $n_{\mathrm{dof}}\gg 1$, this quantity should be
distributed approximately as a Gaussian distribution with zero mean
and unit standard deviation. For $q=10^2\muK_{\mathrm{RJ}}^2$ and $\beta=0$, we see that
$\bar{\chi}^2\approx 1$, which indicates a clear excess
residual. However, for $q=10^4\muK_{\mathrm{RJ}}^2$ and $\beta=0$, this excess is greatly
diminished, while there still is some effect from the prior. In the
following, we adopt this latter combination as our default AME prior. 

For free-free emission, there are no corresponding full-sky
independent spatial templates available in the
literature. Observations of H$\alpha$ \citep{finkbeiner:2003} or radio
recombination lines (RRL; \citealp{alves:2015}) might serve useful
roles, but both are associated with significant short-comings for the
purposes of the current analysis: the H$\alpha$ observations lack most
of the Galactic plane signal due to dust absorption, while the RRL
observations only cover a part of the Galactic plane. For now, the
best available full-sky free-free tracer is in fact the \Planck\ 2015
free-free map \citep{planck2014-a12}, which is based on the same data
set as studied in the current paper, but additionally (and
critically) the \Planck\ HFI observations as well. We therefore adopted
this map as a spatial prior in the current analysis, while recognizing
that this is strictly speaking not admissible in the Bayesian
framework; some data (i.e., LFI, \WMAP, and Haslam) are used twice to
constrain free-free emission and the resulting uncertainties will
therefore be underestimated. In practice, this solution is a way of
integrating HFI observations into the analysis without directly
affecting the CMB component. A critical goal for near-future work is
to integrate HFI observations directly into the analysis in the form
of frequency maps and at that point, this informative free-free prior will
be removed.

We adopted the same parametric function for free-free emission as for
AME, defined by Eq.~\eqref{eq:prior_AME} and adjusted the free
parameters in the same way. The results from this optimization are
shown in the bottom panel of Fig.~\ref{fig:chisq_prior} and we
adopted $q=10^3\muK_{\mathrm{RJ}}^2$ and $\beta=2$ in this case. A comparison of
different prior choices with the actual input prior map is shown in
Fig.~\ref{fig:ff_amp_priors}.

The only algorithmic difference with respect to AME is that we
additionally imposed a Gaussian smoothing for free-free emission, as
per Eq.~\eqref{eq:gauss_smooth_prior} with $\theta=30\arcm$ FWHM. This
is done to account for the fact that the distribution of free-free
emission is highly localized on the sky and therefore requires a high
maximum multipole moment to capture all significant structures; the
additional Gaussian smoothing ensures that no ringing emerges from the
high-$\ell$ truncation.

\section{Validation by simulations}
\label{sec:sim_results}

In order to validate the component separation
  implementations, we ran \commander\ on the simulated data as
  described in Sect.~\ref{subsec:sim_data}. In this section, we
  describe the efficacy of the algorithmic developments by comparing
  the input foreground sky maps with the resulting mean of an ensemble
  of 300 component separation samples as produced by \commander. As
  described in Sec.~\ref{subsec:regions}, the only spectral parameter
  which is sampled and constrained by data is the AME peak frequency,
  $\nu_\mathrm{p}$, which is determined as a full sky value.

For the amplitude maps, we defined a new variable to represent the relative difference between the input amplitude map and the results of running the simulation through the \commander\ component separation procedure. We define $\epsilon$ as the relative difference given by:
\begin{equation}
\epsilon_c = \frac{\a_c^\mathrm{in}-\a_c^\mathrm{out}}{\a_c^\mathrm{in}},
\label{eq:rel_diff}
\end{equation}
where, again, $c$ is an astrophysical sky component, $\a^{\mathrm{in}}$ is the input simulation amplitude map, and $\a^{\mathrm{out}}$ is the mean amplitude map of the component separation samples.

\begin{figure*}
\centering
\includegraphics[width=0.3\linewidth]{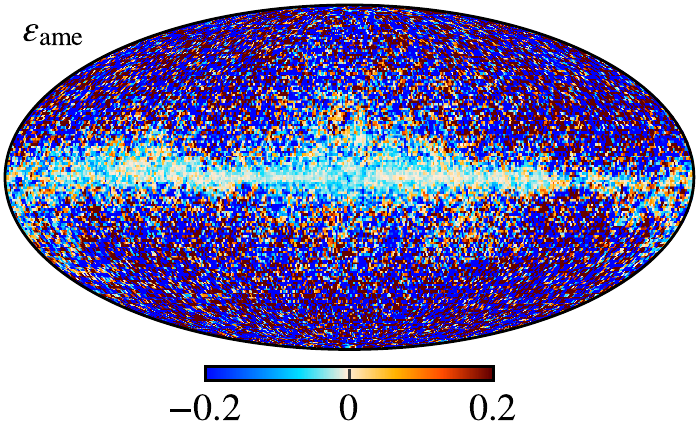}
\includegraphics[width=0.3\linewidth]{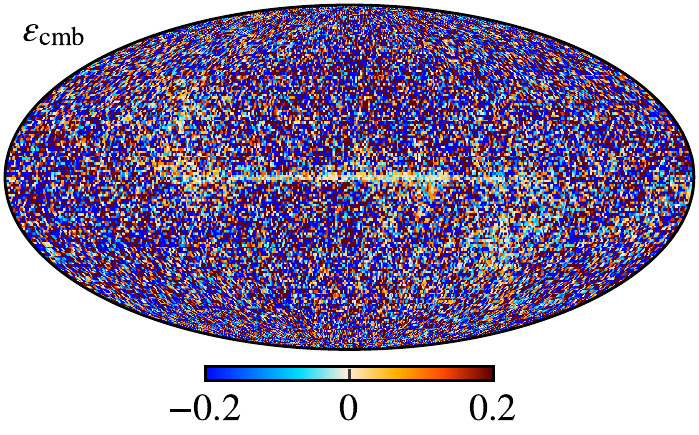}
\includegraphics[width=0.3\linewidth]{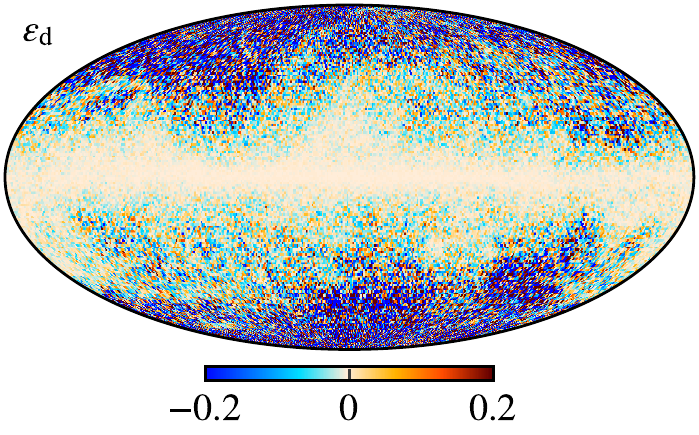}\\
\includegraphics[width=0.3\linewidth]{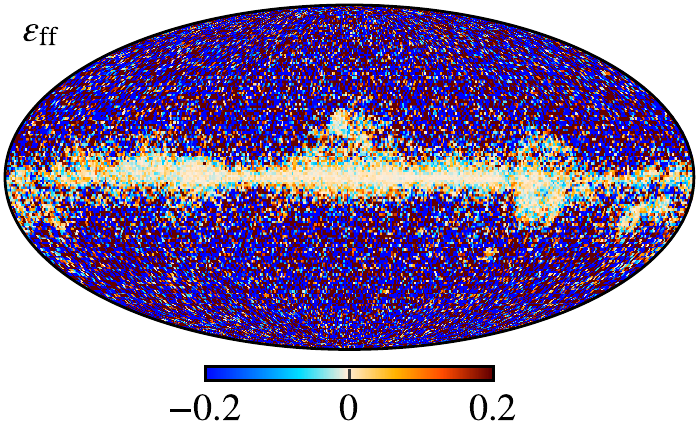}
\includegraphics[width=0.3\linewidth]{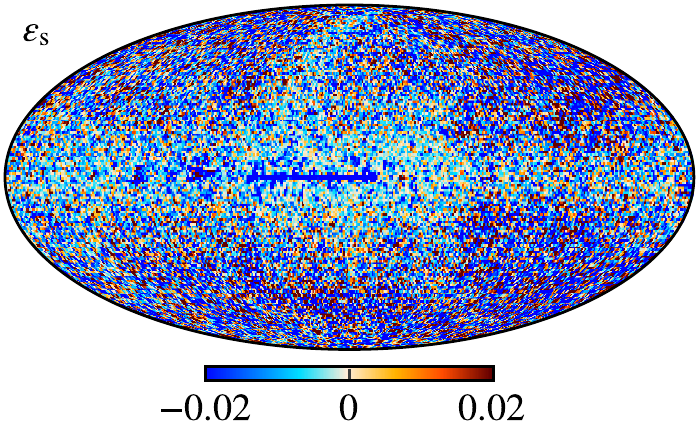}
\caption{Relative difference maps, $\epsilon_i$, for each of the sky components within the simulated \BP\ dataset. Top row: AME, CMB, and thermal dust maps. Bottom row: Free-free and synchrotron maps.}
\label{fig:wspec_results}
\end{figure*}

The results of the amplitude maps, represented in terms of the relative difference maps defined in Eq.~\ref{eq:rel_diff}, are summarized in Fig.~\ref{fig:wspec_results}. Inspecting the difference maps shows that there is good agreement with the input amplitude map, with departures from small relative differences in parts of the sky which are easy to explain. Notably, both the free-free and AME components show high levels of noisy departures off of the Galactic plane. Compared to the relative difference between two samples within component separation, the relative differences seen here are small. We see that in the high S/N observations along the Galactic plane, the agreement is excellent, with the outline of each components amplitudes clearly defined.

For the other three sky components, we see even better agreement. Unsurprisingly, the thermal dust amplitude map shows excellent agreement over the majority of the sky, though with deviations at the 10-20\% level in the low signal-to-noise regions of the sky. Finally, the synchrotron amplitude map shows very small differences in $\epsilon_{\rm s}$, though the imprint of the component is significantly less notable here than in the other components.

\begin{figure}
\includegraphics[width=\linewidth]{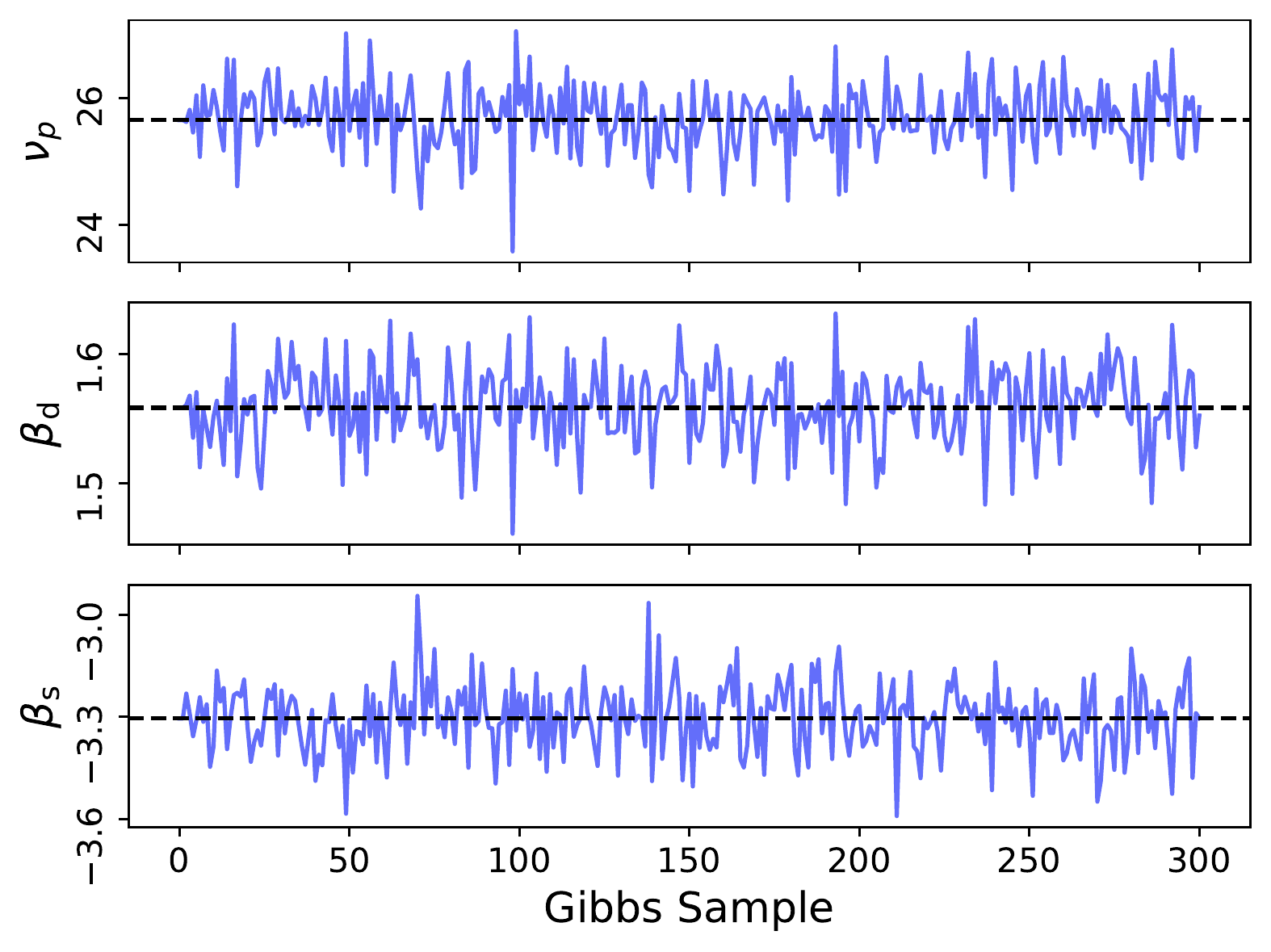}
\caption{Full sky spectral parameters as a function of sample (blue) for the controlled simulation. The nominal input value of the simulation is overlayed as the black dashed line.}
\label{fig:simulated_indices}
\end{figure}

\begin{figure*}
\includegraphics[width=\linewidth]{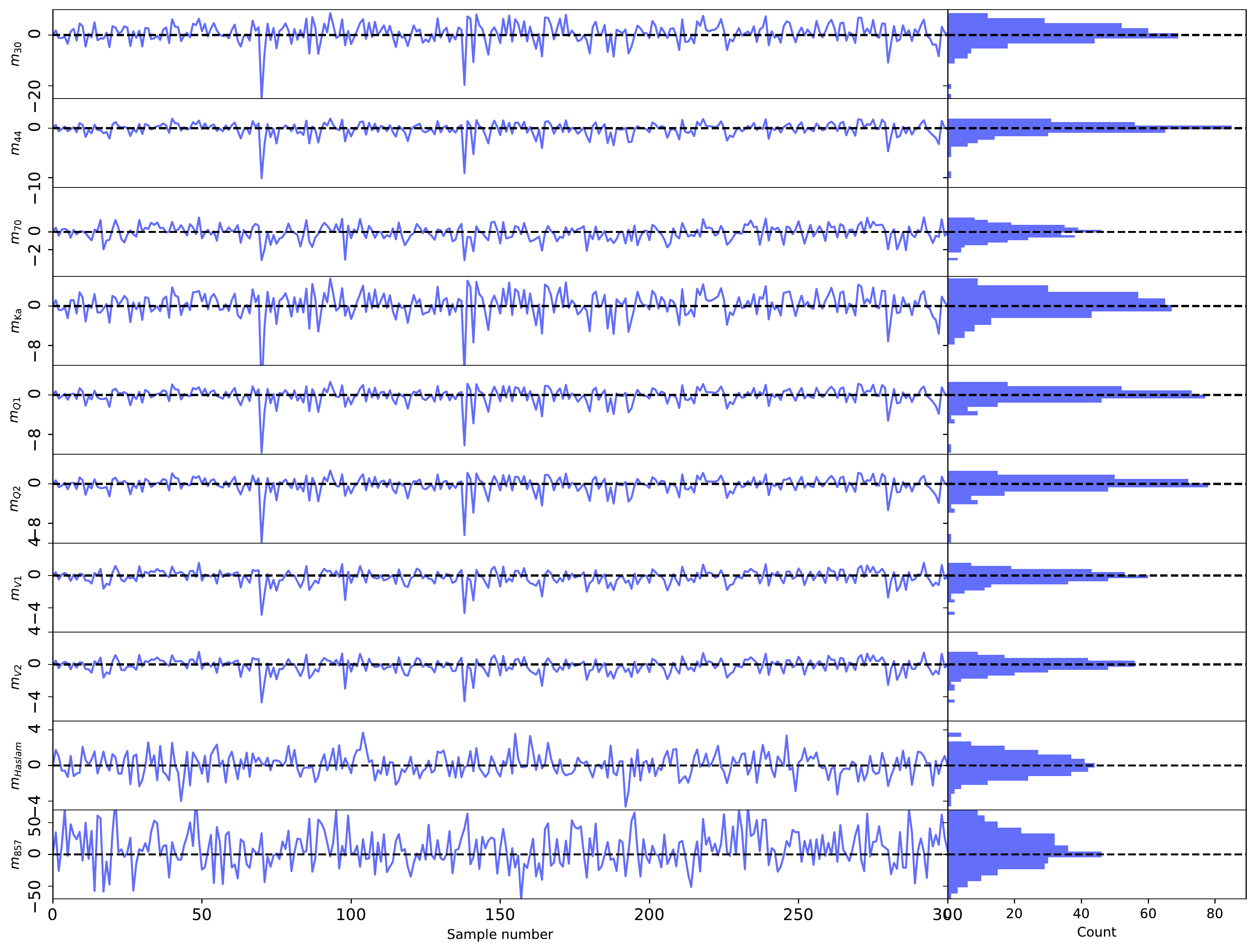}
\caption{Band monopoles for each of the simulated frequency bands as a function of sample. The nominal input value is given by the overlaid black dashed line. The right panel shows the distribution of the samples.}
\label{fig:simulated_monopoles}
\end{figure*}

The results of the 300 samples for the full sky spectral parameters and band monopoles can be seen in Figs.~\ref{fig:simulated_indices} and \ref{fig:simulated_monopoles} respectively. Much like the sky component maps, we see excellent agreement with the input values for both spectral parameters and the band monopoles. We note that both the spectral indices and the band monopoles show a few samples which are significant deviations from the input value. This is to be expected, and is in fact an intention of the algorithms implemented in this work. As described in Sec.~\ref{subsec:monopole_index}, the marginalization over the band monopoles allows for a broader log-likelihood, corresponding to a more complete exploration of the underlying distribution.

\section{\BP\ analysis and posterior distributions}
\label{sec:results}

We now turn our attention to the actual \BP\ analysis and intensity
component posterior distributions derived from the data combination
discussed in Sect.~\ref{sec:data}.
The results described in this section represent the
intensity foreground results of the \BP\ project and are a practical 
demonstration of the algorithms described and tested in Sects.~\ref{sec:algorithms} and \ref{sec:sim_results} respectively. We once again note that the
goal of the \BP\ project is not to derive a new state-of-the-art
astrophysical component model within the relevant CMB frequency range (given that critical data sets such as
\Planck\ HFI and \WMAP\ $K$-band are not included), but rather to lay
the algorithmic groundwork for a statistically robust community-wide
sky model, as will be implemented through the Open Science
\textsc{Cosmoglobe}\footnote{\url{http://cosmoglobe.uio.no}} community
effort. As far as Bayesian intensity sky modeling is concerned,
\citet{planck2014-a12} still represents a state of the art approach.

\subsection{Spectral parameter prior tuning}
\label{subsec:regions}

\begin{figure}
  \center       
  \includegraphics[width=\linewidth]{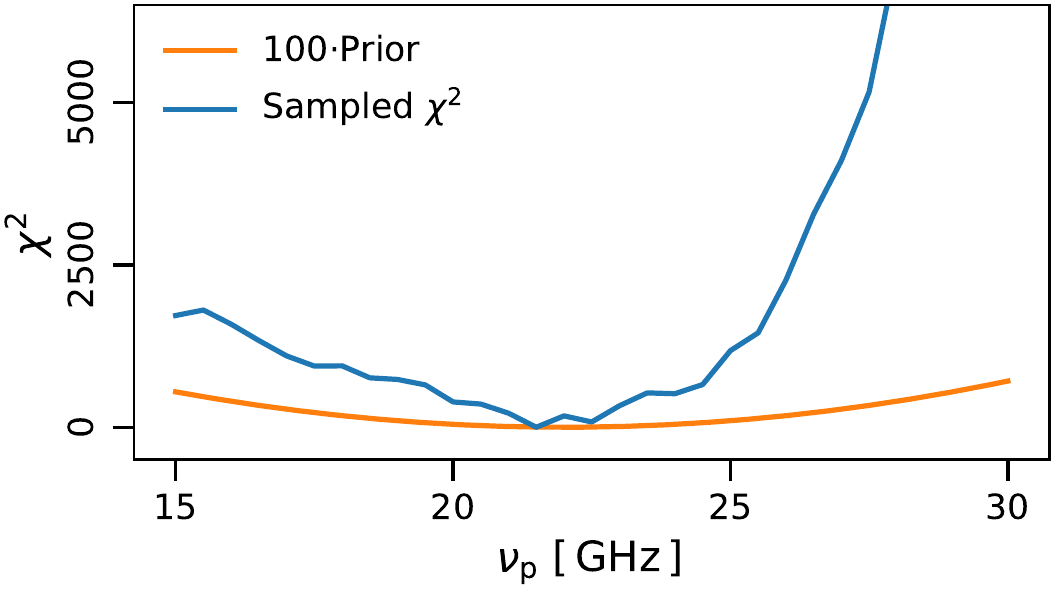}\\
  \includegraphics[width=\linewidth]{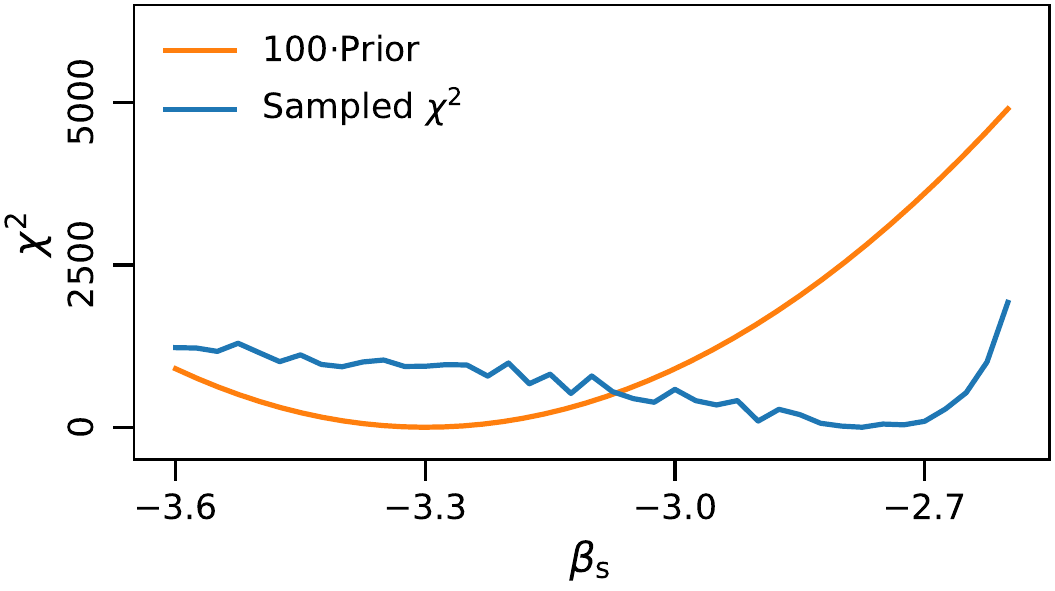}\\
  \includegraphics[width=\linewidth]{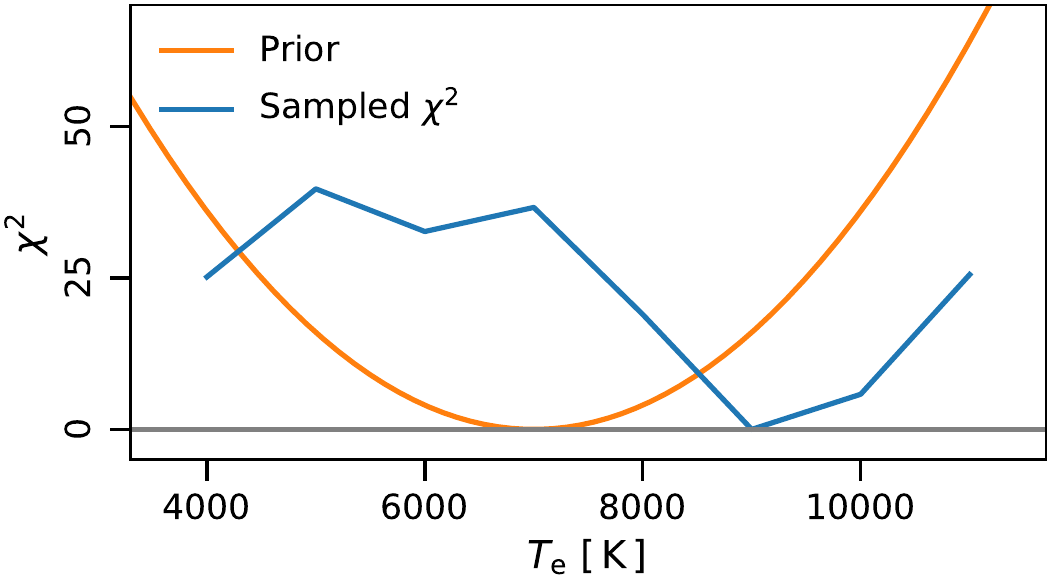}\\
  \caption{$\chi^2$ distributions (blue curves) from coarse grid
    evaluation of each free spectral parameters. From top to bottom,
    the panels show 1) AME peak frequency, $\nup$; 2) synchrotron
    spectral index, \bsynch;
    and 3) free-free electron temperature,
    \Te. The minimum ${\chi}^2$ value for each parameter has been
    subtracted in each case. Orange curves show the priors adopted for
    the given component; see Table~\ref{tab:components}. The prior values for
    \bsynch\ and \nup\ have been scaled by a factor of 100
    to fit in the plot with the derived ${\chi}^2$ values.  }
  \label{fig:spec_param_test}
\end{figure}

Before presenting the \BP\ Markov chains and posterior distributions,
there is still one task that must be completed before the algorithm
described in Sect.~\ref{sec:bp} is carried out to completion, namely, finalizing the
informative spectral parameter priors. With the reduced number of data
sets included in this work, we have reduced constraining power when
sampling spectral parameters, and strong priors are required for most
spectral SED parameters. With this in mind, we can assume that all
free parameters can only be fitted with a single constant value over
the full sky, at least for now. Already at this stage, we fix the
thermal dust temperature, $T_{\mathrm{d}}$, at the sky map derived by
\Planck\ DR4 \citep{npipe}, noting that LFI has no constraining power
for this particular parameter.

Even though LFI should have some constraining power of the thermal dust
spectral index, \bdust, the thermal dust and AME components are
found to be highly degenerate with the limited data set used in this
work. A joint fit of the AME \nup\ and \bdust\ would therefore lead to
unphysical results, with preliminary analyses showing that the
\bdust\ diverged to values $\bdust > 2, $ raising  \nup\ to much higher
frequencies. The uncertainty in the \bdust\ value is important for error
propagation and, thus, we must simply marginalize over the adopted prior,
instead of trying to constrain $\beta_{\mathrm{d}}$ with the current data set.

For each free spectral parameter, we created a dedicated sampling mask,
where we exclude regions on the sky where the other components are
strong in order to reduce potential modeling mismatch errors to
propagate between the various components. These masks are created from
the amplitude maps of the modeled components, evaluated at
44\GHz\ and smoothed to 10\deg\ FWHM. In addition, we masked radio
sources by thresholding the \Planck\ 30\GHz\ compact source map at
three different angular resolutions, namely, at native resolution and
at 1\deg\ and 10\deg\ FWHM. For \bsynch\ and \nup, we excluded regions
of the sky where any other component signals is greater than 40\muKRJ;
while for \Te\ , we excluded the areas where the other
components are greater than 50\muKRJ. For all parameters, we exclude
any pixel for which the smoothed radio sources are stronger than 30\muKRJ.
Finally, we masked out regions of the sky contributing to the largest 15\,\% of a
10\deg\ FWHM \chisq\ map to exclude regions with large known modeling
errors. The accepted sky fractions of the resulting masks are
$f_{\mathrm{synch}} = 0.66$, $f_{\mathrm{ff}} = 0.74$, and
$f_{\mathrm{AME}} = 0.66$,
respectively.

For each free parameter, we adopted a Gaussian informative prior with
some mean and standard deviation, $P(\beta) =
N(\mu_{\mathrm{\beta}},\sigma_{\mathrm{\beta}}^2)$.  The prior
parameters are informed by literature results and listed in
Table~\ref{tab:components}. For synchrotron emission, we note that few
intensity-based constraints are available for frequencies higher than
30\,GHz in the literature and we therefore adopted
$\beta_{\mathrm{s}}=-3.3\pm0.1,$ as derived from polarization
measurements in \citet{planck2016-l05}. We also note that we have
attempted to use flatter mean values of $\beta_{\mathrm{s}}=-3.1$ and
$-3.0$, as suggested from low-frequency surveys, but these result in
obvious artefacts in the current analysis in the form of a
significantly overestimated synchrotron amplitude at 30\,GHz. As an
example, Fig.~\ref{fig:AME_vs_synch_beta} compares the AME and
free-free amplitude maps derived for two different values of
$\beta_{\mathrm{s}}$, namely, $\beta_{\mathrm{s}}=-2.8$ and
$-3.3$. While neither of these solutions produce an excess $\chi^2$
and, therefore, a free likelihood-driven fit is unable to distinguish
between them, it is obvious from visual inspection that the former
spectral index leads to clear synchrotron leakage into both the AME
and free-free components. With a prior of
$\beta_{\mathrm{s}}=-3.3\pm0.1$, the nonphysical flat-index solutions
are largely excluded, while some parameter space is still allowed
toward the steeper end to explore degeneracies. Another approach of
reducing the predicted amplitude at CMB frequencies is by introducing
a negative curvature in the synchrotron SED (as discussed in
Sect.~\ref{subsec:sky_model}) and when comparing the results from this paper
with previous results, it is important to ensure that the models are
compatible.

Given the listed priors, we performed a coarse $\chi^2$ grid evaluation
for each free parameter, conditioning on all other spectral parameter
means, but allowing for amplitudes and band monopoles to adjust to the
given spectral parameter. The resulting \chisq's (evaluated at
a \healpix\ resolution of $\nside=16$) are shown in
Fig.~\ref{fig:spec_param_test}.

Starting with AME $\nup$, shown in the top panel, we see that the
$\chi^2$ is well-defined with a typical best-fit value around
22\,GHz. The actual $\chi^2$ values show
rapid increases at both lower and higher values with variations
ranging in the thousands, and correspondingly, the prior (which is on the
order of unity) is therefore largely irrelevant. It is clear that the
current data combination has significant constraining power for
$\nup$.

The second panel shows similar results for $\beta_{\mathrm{s}}$. In
this case, we see a rapid $\chi^2$ increase for $\beta\lesssim-2.7$,
but  it is otherwise slowly increasing for smaller values of $\beta_{\mathrm{s}}$
in the region $\beta_{\mathrm{s}} < -3.2$, becoming almost flat.
Additionally, we already know from
Fig.~\ref{fig:AME_vs_synch_beta} that spectral indices flatter than
$\beta_{\mathrm{s}}\lesssim-2.8$ lead to clearly contaminated AME
component maps, even if the $\chi^2$ is not able to identify this. At
the same time, the actual $\chi^2$ variations are indeed larger than
the prior, and this typically indicates that the algorithm prefers to
use this unconstrained degree of freedom to fit other degrees of
freedom, for instance, modeling errors in the thermal dust model. To
avoid pathological solutions, we instead chose to
marginalize explicitly over the prior and disable the likelihood term
entirely when sampling this parameter. In other words, we simply
marginalized over the adopted prior, but we did not attempt to constrain
$\beta_{\mathrm{s}}$ with the current data set.

The same considerations hold to an even greater extent for the last parameter.
For the electron temperature, \Te, the $\chi^2$ variations are
entirely spurious and we therefore disable the likelihood term for $T_e$. 

In summary, the only spectral parameter that the current data set is
able to robustly constrain in intensity is the AME peak frequency,
$\nup$. All others are either drawn from their corresponding priors in the
current analysis or frozen. We note that introducing additional data sets to
constrain these parameters is a critically important next step for
future works. 

With \bsynch\ and \bdust\ drawn from priors and \Te\ frozen,
we find that the mask used in the coarse grid sampling of \nup\
is too conservative when sampling \nup, masking out too much of the
galactic plane and leading to large dust-like residual signals in the
LFI 30\GHz\ and the \WMAP\ $Ka$ channels. Additionally, a more dust-like
signal was found to be leaking into the free-free component amplitude.
In order to limit these effects, a less conservative mask had to be used.
The mask used to sample \nup\ in the final \BP\ production is
generated by excluding all pixels with values above 200\muKRJ\ of the
free-free amplitude at both 30\arcm\ and 2\deg\ FWHM angular
resolution evaluated at 40\GHz; all pixels with values above
100\muKRJ\ of the point source amplitudes smoothed with a 1\deg\ FWHM
beam evaluated at the LFI 30\GHz\ band frequency; and the regions of the
sky contributing to the largest 2.5\,\% of the \chisq.  The accepted
sky fraction of the resulting mask is $f_{\mathrm{AME}} = 0.91$.

\subsection{Markov chain trace plots and correlations}
\label{sec:traceplots}

\begin{figure*}
  \center       
  \includegraphics[width=0.78\linewidth]{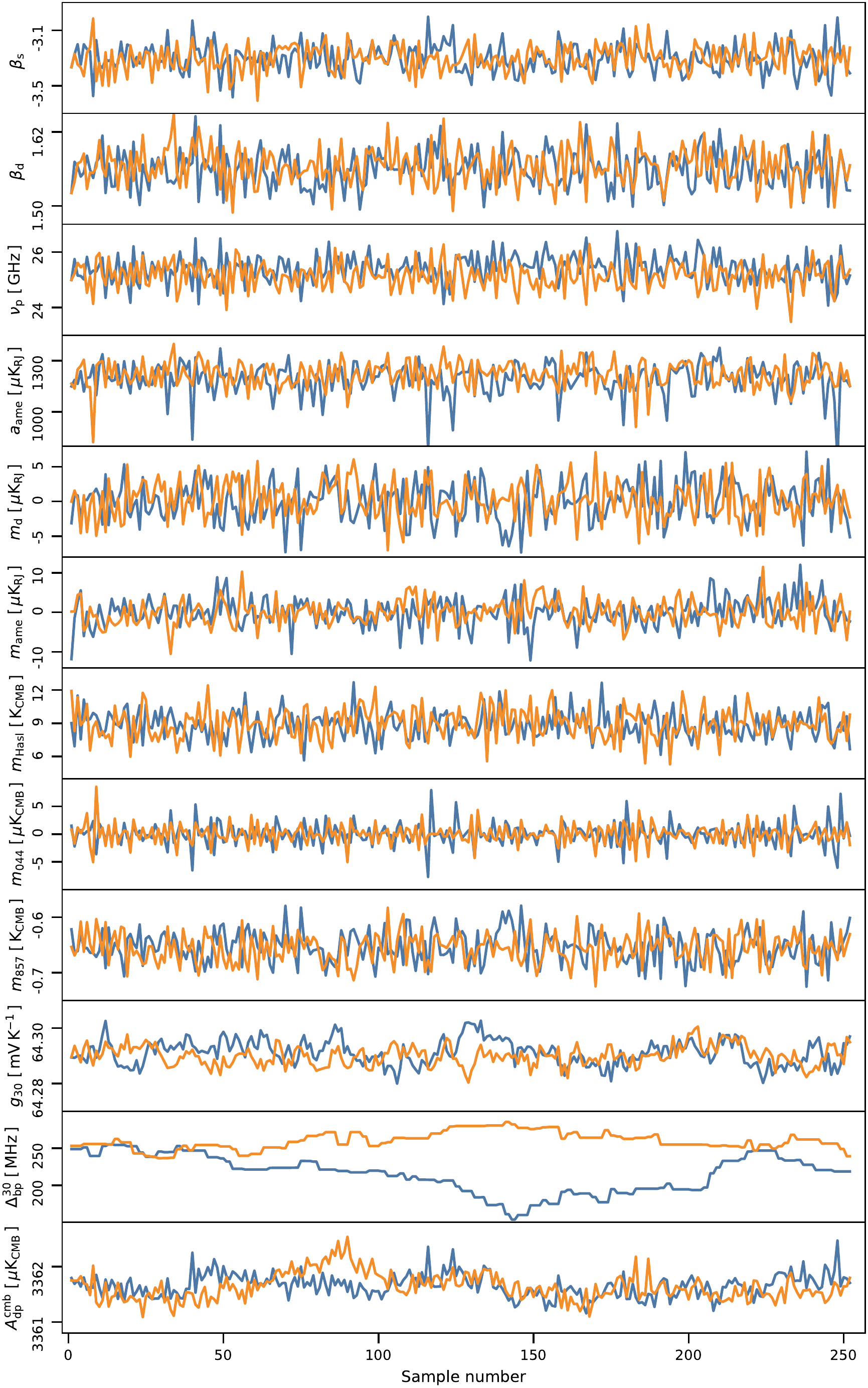}
  \caption{Trace plots of a set of sampled component separation parameters and selected instrument parameters, including absolute calibration ($g$), bandpass shift $(\Delta_{\mathrm{bp}})$, and the Solar CMB dipole ($A_{\mathrm{dp}}^{\mathrm{cmb}}$) from a naive mono- and dipole estimate using the band monopole mask, as described in Fig.~\ref{fig:monopole_masks}. The parameter $a_{\mathrm{ame}}$ is the AME amplitude evaluated at 10\deg\ FWHM and $\nside=16$ of the pixel at Galactic coordinates $(l,b)=(31\deg,5\deg)$. Parameters from chain 1 are plotted in blue and chain 2 in orange. 
    }
  \label{fig:param_traceplot}
\end{figure*}

\begin{figure*}
  \center       
  \includegraphics[width=0.9\linewidth]{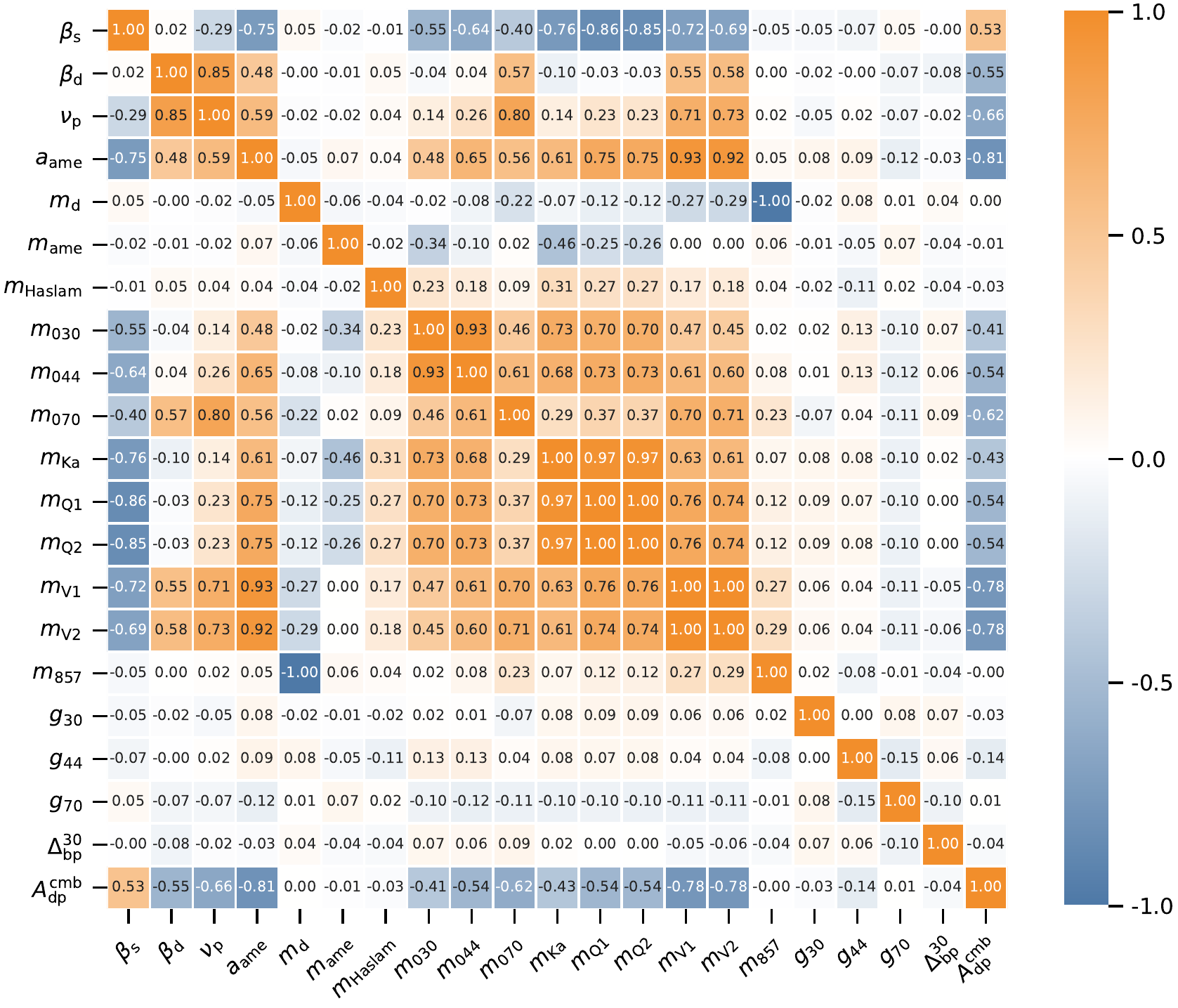}
  \caption{Correlation coefficient plot of local deviation, from a running mean of five prior and five succeeding samples of the sampled spectral parameters, component cross-correlation intersections, frequency band monopoles, absolute gain calibration, LFI 30\GHz\ bandpass shift, and CMB dipole amplitude, all as described in Fig.~\ref{fig:param_traceplot}.
    }
  \label{fig:param_corr_local}
\end{figure*}

With the data, sampling algorithms, and priors in place, we are ready
to consider the actual Markov chain results. As described by
\citet{bp01}, for the final analysis, we produced two independent
chains, each with 750~samples.

Figure~\ref{fig:param_traceplot} shows the first 250 samples for a
selection of relevant parameters. Several points are worth noting in
this figure. First, we note that the burn-in period for most of the
foreground-induced parameters is very short, while it is slightly
longer for some of the global gain and bandpass instrumental
parameters. We removed the first ten samples for burn-in for the
following analysis. This short burn-in is a combination of the novel
sampling algorithms introduced in Sect.~\ref{sec:algorithms} and the
prior choices discussed above. At the same time, we do observe a weak
shift in the average of $\nup$, where the chains split away from each
other around sample 120, which appears to trace some of the more
slowly varying instrumental parameters, primarily the total bandpass
correction of LFI 30 GHz; this is quite typical, as many global
instrumental parameters tend to have long Markov correlation lengths
and these directly affect foreground residuals.

In Fig.~\ref{fig:param_corr_local}, we plot Pearson's correlation
coefficients between the various parameters. For this particular plot,
we have subtracted a running average with a length of ten samples
(five samples prior and succeeding) from each function before
evaluating the correlation coefficient, in order to highlight
sample-by-sample correlations; two parameters may appear to be
spuriously correlated if there are long-term gradients, irrespective
of their origins.

Several interesting observations may be made from this plot. First, we
note that there is a very high correlation between
$\beta_{\mathrm{d}}$ and both the AME peak frequency $\nup$ and the
CMB dipole amplitude $\A_{\mathrm{dp}}^{\mathrm{CMB}}$. This is not
unexpected, given the critically important role of thermal dust
emission at all CMB frequencies. At the same time, this also serves as
a useful reminder that several important \BP\ results depend directly
on official \Planck\ results through the use of the HFI 857\,GHz
frequency band and the assumed thermal dust spectral index prior of
$\beta_{\mathrm{d}}=1.56\pm0.03$
\citep{planck2014-a12,planck2016-l04,npipe}, and the systematic errors
in these are not propagated properly in the current
analysis. Integrating HFI observations into the \BP\ framework at the
TOD level is clearly an important goal of near-term work.

Next, we note that all microwave frequency monopoles are internally
strongly correlated and, notably, anticorrelated with the component
monopoles. Both of these observations make intuitive sense, as if
one increases a given component monopole, the frequency monopoles have
to decrease to result in a net zero change to the overall model. In
addition, all frequency monopoles have to change in the same
direction to a given change in the component monopoles.
Accurately accounting for all of these correlations in terms of MCMC
samples is perhaps the most important advantage of the Bayesian
end-to-end \BP\ framework.

\subsection{Goodness-of-fit statistics}
\label{sec:gof}

\begin{figure*}
  \center       
  \includegraphics[width=0.32\linewidth]{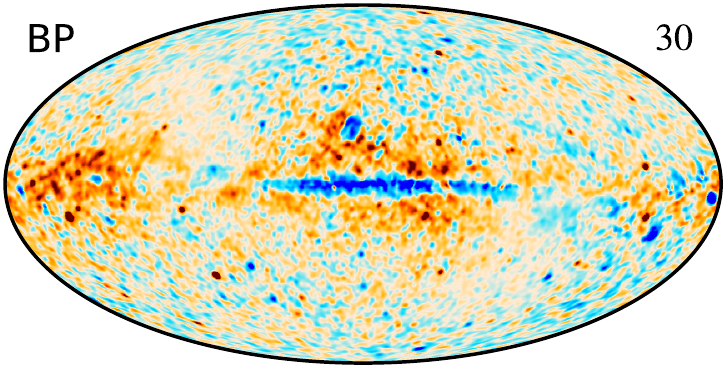}
  \includegraphics[width=0.32\linewidth]{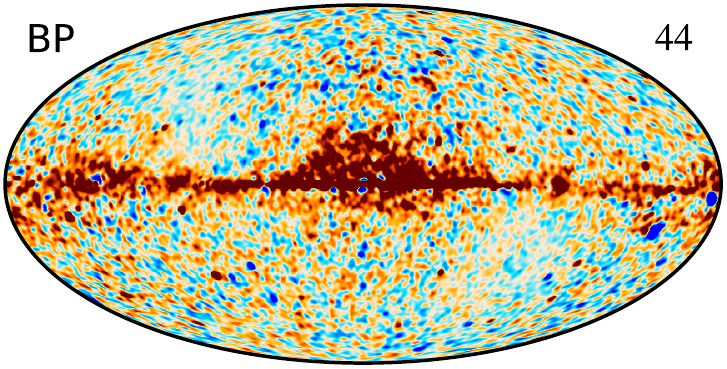}
  \includegraphics[width=0.32\linewidth]{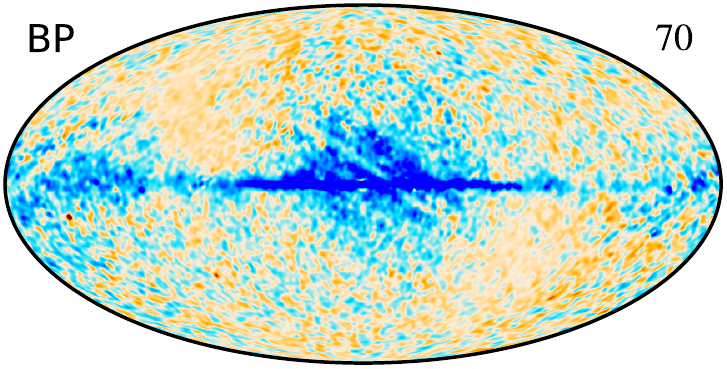}\\
  \includegraphics[width=0.7\linewidth]{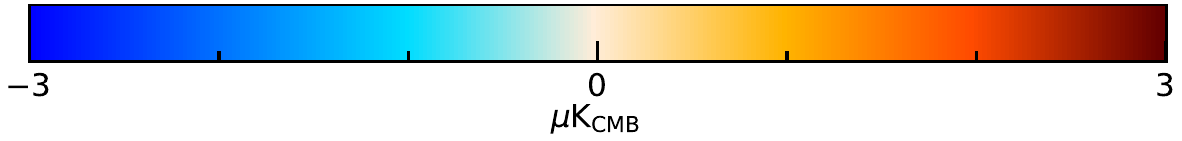}\vspace{0.3cm}\\
  \includegraphics[width=0.32\linewidth]{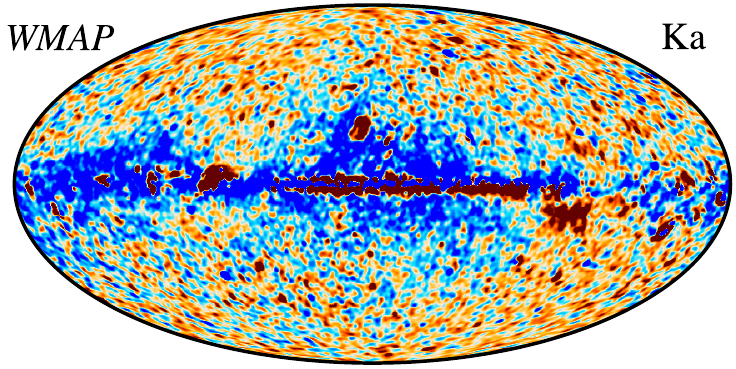}
  \includegraphics[width=0.32\linewidth]{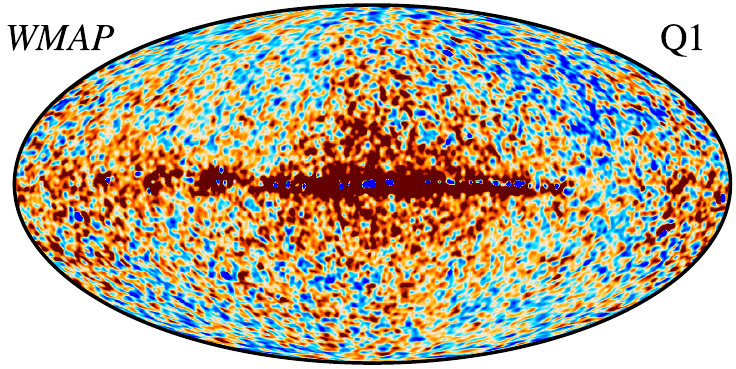}
  \includegraphics[width=0.32\linewidth]{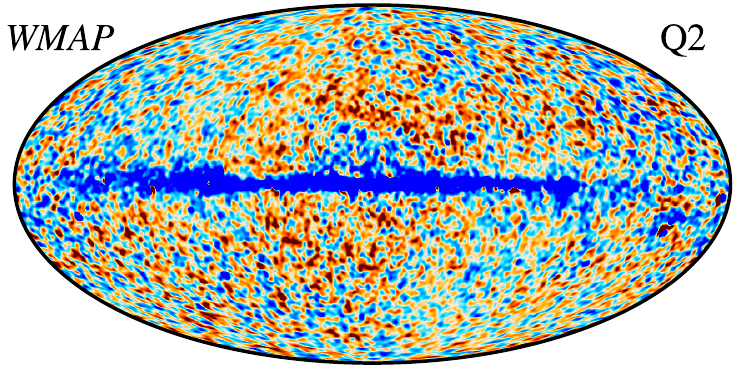}\\
  \includegraphics[width=0.32\linewidth]{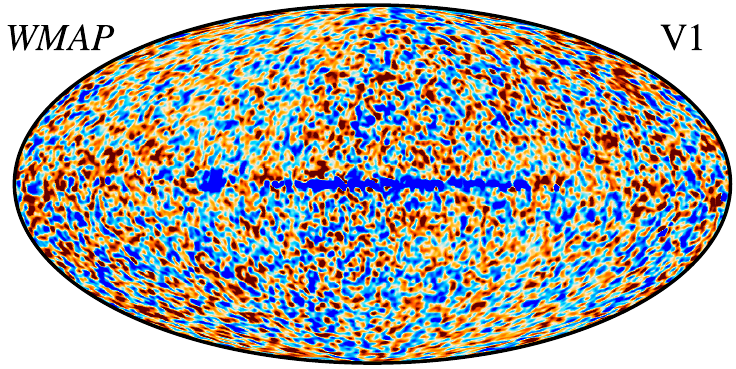}
  \includegraphics[width=0.32\linewidth]{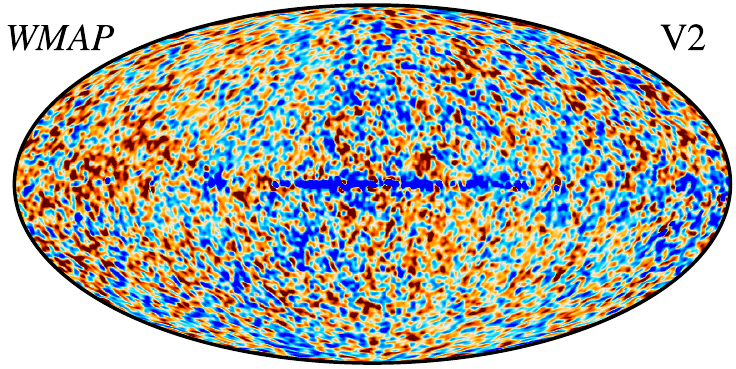}\\
  \includegraphics[width=0.7\linewidth]{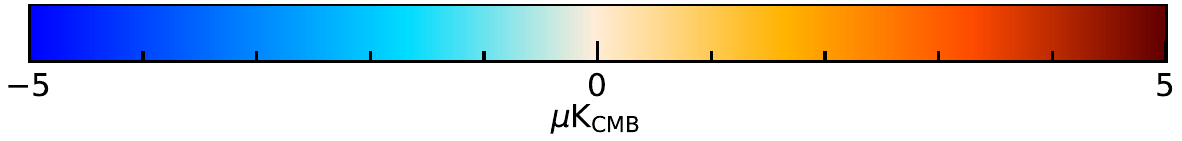}\vspace{0.3cm}\\
  \includegraphics[width=0.33\linewidth]{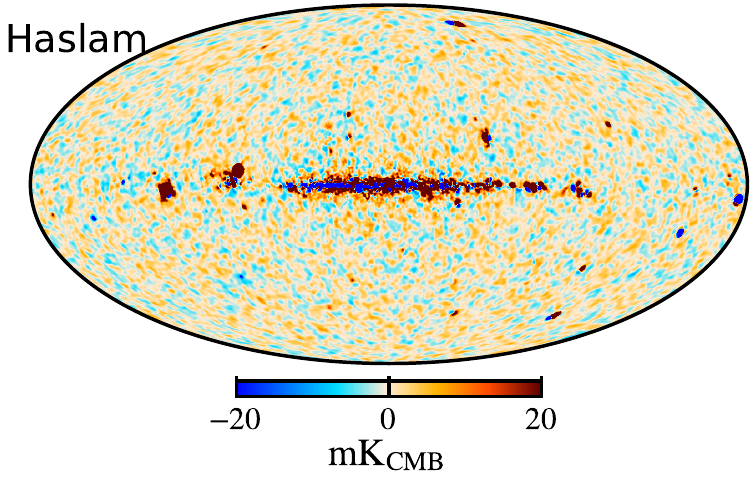}
  \includegraphics[width=0.33\linewidth]{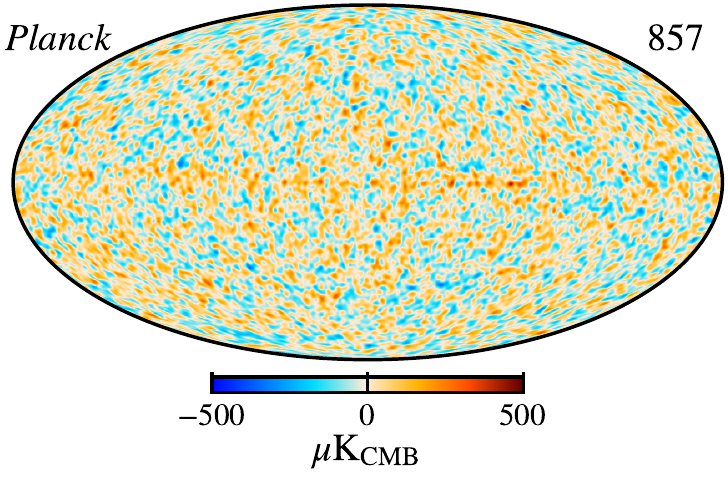}
  \includegraphics[width=0.33\linewidth]{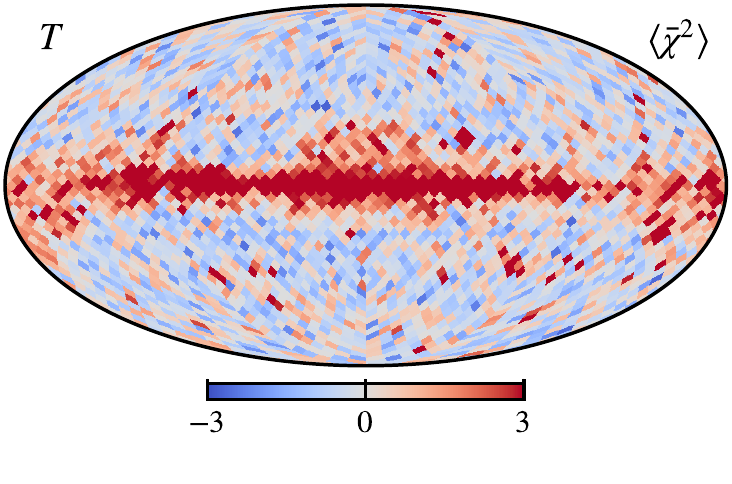}  
  \caption{Mean residual maps for the different frequency channels
    included in the \BP\ analysis. All maps have been smoothed to a
    common angular resolution of 2\deg\ FWHM. The \Planck\ LFI
    residuals are plotted with a range of 3\,\muKCMB, and the
    \WMAP\ residuals are plotted with a range of 5\,\muKCMB. The
    bottom right panel shows the mean reduced chi-squared
    $\bar{\chi}^2$ per $\nside = 16$ pixel of the \BP\ Gibbs chain.}
  \label{fig:BP_residuals_mean}
\end{figure*}

Before presenting the actual posterior distributions, we consider the
goodness-of-fit of the derived model. First,
Fig.~\ref{fig:BP_residuals_mean} shows posterior mean residual maps on
the form $\r_{\nu} = \d_{\nu}-\s_{\nu}$ for each of the ten frequency
bands included in the analysis, where the average is evaluated over
all samples in the Markov chain.

Starting with the LFI channels shown in the top section of the figure,
we see that the residuals at high Galactic latitudes are largely
consistent with instrumental noise, except for scattered point source
residuals at 30\,GHz, while at low Galactic latitudes there is an
obvious dust morphology residual coming from either AME or thermal
dust. At 30\,GHz it is also possible to see some bright negative
free-free residuals. Overall, though, the fits are performing quite
well, with typical residuals smaller than 3\muK\ over most of the sky,
and the remaining artefacts are relatively easy to mask through
Galactic and point source thresholding.

The middle section shows the \WMAP\ channels, plotted on a color range
of $\pm5\muK$. The most striking residual feature in this case is a
strong dust residual in the \textit{Ka} (33\,GHz) channel. This, combined
with the weaker dust residuals observed in the LFI channels, strongly
suggests that the current single-component AME model adopted for the
\BP\ analysis is not a statistically adequate model for the actual AME
sky. In fact, this shortcoming was already pointed out by
\citet{planck2014-a12}, who introduced a second AME component to fit
the full contribution. Doing the same with the current data selection
would lead to a massively increased noise level for all derived
components, and we instead accept the foreground mismodeling here
and we instead simply make sure to mask out the contaminated regions
of the sky in higher-order analyses.

Other notable features in the \WMAP\ residuals are large regions of
low-level residuals at high Galactic latitudes that do not obviously
trace known Galactic components. As discussed by \citet{barnes2003},
an important challenge regarding this data set on large angular scales
is sidelobe modeling and this may also be relevant for the residuals
we see in Fig.~\ref{fig:BP_residuals_mean}. A re-analysis of the
time-ordered \WMAP\ data within the \BP\ framework is already
ongoing \citep{bp17}. 

The last two frequency channels, Haslam 408\,MHz and HFI 857\,GHz,
show very uniform residuals. This is simply due to the fact that their
unique S/N values massively dominate the synchrotron and
thermal dust components, respectively, and any potential mismodeling
would therefore leak directly into the component amplitude maps. A flat
residual should thus not be interpreted as the absence of systematic
errors, but rather simply as an indication that these two channels
have no significant crosscheck in terms of their effect on the signal
model by any other channel. We note that this is not entirely true for the Haslam
408\,MHz map, which does have competitors in terms of free-free
emission near the Galactic center in \Planck\ and \WMAP, and
corresponding residuals may be seen here.

The bottom right panel of Figure~\ref{fig:BP_residuals_mean} shows the
reduced and normalized $\chi^2$ per $N_{\mathrm{side}}=16$ pixel, as
defined by:
\begin{equation}
  \bar{\chi}^2(p) = \frac{\sum_{\nu,p'\in p} \left[\frac{d_{\nu}(p')-s_{\nu}(p')}{\sigma_{\nu}(p')}\right]^2 -n_{\mathrm{dof}}}{\sqrt{2n_{\mathrm{dof}}}},
\end{equation}
where the sum runs over all pixels within a given low-resolution
pixel, and $n_{\mathrm{dof}}=15\,400$ is an estimate of the total
number of degrees of freedom for each low-resolution pixel. Since
$n_{\mathrm{dof}}$ is large, this quantity is expected to be $N(0,1)$
distributed in the ideal case, and Fig.~\ref{fig:BP_residuals_mean} thus
quantifies the agreement between the data and the model in units of
standard deviations per pixel. Overall, we see that the distribution
agrees with the expectation to about $1\,\sigma$ at high Galactic
latitudes, except for some compact sources, while at low Galactic
latitudes there is a strong residual with a clear dust-like
morphology. This $\chi^2$ map serves as an important input for
producing masks for higher order analyses. 

\subsection{Signal component posterior distributions}
\label{sec:diffuse_comp}

\begin{table}[t]
\centering
\caption{Comparison of frequency channel monopole constraints from
  \BP\ and \Planck. For \Planck, all numbers correspond to the
  \Planck\ 2015 analysis \citep{planck2014-a12}, except for the HFI
  857\,GHz, which is taken from \Planck\ DR4 \citep{npipe}. We note that
  the \Planck\ 2015 analysis did not fit for the \WMAP\ \textit{Ka}-band
  monopole, but fixed it at the given value; this parameter therefore
  has no associated uncertainty. Lastly, we note that while the \WMAP\ and
  HFI maps are the same in both columns, the LFI maps are not, and they have
  different zero-levels coming from the TOD processing. } \footnotesize
\renewcommand{\arraystretch}{1.4}
 \begin{tabular}{l c c l}
  \hline 
  \hline
  Channel & \BP\ & \Planck\ & Unit   \\
  \hline
  LFI 30\,GHz & $0\pm6$ & $-17\pm1\phantom{-0}$ & $\mu\mathrm{K_{CMB}}$   \\
  LFI 44\,GHz & $0\pm2$ & $\phm11\pm1\phantom{-0}$& $\mu\mathrm{K_{CMB}}$   \\
  LFI 70\,GHz & $0\pm1$ & $\phm16\pm1\phantom{-0}$& $\mu\mathrm{K_{CMB}}$   \\
  \WMAP\ \textit{Ka} & $16\pm 3\phm$ & $3$& $\mu\mathrm{K_{CMB}}$   \\
  \WMAP\ \textit{Q}1 & $12\pm 2\phm$ & $\phm2\pm1\phantom{-0}$& $\mu\mathrm{K_{CMB}}$   \\
  \WMAP\ \textit{Q}2 & $11\pm 2\phm$ & $\phm2\pm1\phantom{-0}$& $\mu\mathrm{K_{CMB}}$   \\
  \WMAP\ \textit{V}1 & $6\pm 1$  &$\phm1\pm1\phantom{-0}$& $\mu\mathrm{K_{CMB}}$   \\
  \WMAP\ \textit{V}2 & $6\pm 1$ & $\phm1\pm1\phantom{-0}$& $\mu\mathrm{K_{CMB}}$   \\
  HFI 857\,GHz & $-0.65\pm 0.03\phm$ & $-0.72\pm0.01\phantom{-0}$& $\phantom{\mu}\mathrm{K_{CMB}}$   \\
  \hline
  \end{tabular}
 \label{tab:spec_par_dist}
 \endPlancktable

\end{table}

\begin{figure*}[p]
  \center       
  \includegraphics[width=0.47\textwidth]{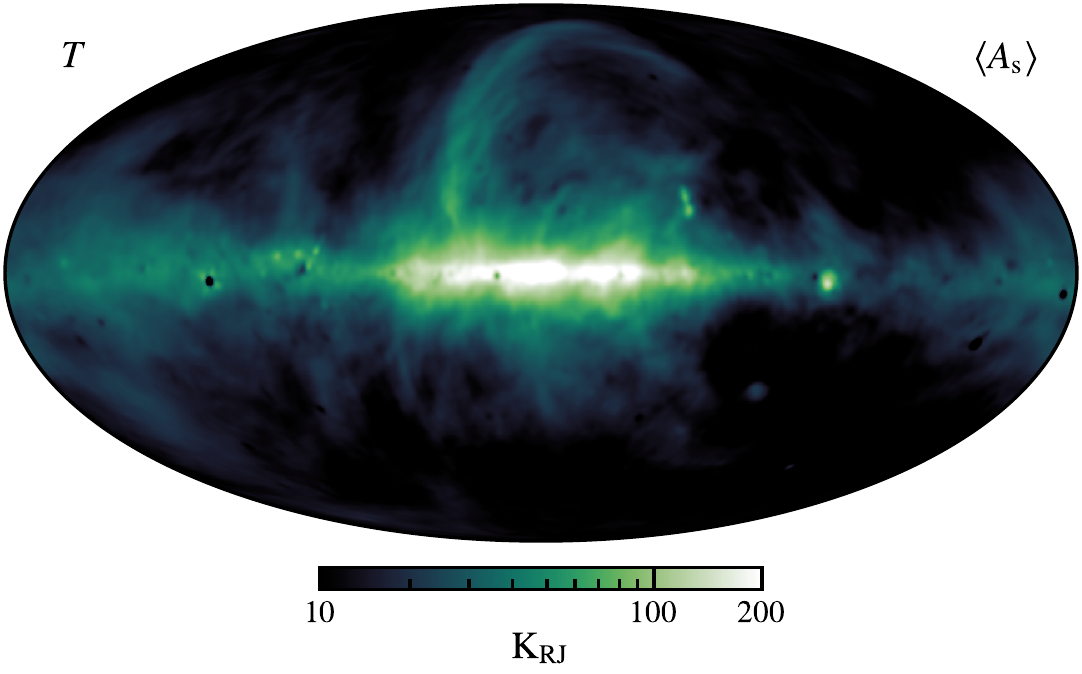}
  \includegraphics[width=0.47\textwidth]{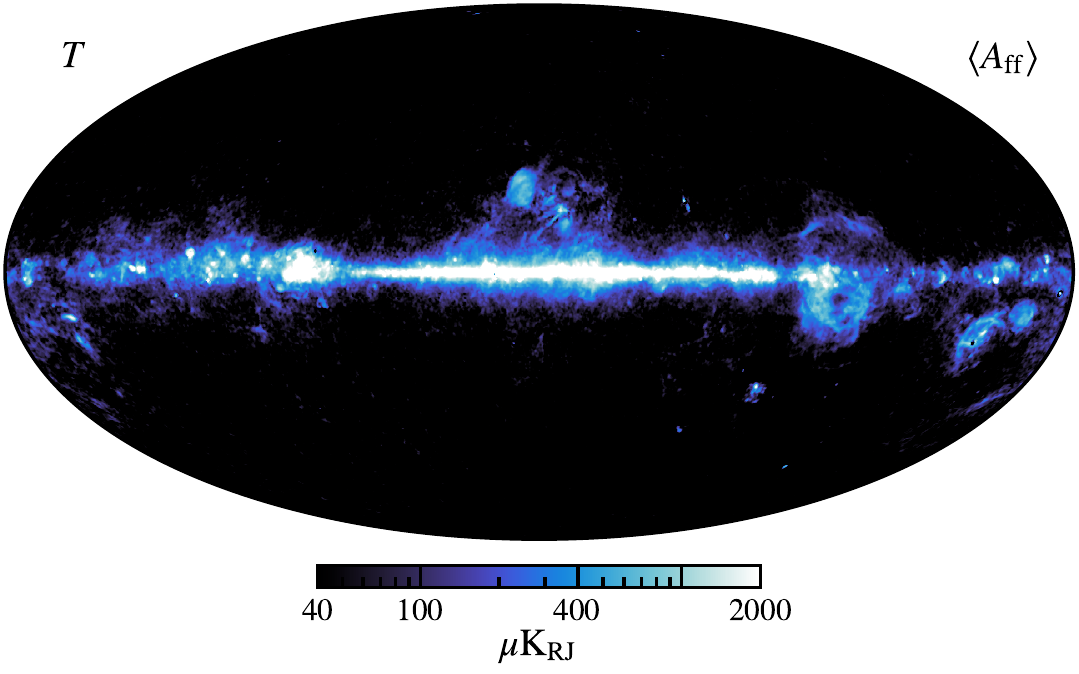}\\
  \includegraphics[width=0.47\textwidth]{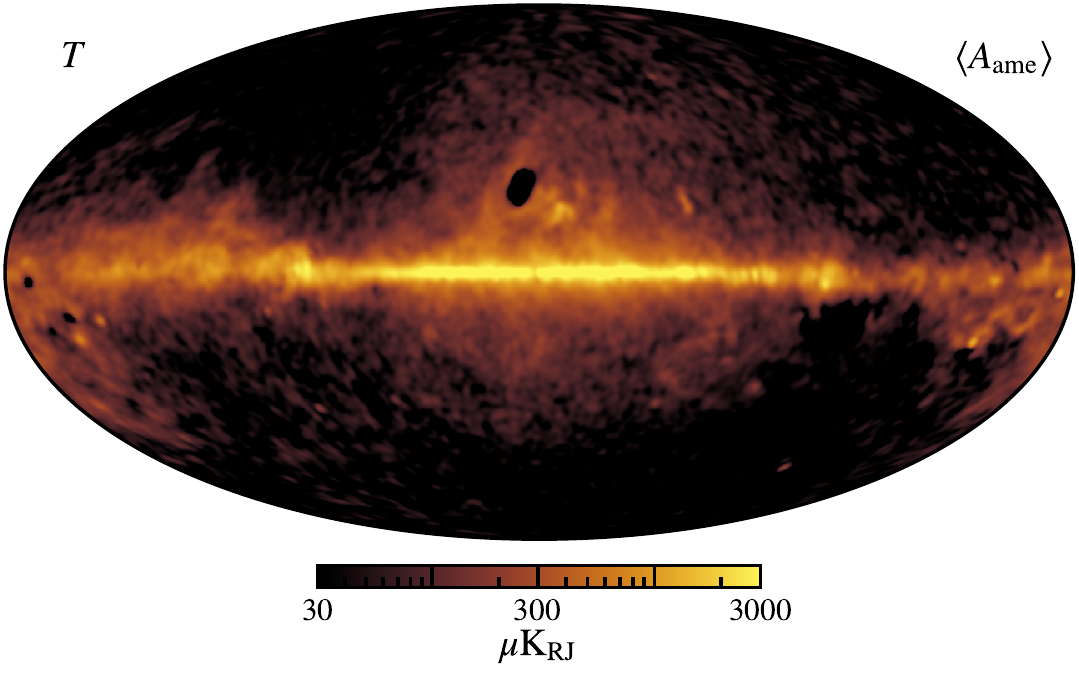}
  \includegraphics[width=0.47\textwidth]{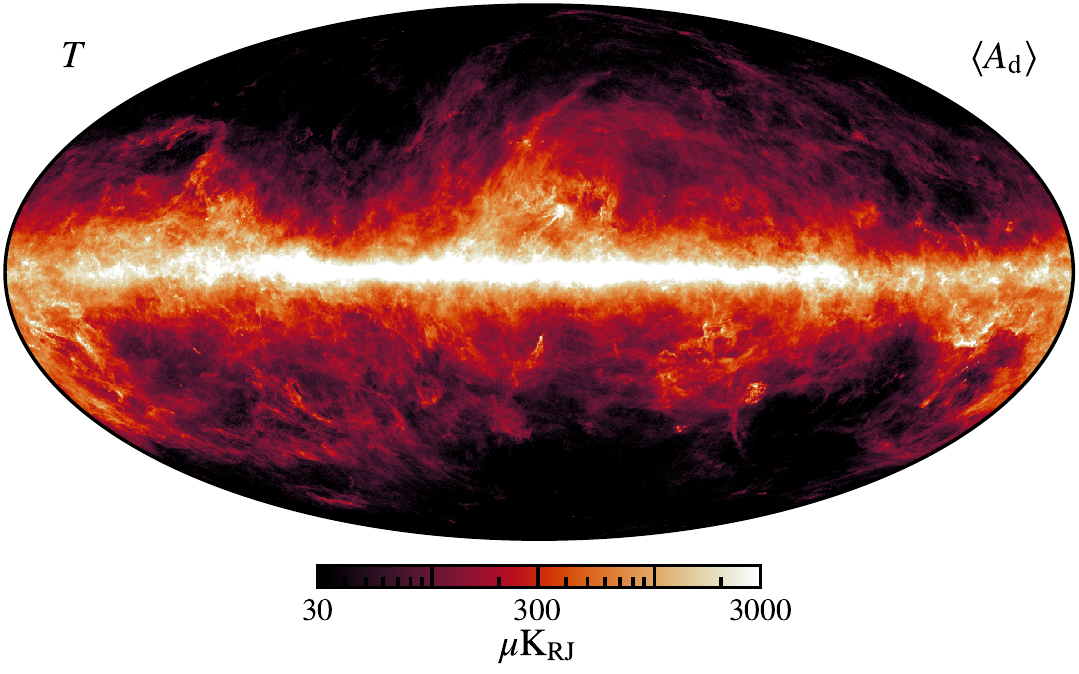}
  \caption{Posterior mean amplitude maps for each of the four fitted
    foreground component; synchrotron (top-left), free-free
    (top-right), AME (bottom-left), and thermal dust
    emission (bottom-right). The angular resolutions of the
    four maps are 120, 30, 120, and 10\arcm\ FWHM, respectively.
  }
  \label{fig:diff_mean}
\vspace*{3mm}
  \center       
  \includegraphics[width=0.47\textwidth]{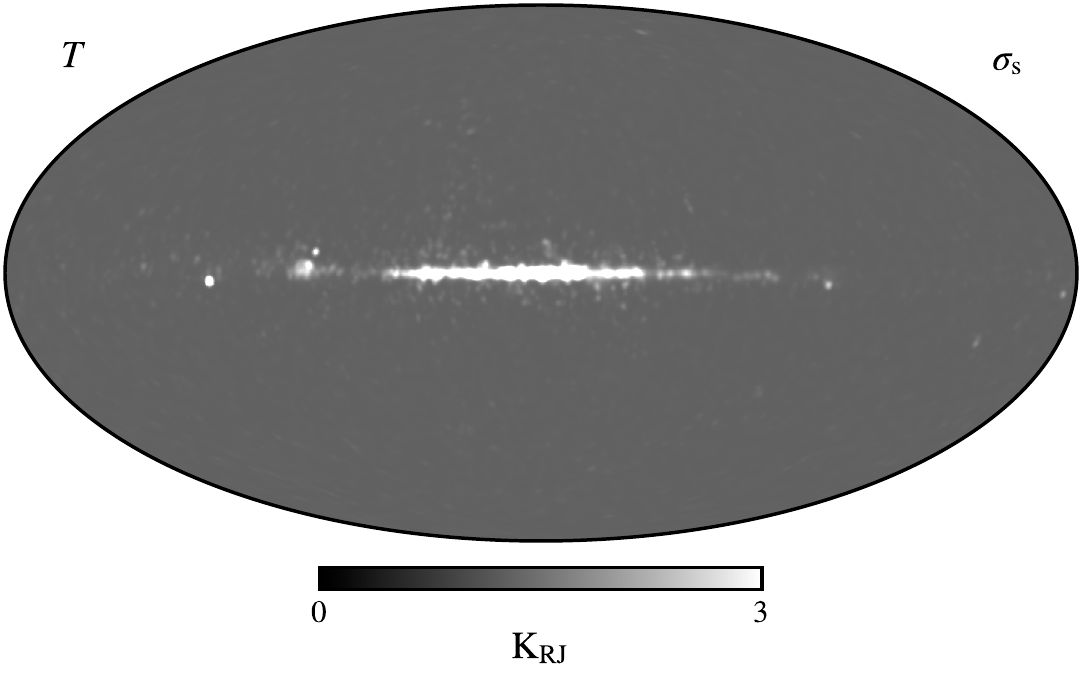}
  \includegraphics[width=0.47\textwidth]{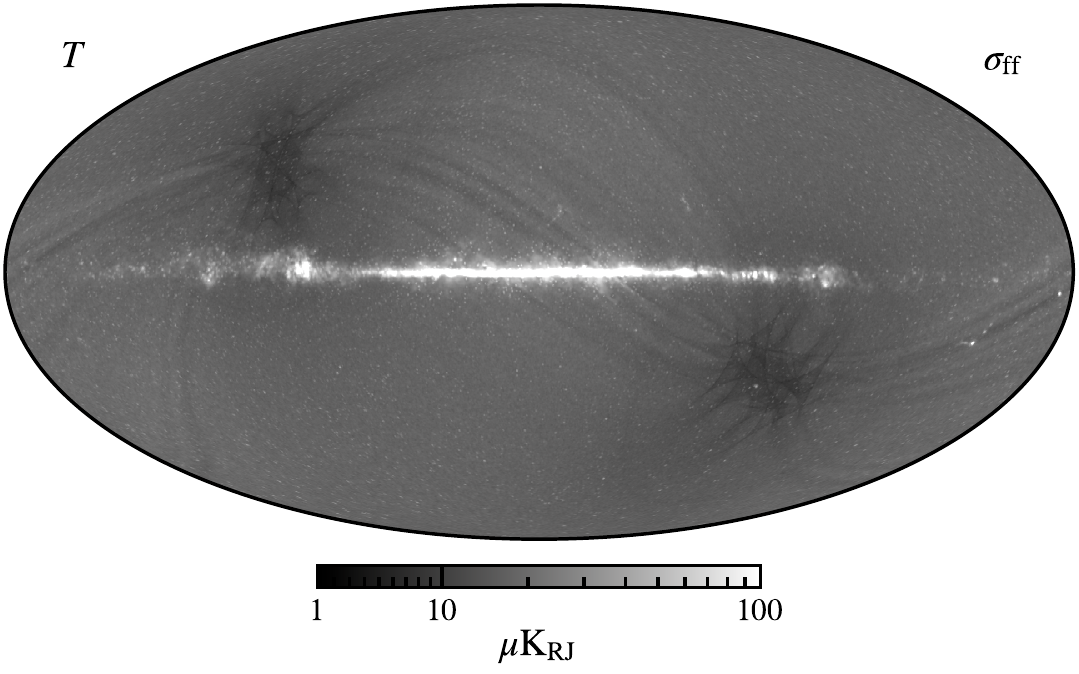}\\
  \includegraphics[width=0.47\textwidth]{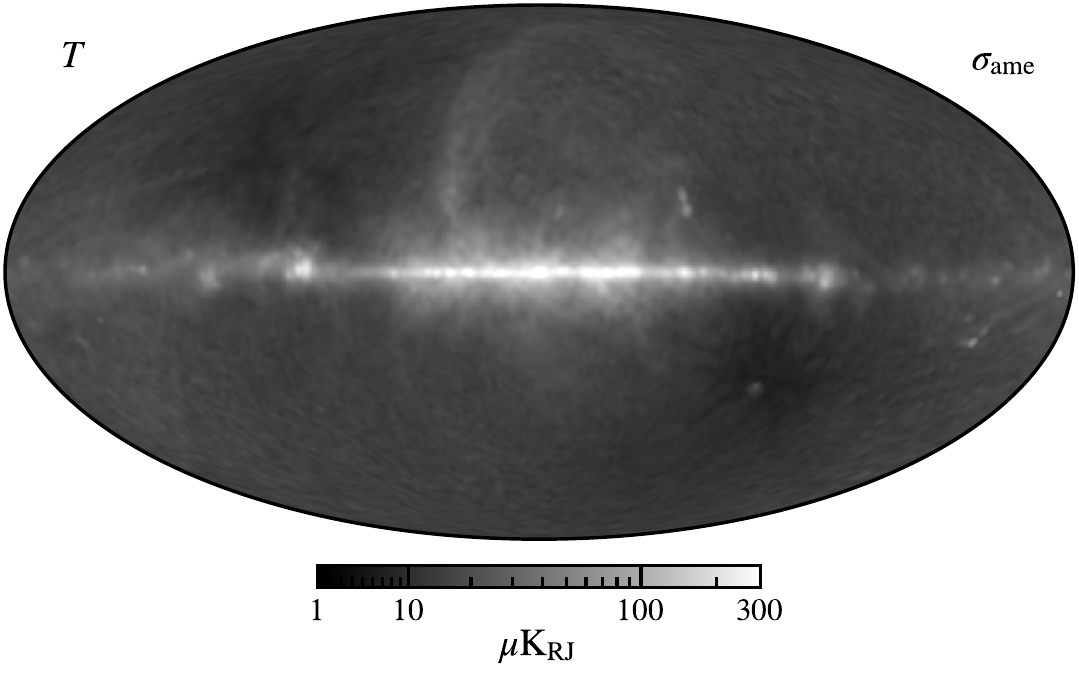}
  \includegraphics[width=0.47\textwidth]{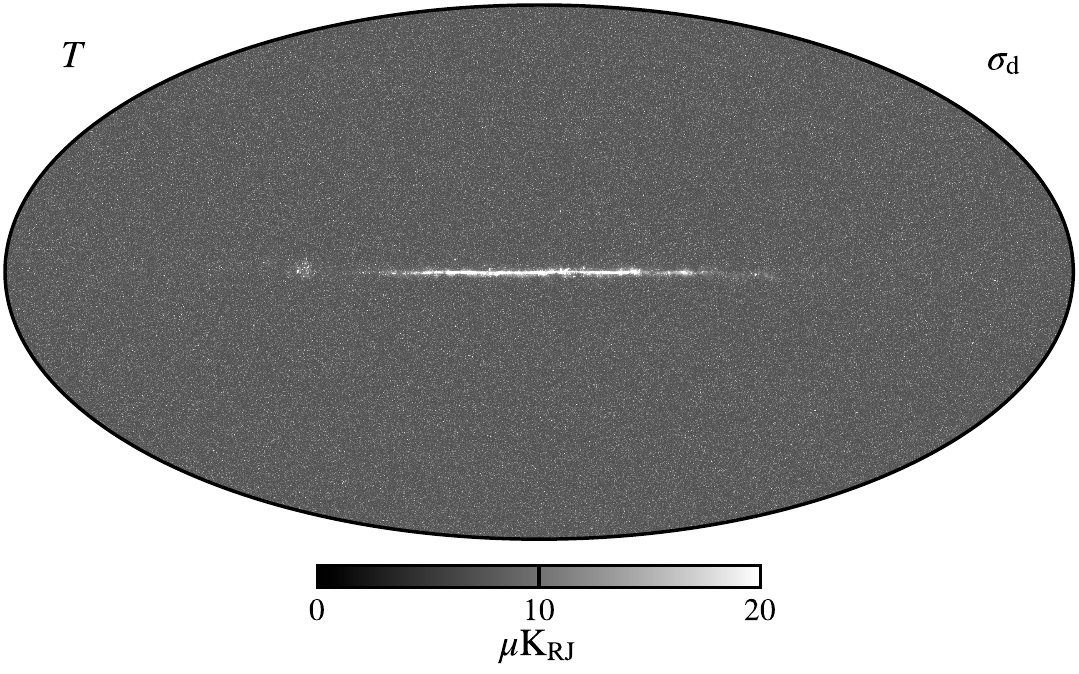}
  \caption{Posterior standard deviations for the same maps as shown in
    Fig.~\ref{fig:diff_mean}. We note that synchrotron and dust emission are potted
    with linear scaling.
  }
  \label{fig:diff_rms}
\end{figure*}

\begin{figure*}
  \center       
  \includegraphics[width=0.32\linewidth]{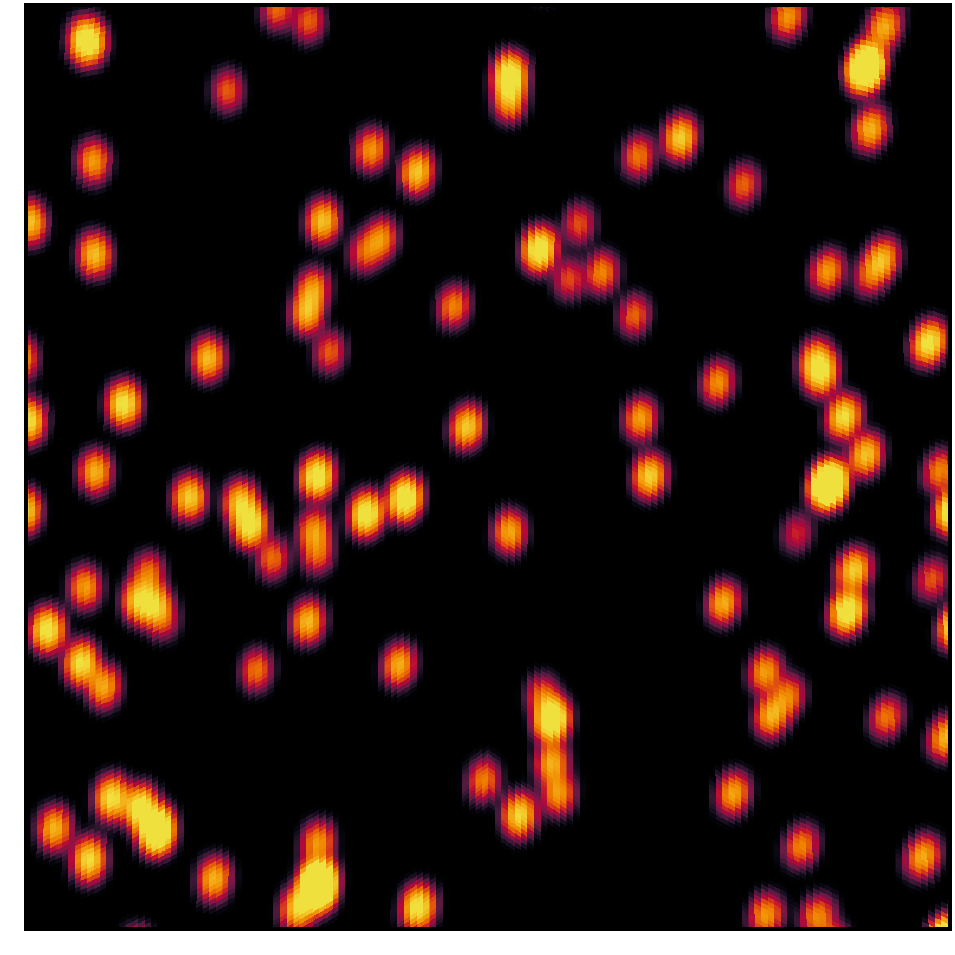}
  \includegraphics[width=0.32\linewidth]{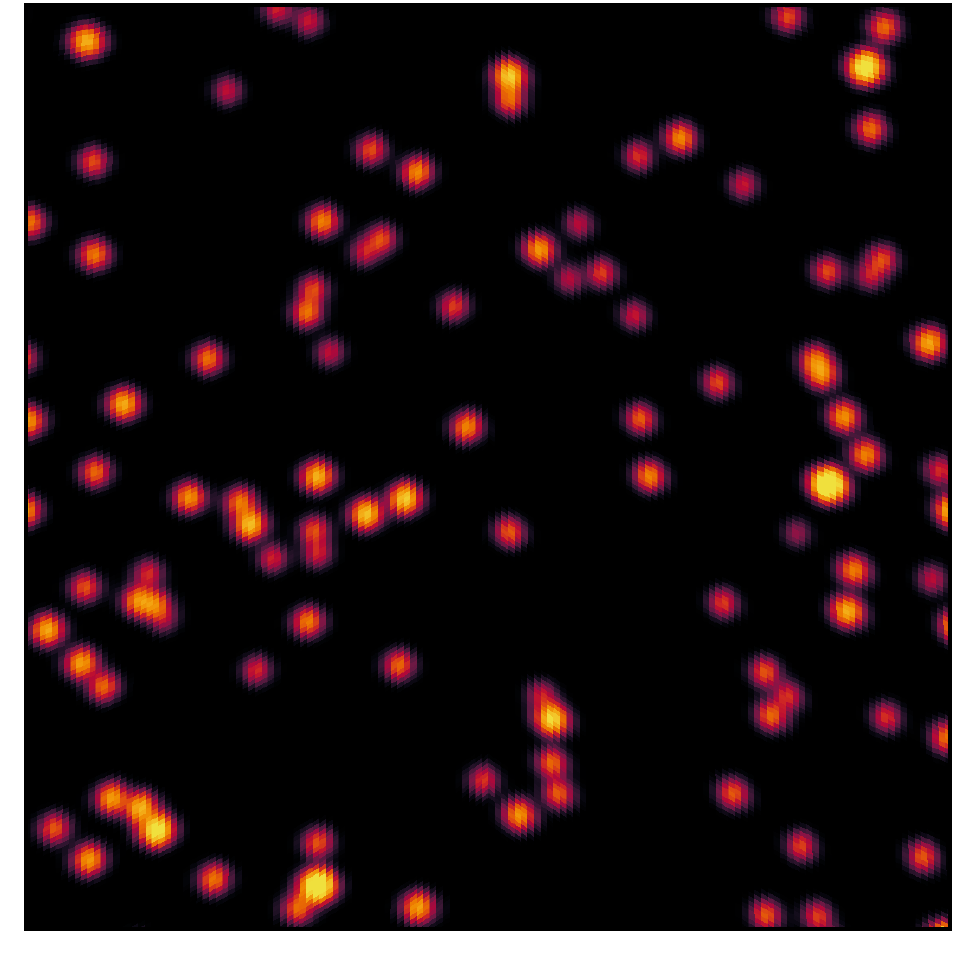}
  \includegraphics[width=0.32\linewidth]{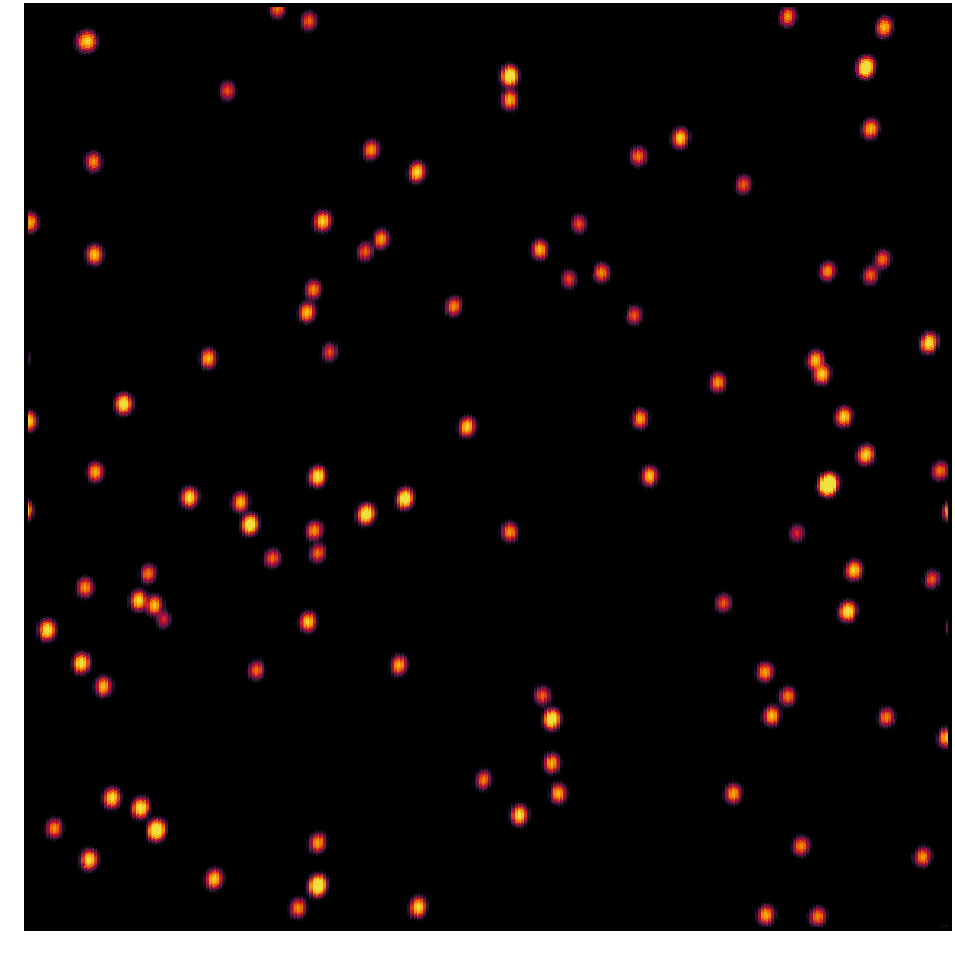}\\
  \includegraphics[width=0.7\linewidth]{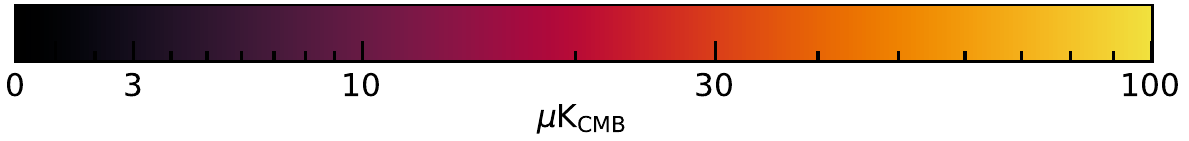}\vspace{0.3cm}\\
  \includegraphics[width=0.32\linewidth]{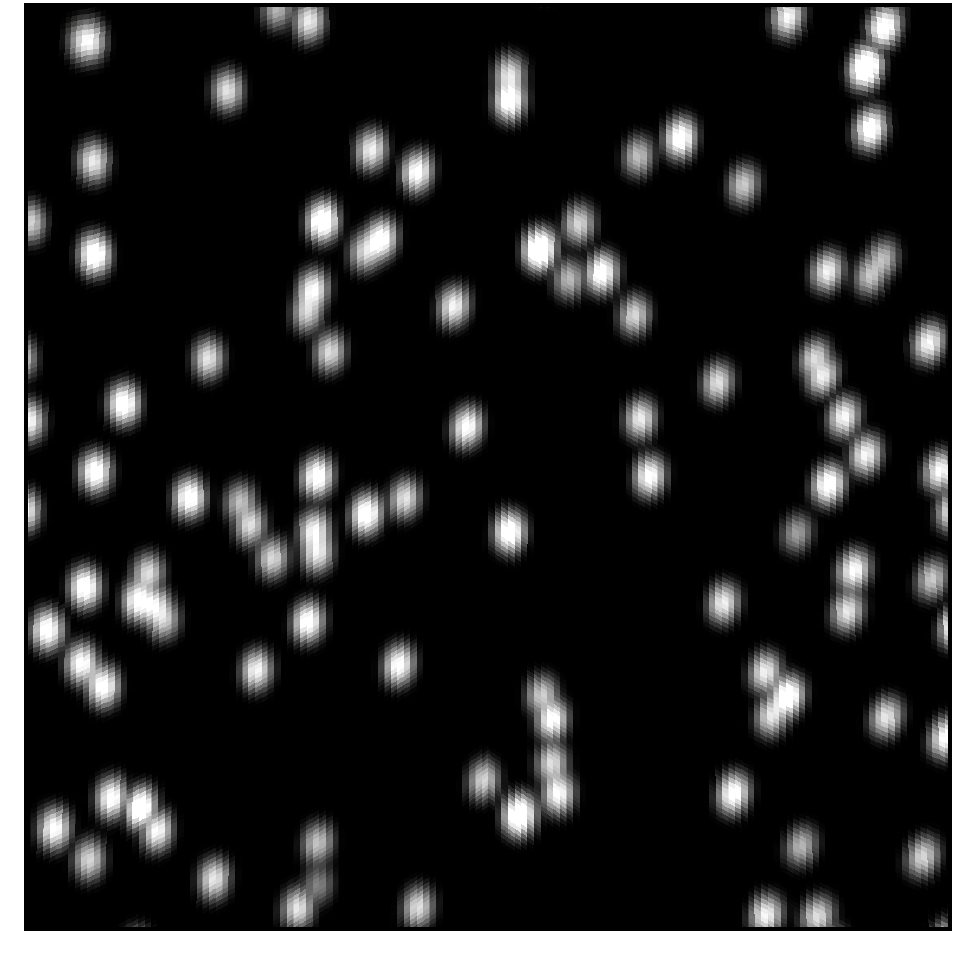}
  \includegraphics[width=0.32\linewidth]{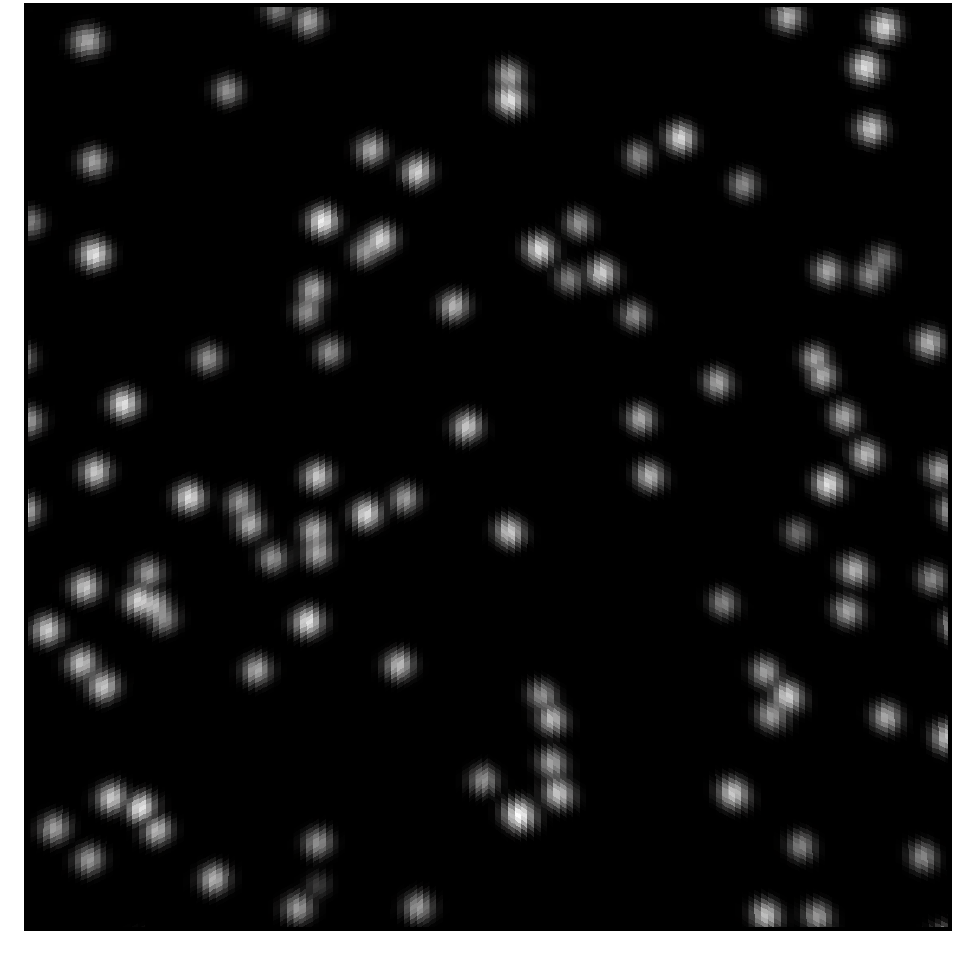}
  \includegraphics[width=0.32\linewidth]{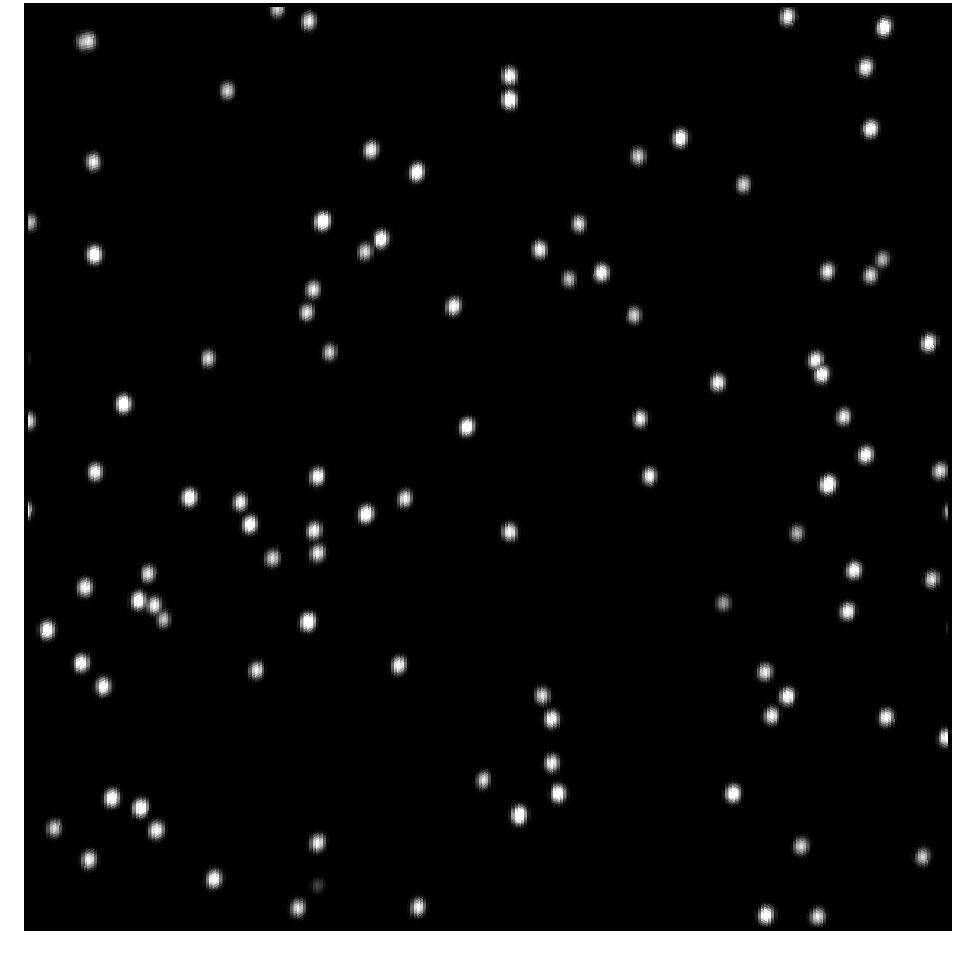}\\
  \includegraphics[width=0.7\linewidth]{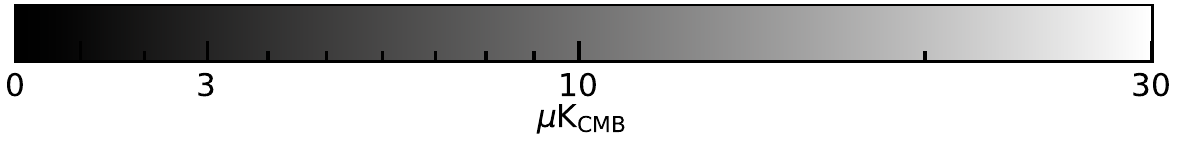}

  \caption{Partial sky plots of the mean (top) and
    standard deviation (bottom) amplitude of the
    fitted compact sources (point sources) as seen by
    the three \Planck\ LFI detector bands; 30\,GHz (\emph{left}),
    44\,GHz (middle), and 70\,GHz (right).
    The plots show gnomonic projections of a $20\times 20$
    degree patch of the sky centered on 90\deg\ longitude and
    70\deg\ latitude, with the north galactic pole located towards
    the top-center of the plots. All plots are at native angular
    resolution and pixelization, see Table~\ref{tab:data_survey_char}
    for details. }
  \label{fig:BP_radio}
\end{figure*}

\begin{figure*}
  \center       
  \includegraphics[width=0.49\linewidth]{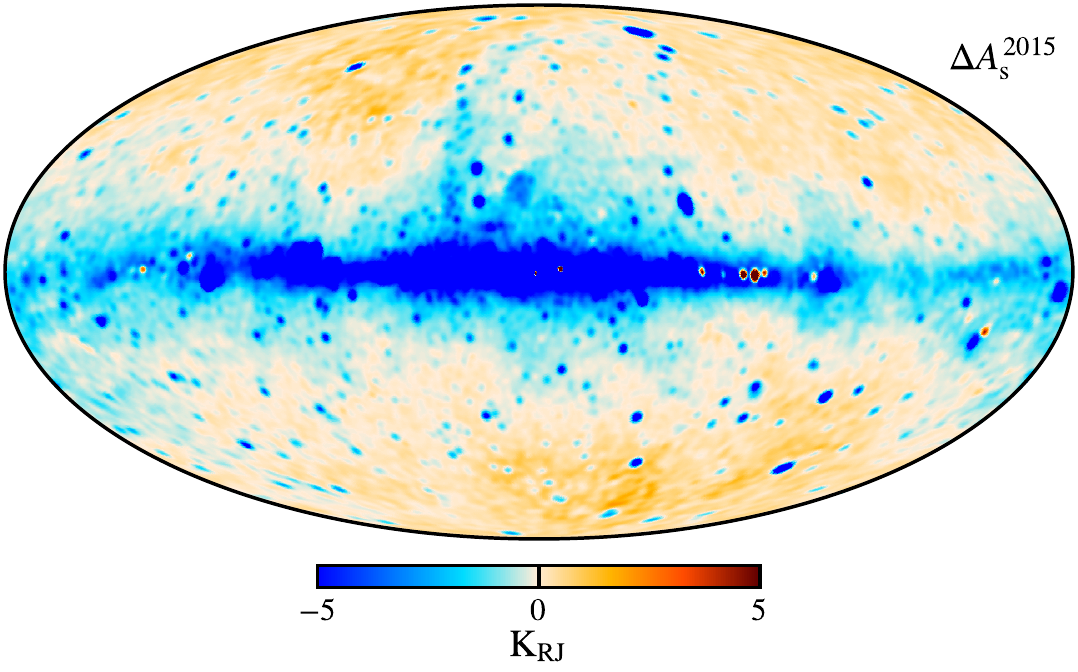}
  \includegraphics[width=0.49\linewidth]{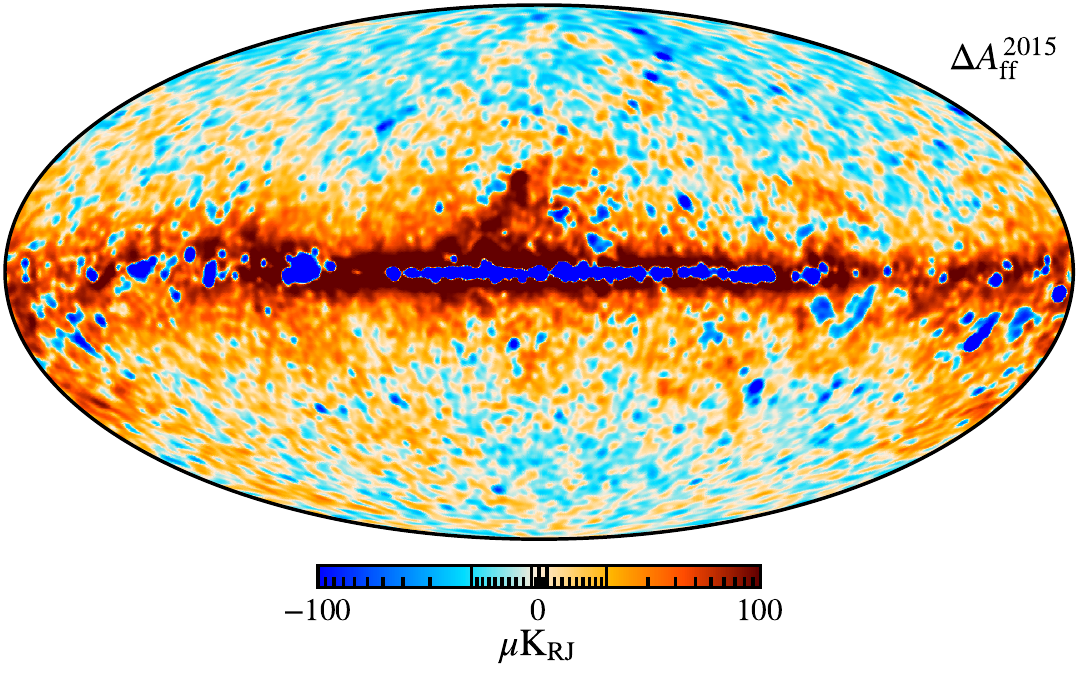} \\
  \includegraphics[width=0.49\linewidth]{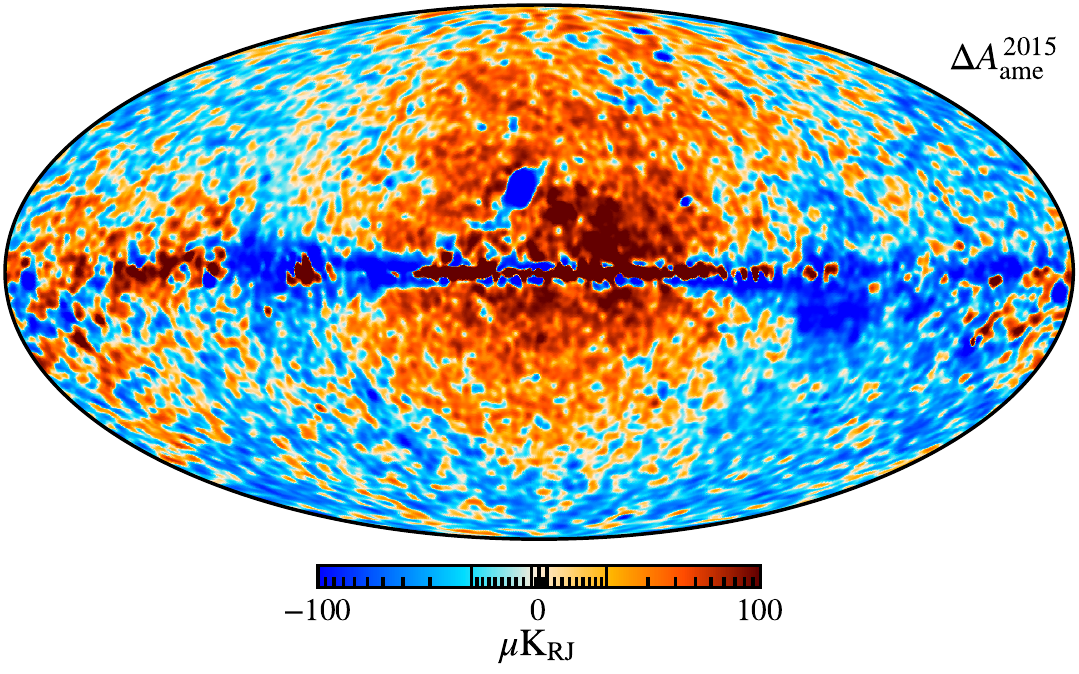} 
  \includegraphics[width=0.49\linewidth]{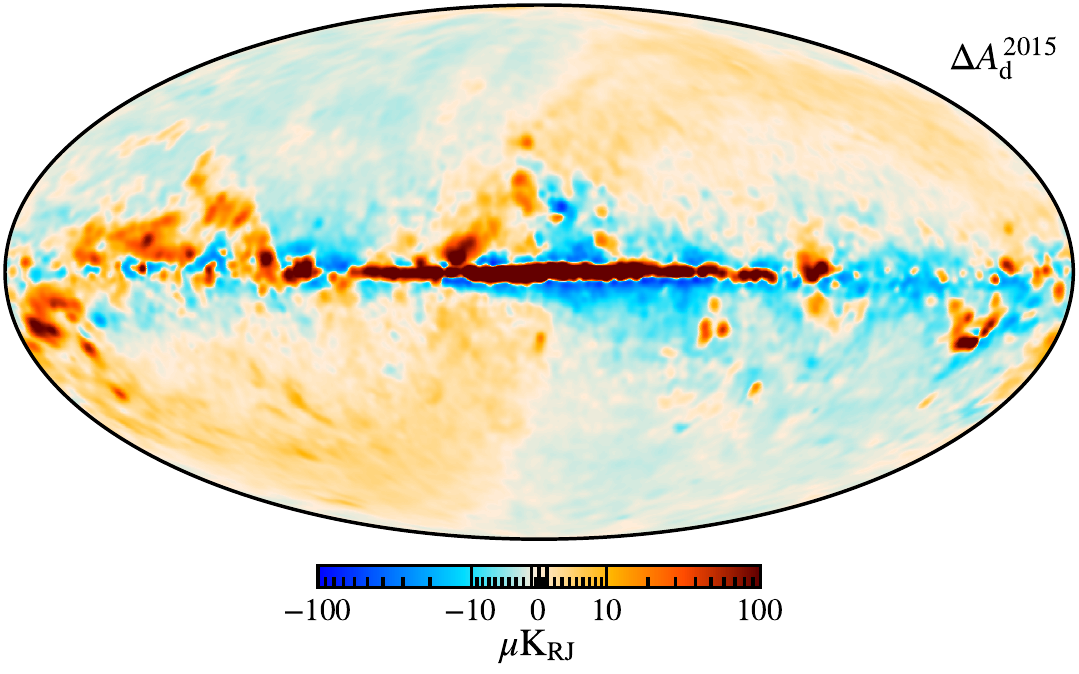}  \\
  \caption{ Difference maps between the component amplitude maps
    derived in \BP\ and the \Planck\ 2015 analysis, for synchrotron
    (top-left), free-free (top-right), AME (bottom-left), and thermal dust emission (bottom-right),
    respectively. All maps have been smoothed to a common angular
    resolution of 2\deg\ FWHM, a relative offset has been fitted and subtracted
    using the frequency-band monopole mask discussed in
    Sect.~\ref{subsec:monopole_index} and differences in reference
    frequencies (where relevant) have been accounted for by a single
    multiplicative scaling factor.  }
  \label{fig:diff_maps}
\end{figure*}

\begin{figure*}
  \center       
  \includegraphics[width=\linewidth]{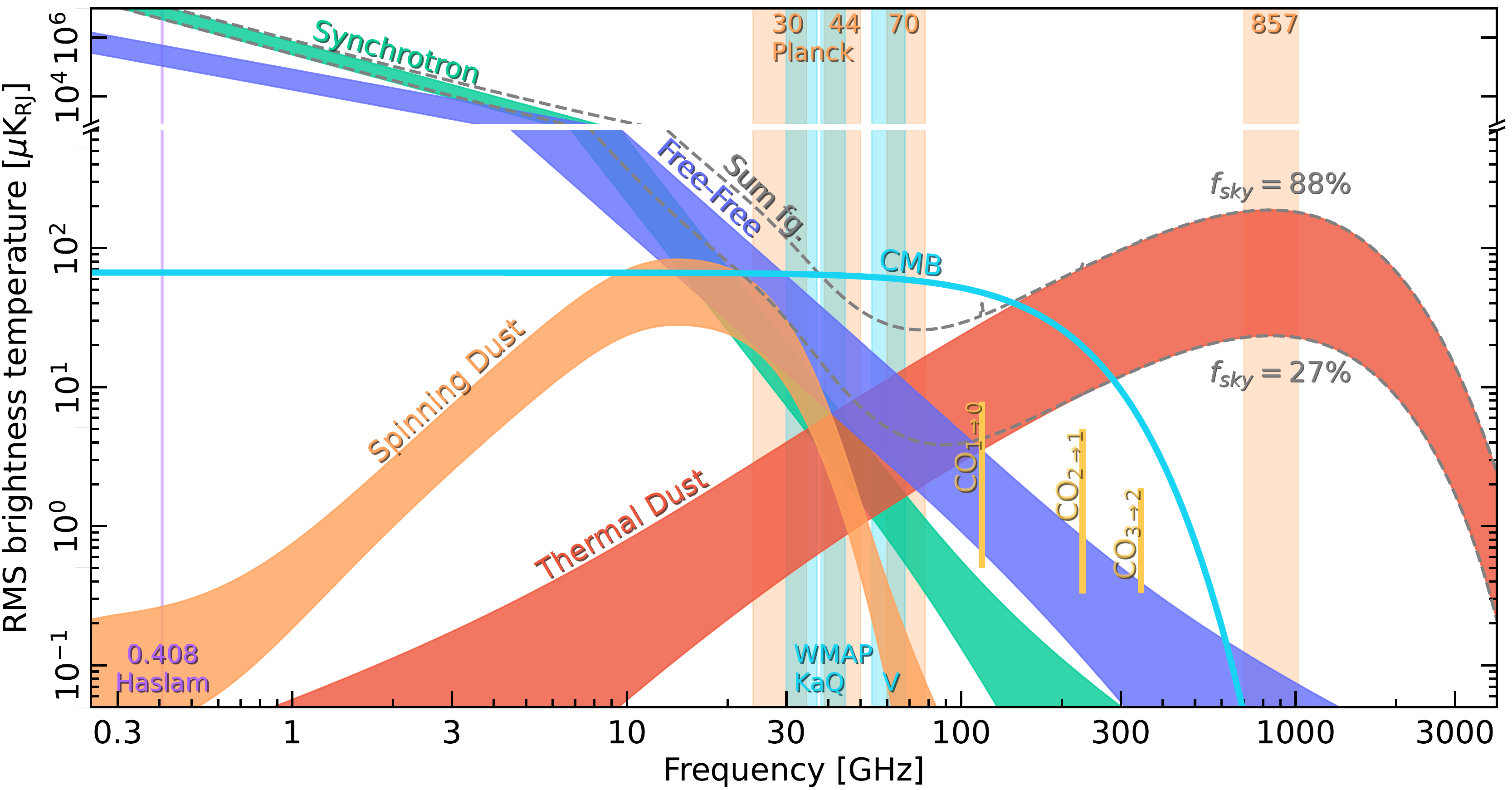}
  \includegraphics[width=\linewidth]{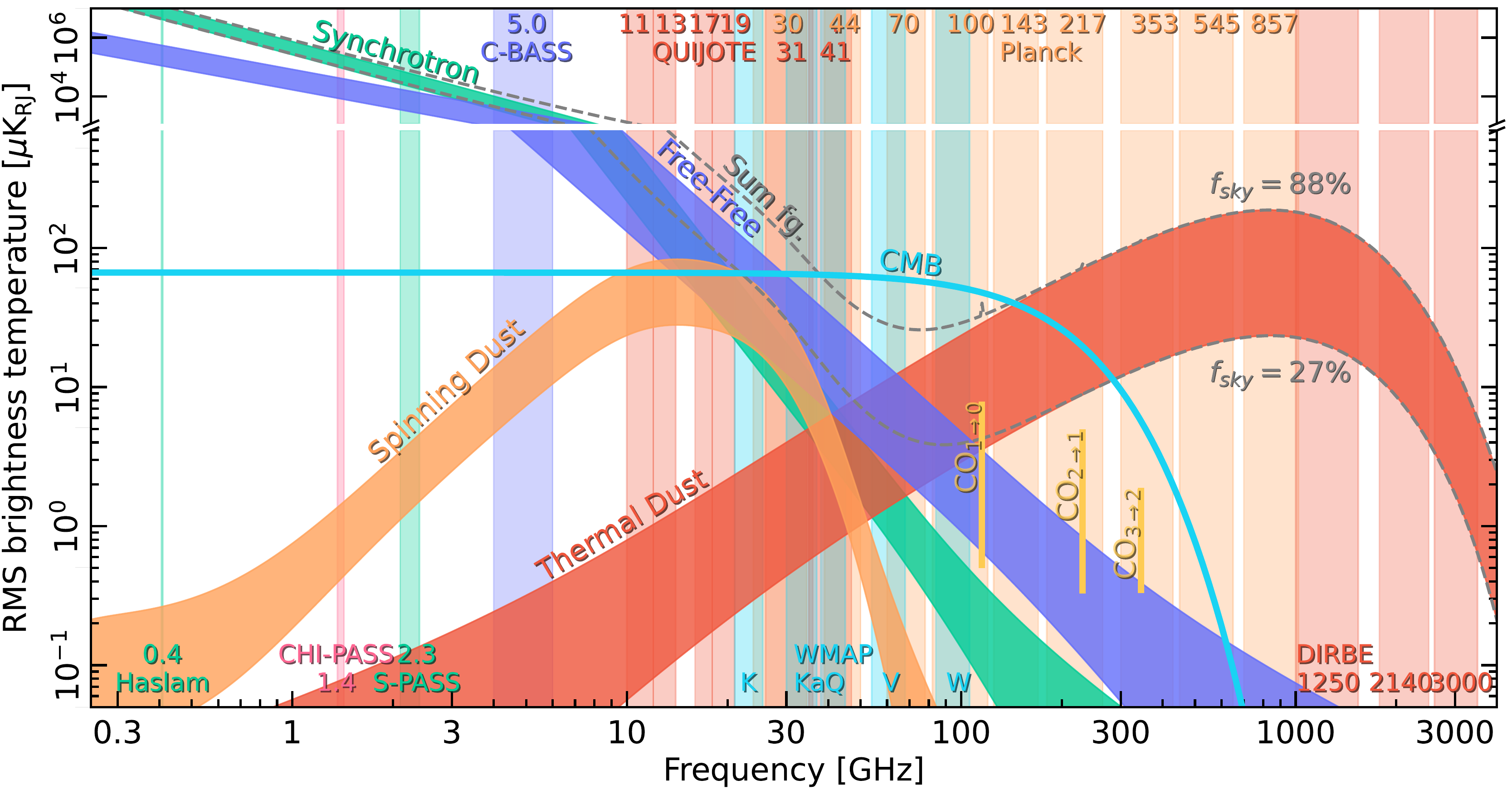}  
  \caption{Brightness temperature RMS as a function of frequency and astrophysical
    component for temperature. Each component is smoothed to an angular resolution
    of 1\deg\ FWHM, and the lower and upper edges of each band are defined by
    masks covering 27 and 88\,\% of the sky, respectively. We note that foreground RMS values decrease nearly monotonically with sky fraction, whereas the CMB RMS is independent of sky
    fraction, up to random variations. The vertical bands represent the frequency range of detector data, where the top panel shows the data employed in this paper and the bottom panel shows some data available for future analysis. 
    }
  \label{fig:fg_comp_spectrum}
\end{figure*}

Next, we consider the signal component posterior distributions
and we start with the frequency monopoles, as summarized in terms of
posterior mean and standard deviations in the second column of
Table~\ref{tab:spec_par_dist}. For comparison purposes, the third
column shows corresponding results derived by the \Planck\ team; all
results except 857\,GHz are reproduced from the \Planck\ 2015 analysis
\citep{planck2014-a12}, while the 857\,GHz result is taken from
\Planck\ DR4 \citep{npipe}.

Several important differences between the two sets of results can be
identified. First, we note that the \BP\ LFI mean monopoles are all
zero; this happens by construction during the mapmaking process, as
the frequency band monopoles are determined directly from the sky
model and any deviation from this is assigned to the correlated noise
component \citep{bp06}. On the other hand, we do see that the
uncertainties of the LFI zero-levels are larger at both 30 and
44\,GHz, reflecting the difficulty of uniquely determining the AME
offset, as discussed in Sect.~\ref{subsec:monopole_sampler}.
In addition, the uncertainty of the Haslam, and thereby the synchrotron monopole
propagates down to the lower frequency \Planck\ and \WMAP\ channels and
contributes to increased uncertainties of these monopoles. We argue
that the uncertainties determined through component-based monopole
determination are more realistic than those obtained through
morphologically based frequency monopole determination.

For \WMAP, we note that the our monopole corrections are considerably
larger compared to those determined in the \Planck\ 2015
estimates. This is caused by two main differences. First, the 2015
analysis adopted the \WMAP\ \textit{Ka}-band explicitly as a fixed ``anchor
channel'' \citep{planck2014-a12} that was not allowed to vary in the
analysis. In the current analysis, for which we instead impose
constraints directly on the component monopoles, this channel is
instead associated with a $16\muKCMB$ correction, which is then accounted
for by an offset of $-17\muKCMB$ at LFI 30\,GHz in the 2015
analysis. Second, we have changed the average synchrotron index from
$\beta_{\mathrm{s}}=-3.1$ in the \Planck\ 2015 analysis
\citep{planck2014-a12} to $\beta_{\mathrm{s}}=-3.3$ in the current
analysis. Since the synchrotron monopole is set to 8.9\,\KRJ\ at
408\,MHz (see Sect.~\ref{sec:data}), the predicted monopole at
\WMAP\ \textit{Ka}-band is 11\muKCMB\ with $\bsynch=-3.1$ and 4\muKCMB\ with
$\bsynch=-3.3$, a difference of 7\muKCMB. The frequency band monopole
has to adjust for this difference. Similarly, differences in the
zero-level of the other components, especially the AME, will
contribute to frequency-band monopole differences in equal
fashion. Regarding the offsets for the 857\,GHz channel, we see that
these agree within $2\,\sigma$ as estimated by \BP and the
uncertainties are significantly larger (and, we believe, more
realistic) in the new approach, which is a reflection of the fact that
we are now propagating uncertainties in the thermal dust zero-level as
discussed in Sect.~\ref{subsec:monopole_sampler}.

Regarding spectral parameters, the AME peak frequency is the only
spectral parameter that is fully sampled from the data in this
component separation work. For this, we find a posterior mean and
standard deviation of $\nup = 25.3\pm0.5\GHz$.  From the high
correlation coefficient between \nup\ and the thermal dust
\bdust\ seen in Fig.~\ref{fig:param_corr_local}, it is clear that this
result is highly dependent on the assumptions made regarding thermal
dust modeling in this paper, and future work that includes
\Planck\ HFI observations will be critically important to make the AME
model more robust.

Posterior mean and standard deviation maps for each of the four
component maps (synchrotron, free-free, AME, and thermal dust
emission) are shown in Figs.~\ref{fig:diff_mean} and
\ref{fig:diff_rms}, each at their own angular resolution (120, 30, 120,
and 10\arcm) and reference frequency (408\,MHz, 40\,GHz, 22\,GHz, and
857\,GHz), and all shown in terms of brightness
temperature. In Figure~\ref{fig:diff_rms}, we notice the dark stripes
following the \Planck\ scanning pattern at high galactic latitudes
in the standard deviation map of the free-free component.
These stripes are barely visible in the AME standard deviation map
while being completely absent from the synchrotron and thermal dust maps.
This can be explained by the sampling of the amplitude zero-levels,
where the free-free zero-level is the only one not marginalized over
in the \BP\ Gibbs chain. The base value at higher latitudes of the
amplitude RMS is equivalent to the RMS of the sampled zero-levels.  

Figure~\ref{fig:BP_radio} shows mean and standard distribution
plots for the compact source component for a $20\deg\times20\deg$ field of the sky
in the Northern Galactic hemisphere for each of the three LFI
frequency channels.
The four-component posterior mean maps may be compared to similar
products from the \Planck\ 2015 analysis \citep{planck2014-a12} and
corresponding difference maps are shown in
Fig.~\ref{fig:diff_maps}. All maps are smoothed to a common angular
resolution of $2\deg$ FWHM before differencing, and a free offset has
been fitted and subtracted using the frequency-band monopole mask discussed in
Sect.~\ref{subsec:monopole_index}. In addition, for the components
that have different reference frequencies in the two analyses (i.e..,
free-free, AME, and thermal dust emission), a single multiplicative
factor has been fitted to take into account SED scaling differences.

Starting with the synchrotron case, we note that typical high-latitude
differences are small compared to the overall amplitude of the Haslam
408\,MHz map, typically less than $1\,\mathrm{K}_{\mathrm{RJ}}$. This
is of course entirely expected, since the two analyses are both
dominated by the same map. However, we do see differences at both low
and high Galactic latitudes; for high latitudes we note that the
current analysis explicitly models individual point sources, while in
the \Planck\ 2015 analysis there was no such component and compact
sources were therefore part of the diffuse components. At low
latitudes, the differences are dominated by differences in the
free-free, AME, and thermal dust emission models.

For free-free emission, we see clear negative imprints of the free-free
amplitude map itself at low Galactic latitudes. In this case, we note
that while the \Planck\ 2015 analysis fit the electron temperature
pixel-by-pixel, we adopt a single constant value of $T_e =
7000\,\mathrm{K}$ in the current analysis. Second, we see clear positive
imprints of AME or dust around the negative free-free imprint at low
Galactic latitudes, resulting from degeneracies between the component
signals. Lastly, we also note that
the current analysis suffers significantly with respect to free-free
emission due to the absence of the \Planck\ HFI channels, which
provide both angular resolution and sensitivity to this component.

The AME component residual map is largely dominated by a dipole
component aligned with the direction of the Galactic center. In this
case, we note that the \BP\ analysis estimates the absolute
calibration of each \Planck\ LFI frequency channel through a joint fit
with all frequencies and this provides more robust estimates of the
CMB dipole. We also see a negative free-free imprint at low Galactic
latitudes similar to what we see in the free-free difference map.
One key feature is the almost circular blob just above and slightly to
the left of the Galactic center, which is also clearly seen as a dark
blob in the mean map in Fig.~\ref{fig:diff_mean}. This coincides with
a similarly bright circular blob in the free-free amplitude. A similar
signal can be seen in the \Planck\ 2015 free-free signal \citep{planck2014-a12},
although it did not affect the AME amplitude in the same way.
If we look at the residuals in Fig.~\ref{fig:BP_residuals_mean}, we
notice the same blob in the LFI 30\GHz\ and the \WMAP\ $Ka$ maps,
indicating a degeneracy between the modeled AME and free-free components.
To break this degeneracy, either more data sets needs to be introduced or
the free-free \Te\ and the AME \nup\ needs to be sampled with spatial
variance, or both. This is, however, not possible with the limited data set
in this analysis and therefore needs to be left for future analyses.

Finally, for the thermal dust emission amplitude map, which is
essentially defined by the \Planck\ 857\,GHz frequency map, the
differences are explained very closely by the morphology of the
thermal dust spectral index map $\beta_{\mathrm{d}}(p)$ presented by
\citet{planck2014-a12}. This is also entirely expected, given the fact
that we only model $\beta_{\mathrm{d}}$ in terms of a single spatial
constant over the full sky.

\section{Summary and outlook}
\label{sec:conclusions}

The main goal of the current paper is to establish an efficient Monte
Carlo sampling scheme for intensity foregrounds within an end-to-end
Bayesian CMB analysis pipeline, as implemented in the
\BP\ framework. This sampling scheme must be able to operate in both
low and high signal-to-noise regimes, and it must be able to
incorporate both algorithmic and informative priors in a controlled
and transparent manner. Degeneracies between different parameters must
be explored properly and it must be possible to propagate
corresponding uncertainties to higher level products.

Most of the algorithmic elements in the machinery used in this
paper were developed and applied within the context of the official
\Planck\ project, as described in, for instance,
\citet{planck2014-a12}, \citet{planck2016-l04}, and \citet{npipe}. In this paper, we have
added four new algorithmic components to this machinery, namely: 1) joint amplitude
and spectral index sampling, borrowing heavily from ideas already
introduced by \citet{2009MNRAS.392..216S} and \citet{stivoli:2010}; 2)
component-based monopole determination; 3) joint spectral index and
monopole sampling; and 4) the application of informative spatial
priors. Each of these steps significantly improve the computational
efficiency and robustness of the Gibbs sampling-based
\commander\ approach.

We stress that the current \BP\ results are not intended to define a
new state-of-the-art model of the astrophysical sky. Rather, the
current framework and analysis constitute a ``skeleton'' to which
additional data sets, both legacy and future, may be added in a
controlled fashion, keeping track of both instrumental and
astrophysical modeling errors. As more and more data sets are added,
the dependency on external priors may be gradually lifted until,
hopefully, all key parameters of the model become data
driven. Obviously, the single most important step towards realizing
this goal is the introduction of \Planck\ HFI TOD observations, which
still define the state-of-the-art in terms of full-sky CMB sensitivity
to date. On a longer term, we argue that all key data sets in the
community should be integrated into the model, allowing one set to
break the degeneracies of others. This is the goal of the Open Source
\textsc{Cosmoglobe} project, and the transition from \BP\ to
\textsc{Cosmoglobe} may be illustrated in
Fig.~\ref{fig:fg_comp_spectrum}: the top panel of this figure shows
the standard deviation brightness temperature as a function of
frequency for each primary intensity CMB foreground as colored
bands. The \BP\ frequency bands are indicated by vertical
bars. Looking at this figure, noting the large unexplored frequency
ranges, it is strikingly obvious why the current data model is
significantly prior dominated; thus, more data are desperately needed. The
bottom panel of Fig.~\ref{fig:fg_comp_spectrum} shows an alternative
scenario, in which almost the entire frequency range is
covered by including data from several past and planned sky surveys.
Providing a computationally and organizationally efficient
platform to make this happen is the goal of \textsc{Cosmoglobe} and
the \BP\ project represents an important step towards realizing this
promise. 

\begin{acknowledgements}
  We thank Prof.\ Pedro Ferreira and Dr.\ Charles Lawrence for useful suggestions, comments and 
  discussions. We also thank the entire \Planck\ and \WMAP\ teams for
  invaluable support and discussions, and for their dedicated efforts
  through several decades without which this work would not be
  possible. The current work has received funding from the European
  Union’s Horizon 2020 research and innovation programme under grant
  agreement numbers 776282 (COMPET-4; \BP), 772253 (ERC;
  \textsc{bits2cosmology}), and 819478 (ERC; \textsc{Cosmoglobe}). In
  addition, the collaboration acknowledges support from ESA; ASI and
  INAF (Italy); NASA and DoE (USA); Tekes, Academy of Finland (grant
   no.\ 295113), CSC, and Magnus Ehrnrooth foundation (Finland); RCN
  (Norway; grant nos.\ 263011, 274990); and PRACE (EU).
\end{acknowledgements}

\bibliographystyle{aa}

\bibliography{Planck_bib,BP_bibliography}

\end{document}